# MONTE CARLO METHODS FOR THE SELF-AVOIDING WALK

ALAN D. SOKAL
*Department of Physics*
*4 Washington Place*
*New York University*
*New York NY 10003, USA*
SOKAL@NYU.EDU

**Abstract**

This article is a pedagogical review of Monte Carlo methods for the self-avoiding walk, with emphasis on the extraordinarily efficient algorithms developed over the past decade.

# Contents

# 1 Introduction

## 1.1 Why is the SAW a Sensible Model?

The self-avoiding walk (SAW) was first proposed nearly half a century ago as a model of a linear polymer molecule in a good solvent [1, 2]. At first glance it seems to be a ridiculously crude model, almost a caricature: Real polymer molecules[1] live in continuous space and have tetrahedral (109.47°) bond angles, a non-trivial energy surface for the bond rotation angles, and a rather complicated monomer-monomer interaction potential. By contrast, the self-avoiding walk lives on a discrete lattice and has non-tetrahedral bond angles (e.g. 90° and 180° on the simple cubic lattice), an energy independent of the bond rotation angles, and a repulsive hard-core monomer-monomer potential.

In spite of these rather extreme simplifications, there is now little doubt that the self-avoiding walk is not merely an excellent but in fact a *perfect* model for some (but not all!) aspects of the behavior of linear polymers in a good solvent.[2] This apparent miracle arises from **universality**, which plays a central role in the modern theory of critical phenomena [3, 4]. In brief, critical statistical-mechanical systems are divided into a small number of **universality classes**, which are typically characterized by spatial dimensionality, symmetries and other rather general properties. In the vicinity of a critical point (and only there), the leading asymptotic behavior is *exactly* the same (modulo some system-dependent scale factors) for all systems of a given universality class; the details of chemical structure, interaction energies and so forth are completely irrelevant (except for setting the nonuniversal scale factors). Moreover, this universal behavior is given by simple **scaling laws**, in which the dependent variables are generalized homogeneous functions of the parameters which measure the deviation from criticality.

The key question, therefore, is to determine for each physical problem which quantities are universal and which are nonuniversal.

To compute the nonuniversal quantities, one employs the traditional methods of theoretical physics and chemistry: semi-realistic models followed by a process of successive refinement. All predictions from such models must be expected to be only approximate, *even if the mathematical model is solved exactly*, because the mathematical model is itself only a crude approximation to reality.

To compute the universal quantities, by contrast, a very different approach is available: one may choose *any* mathematical model (the simpler the better) belonging to the same universality class as the system under study, and by solving it determine the *exact* values of universal quantities. Of course, it may not be feasible to solve this mathematical model exactly, so further approximations (or numerical simulations) may be required in practice; but these latter approximations are the *only* sources of error in the computation of universal quantities. At a subsequent stage it is prudent to test variants and refinements of the original model, but solely for the purpose of determining the boundaries of the universality class: if the refined model belongs to the same universality class as the original model, then the refinement has *zero* effect on the universal quantities.

[1] More precisely, linear polymers whose backbones consist solely of carbon-carbon single bonds.

[2] Here "good solvent" means "at any temperature strictly above the theta temperature for the given polymer-solvent pair".

The behavior of polymer molecules as the chain length tends to infinity is, it turns out, a critical phenomenon in the above sense [5]. Thus, it is found empirically — although the existing experimental evidence is admittedly far from perfect [6, 7, 8, 9, 10] — that the mean-square radius of gyration $\langle R_g^2 \rangle$ of a linear polymer molecule consisting of $N$ monomer units has the leading asymptotic behavior

$$\langle R_g^2 \rangle \;=\; AN^{2\nu} \left[ 1 + O(N^{-\Delta}) \right] \tag{1.1}$$

as $N \to \infty$, where the **critical exponent** $\nu \approx 0.588$ is universal, i.e. exactly the same for all polymers, solvents and temperatures (provided only that the temperature is above the theta temperature for the given polymer-solvent pair). The **critical amplitude** $A$ is nonuniversal, i.e. it depends on the polymer, solvent and temperature, and this dependence is *not* expected to be simple.

There is therefore good reason to believe that any (real or mathematical) linear polymer chain which exhibits some flexibility and has short-range[3], predominantly repulsive[4] monomer-monomer interactions lies in the same universality class as the self-avoiding walk. This belief should, of course, be checked carefully by both numerical simulations and laboratory experiments; but at present there is, to my knowledge, no credible numerical or experimental evidence that would call it into question.

## 1.2 Numerical Methods for the Self-Avoiding Walk

Over the decades, the SAW has been studied extensively by a variety of methods. *Rigorous methods* have thus far yielded only fairly weak results [11]; the SAW is, to put it mildly, an extremely difficult mathematical problem. *Non-rigorous analytical methods*, such as perturbation theory and self-consistent-field theory, typically break down in precisely the region of interest, namely long chains [12]. The exceptions are methods based on the renormalization group (RG) [13, 14, 15], which have yielded reasonably accurate estimates for critical exponents and for some universal amplitude ratios [16, 17, 18, 19, 20, 21, 22, 23, 24]. However, the conceptual foundations of the renormalization-group methods have not yet been completely elucidated [25, 26]; and high-precision RG calculations are not always feasible. Thus, considerable work has been devoted to developing *numerical methods* for the study of long SAWs. These methods fall essentially into two categories: exact enumeration and Monte Carlo.

In an **exact-enumeration** study, one first generates a complete list of all SAWs up to a certain length (usually $N \approx$ 15–35 steps), keeping track of the properties of interest such as the number of such walks or their squared end-to-end distances [27]. One then performs an extrapolation to the limit $N \to \infty$, using techniques such as the ratio method, Padé approximants or differential approximants [28, 29, 30]. Inherent in any such extrapolation is an assumption about the behavior of the coefficients beyond those actually computed. Sometimes this assumption is fairly explicit; other times it is hidden in the details of the extrapolation method. In either case, the assumptions made have a profound effect on the numerical results obtained [25]. For this reason, the claimed error bars in exact-enumeration/extrapolation studies should be viewed with a healthy skepticism.

[3]Here I mean that the potential is short-range *in physical space*. It is of course — and this is a crucial point — long-range *along the polymer chain*, in the sense that the interaction between two monomers depends only on their positions in physical space and is essentially independent of the locations of those monomers along the chain.

[4]Here "predominantly repulsive" means "repulsive enough so that the temperature is strictly above the theta temperature for the given polymer-solvent pair".

In a **Monte Carlo** study, by contrast, one aims to probe directly the regime of fairly long SAWs (usually $N \approx 10^2$–$10^5$ steps). Complete enumeration is unfeasible, so one generates instead a random sample. The raw data then contain statistical errors, just as in a laboratory experiment. These errors can, however be estimated — sometimes even *a priori* (see Section 7.3) — and they can in principle be reduced to an arbitrarily low level by the use of sufficient computer time. An extrapolation to the regime of extremely long SAWs is still required, but this extrapolation is much less severe than in the case of exact-enumeration studies, because the point of departure is already much closer to the asymptotic regime.

Monte Carlo studies of the self-avoiding walk go back to the early 1950's [31, 32], and indeed these simulations were among the first applications of a new invention — the "high-speed electronic digital computer" — to pure science.[5] These studies continued throughout the 1960's and 1970's, and benefited from the increasingly powerful computers that became available. However, progress was slowed by the high computational complexity of the algorithms then being employed, which typically required a CPU time of order at least $N^{2+2\nu} = N^{\approx 3.2}$ to produce one "effectively independent" $N$-step SAW. This rapid growth with $N$ of the autocorrelation time — called **critical slowing-down**[6] — made it difficult in practice to do high-precision simulations with $N$ greater than about 30–100.

In the past decade — since 1981 or so — vast progress has been made in the development of *new and more efficient algorithms* for simulating the self-avoiding walk. These new algorithms reduce the CPU time for generating an "effectively independent" $N$-step SAW from $\sim N^{\approx 3.2}$ to $\sim N^{\approx 2}$ or even $\sim N$. The latter is quite impressive, and indeed is the best possible order of magnitude, since it takes a time of order $N$ merely to *write down* an $N$-step walk! As a practical matter, the new algorithms have made possible high-precision simulations at chain lengths $N$ up to nearly $10^5$ [39].

The purpose of this article is thus to give a comprehensive overview of Monte Carlo methods for the self-avoiding walk, with emphasis on the extraordinarily efficient algorithms developed since 1981. I shall also discuss briefly some of the physical results which have been obtained from this work.

[5] Here "pure" means "not useful in the sense of Hardy": "a science is said to be useful if its development tends to accentuate the existing inequalities in the distribution of wealth, or more directly promotes the destruction of human life" [33, p. 120*n*].

[6] For a general introduction to critical slowing-down in Monte Carlo simulations, see [34, 35, 36, 37]. See also [38] for a pioneering treatment of critical slowing-down in the context of dynamic critical phenomena.

The plan of this article is as follows: I begin by presenting background material on the self-avoiding walk (Section 2) and on Monte Carlo methods (Section 3). In Section 4 I discuss *static* Monte Carlo methods for the generation of SAWs: simple sampling and its variants, inversely restricted sampling (Rosenbluth-Rosenbluth algorithm) and its variants, and dimerization. In Section 5 I discuss *quasi-static* Monte Carlo methods: enrichment and incomplete enumeration (Redner-Reynolds algorithm). In Section 6 I discuss *dynamic* Monte Carlo methods: the methods are classified according to whether they are local or non-local, whether they are $N$-conserving or $N$-changing, and whether they are endpoint-conserving or endpoint-changing. In Section 7 I discuss some miscellaneous algorithmic and statistical issues. In Section 8 I review some preliminary physical results which have been obtained using these new algorithms. I conclude (Section 9) with a brief summary of practical recommendations and a listing of open problems.

For previous reviews of Monte Carlo methods for the self-avoiding walk, see Kremer and Binder [40] and Madras and Slade [11, Chapter 9].

# 2 The Self-Avoiding Walk (SAW)

## 2.1 Background and Notation

In this section we review briefly the basic facts and conjectures about the SAW that will be used in the remainder of this article. A comprehensive survey of the SAW, with emphasis on rigorous mathematical results, can be found in the excellent new book by Madras and Slade [11].

Real polymers live in spatial dimension $d = 3$ (ordinary polymer solutions) or in some cases in $d = 2$ (polymer monolayers confined to an interface [41, 42]). Nevertheless, it is of great conceptual value to define and study the mathematical models — in particular, the SAW — in a general dimension $d$. This permits us to distinguish clearly between the *general* features of polymer behavior (in any dimension) and the *special* features of polymers in dimension $d = 3$.[7] The use of arbitrary dimensionality also makes available to theorists some useful technical tools (e.g. dimensional regularization) and some valuable approximation schemes (e.g. expansion in $d = 4 - \epsilon$) [15].

So let $\mathcal{L}$ be some regular $d$-dimensional lattice. Then an **$N$-step self-avoiding walk**[8] (SAW) $\omega$ on $\mathcal{L}$ is a sequence of *distinct* points $\omega_0, \omega_1, ..., \omega_N$ in $\mathcal{L}$ such that each point is a nearest neighbor of its predecessor.[9] We denote by $|\omega| \equiv N$ the number of steps in $\omega$. For simplicity we shall restrict attention to the simple (hyper-)cubic lattice $\mathbb{Z}^d$; similar ideas apply with minor alterations to other regular lattices. We assume all walks to begin at the origin ($\omega_0 = 0$) unless stated otherwise, and we let $\mathcal{S}_N$ [resp. $\mathcal{S}_N(x)$] be the set of all $N$-step SAWs starting at the origin and ending anywhere [resp. ending at $x$].

[7]It turns out that $d = 3$ is *very* special, because it is the upper critical dimension for tricritical behavior. This is the deep reason underlying the fact that polymers at the theta point in $d = 3$ are "quasi-ideal" (i.e. have size exponent $\nu = \frac{1}{2}$ and have all dimensionless virial coefficients vanishing in the limit of infinite chain length). In dimension $d < 3$, polymers at the theta point are *not* quasi-ideal [43, 44, 45, 46, 47].

[8]The term "walk" is a misnomer. The SAW should *not* be thought of as the path of a particle which "walks" (over time). Rather, it should be thought of as the configuration of a polymer chain

First we define the quantities relating to the *number* (or "entropy") of SAWs: Let $c_N$ [resp. $c_N(x)$] be the number of $N$-step SAWs on $\mathbb{Z}^d$ starting at the origin and ending anywhere [resp. ending at $x$]. Then $c_N$ and $c_N(x)$ are believed to have the asymptotic behavior

$$c_N \sim \mu^N N^{\gamma-1} \tag{2.1}$$

$$c_N(x) \sim \mu^N N^{\alpha_{sing}-2} \qquad (x \text{ fixed} \neq 0) \tag{2.2}$$

as $N \to \infty$; here $\mu$ is called the **connective constant** of the lattice, and $\gamma$ and $\alpha_{sing}$ are **critical exponents**. The connective constant is definitely lattice-dependent, while the critical exponents are believed to be universal among lattices of a given dimension $d$. For rigorous results concerning the asymptotic behavior of $c_N$ and $c_N(x)$, see [11, 48, 49, 50, 51].

Next we define several measures of the *size* of an $N$-step SAW:

- The **squared end-to-end distance**

$$R_e^2 = (\omega_N - \omega_0)^2 \ . \tag{2.3}$$

- The **squared radius of gyration**

$$R_g^2 = \frac{1}{N+1} \sum_{i=0}^{N} \left( \omega_i - \frac{1}{N+1} \sum_{j=0}^{N} \omega_j \right)^2 \tag{2.4a}$$

$$= \frac{1}{N+1} \sum_{i=0}^{N} \omega_i^2 - \left( \frac{1}{N+1} \sum_{i=0}^{N} \omega_i \right)^2 \tag{2.4b}$$

$$= \frac{1}{2(N+1)^2} \sum_{i,j=0}^{N} (\omega_i - \omega_j)^2 \ . \tag{2.4c}$$

- The **mean-square distance of a monomer from the endpoints**

$$R_m^2 = \frac{1}{2(N+1)} \sum_{i=0}^{N} \left[ (\omega_i - \omega_0)^2 + (\omega_i - \omega_N)^2 \right] \ . \tag{2.5}$$

---

*at one instant of time*. [In mathematical terms, the SAW is not a stochastic process (not even a non-Markovian one): the trouble is that the equal-weight distributions on $N$-step and $(N+1)$-step SAWs are not consistent.]

[9] Note that a SAW is an *oriented* object, i.e. we distinguish between the starting point ($\omega_0$) and the ending point ($\omega_N$). However, all probability distributions and all observables that we shall consider are invariant under reversal of orientation ($\tilde{\omega}_i \equiv \omega_{N-i}$). This is necessary if the SAW is to be a sensible model of a real homopolymer molecule, which is of course (neglecting end-group effects) *unoriented*.

We then consider the mean values $\langle R_e^2 \rangle_N$, $\langle R_g^2 \rangle_N$ and $\langle R_m^2 \rangle_N$ in the probability distribution which gives equal weight to each $N$-step SAW. Very little has been proven rigorously about these mean values, but they are believed to have the asymptotic behavior

$$\langle R_e^2 \rangle_N,\ \langle R_g^2 \rangle_N,\ \langle R_m^2 \rangle_N \ \sim\ N^{2\nu} \tag{2.6}$$

as $N \to \infty$, where $\nu$ is another (universal) critical exponent. Moreover, the amplitude ratios

$$A_N \ =\ \frac{\langle R_g^2 \rangle_N}{\langle R_e^2 \rangle_N} \tag{2.7}$$

$$B_N \ =\ \frac{\langle R_m^2 \rangle_N}{\langle R_e^2 \rangle_N} \tag{2.8}$$

are expected to approach universal values in the limit $N \to \infty$.[10,11]

Finally, let $c_{N_1,N_2}$ be the number of pairs $(\omega^{(1)}, \omega^{(2)})$ such that $\omega^{(1)}$ is an $N_1$-step SAW starting at the origin, $\omega^{(2)}$ is an $N_2$-step SAW starting *anywhere*, and $\omega^{(1)}$ and $\omega^{(2)}$ have at least one point in common (i.e. $\omega^{(1)} \cap \omega^{(2)} \neq \emptyset$). Then it is believed that

$$c_{N_1,N_2} \ \sim\ \mu^{N_1+N_2} (N_1 N_2)^{(2\Delta_4+\gamma-2)/2} g(N_1/N_2) \tag{2.9}$$

as $N_1, N_2 \to \infty$, where $\Delta_4$ is yet another (universal) critical exponent and $g$ is a (universal) scaling function.

The quantity $c_{N_1,N_2}$ is closely related to the second virial coefficient. To see this, consider a rather general theory in which "molecules" of various types interact. Let the molecules of type $i$ have a set $S_i$ of "internal states", so that the complete state of such a molecule is given by a pair $(x, s)$ where $x \in \mathbb{Z}^d$ is its position and $s \in S_i$ is its internal state. Let us assign Boltzmann weights (or "fugacities") $W_i(s)$ $[s \in S_i]$ to the internal states, normalized so that $\sum_{s \in S_i} W_i(s) = 1$; and let us assign an interaction energy $\mathcal{V}_{ij}\big((x,s),(x',s')\big)$ $[x, x' \in \mathbb{Z}^d,\ s \in S_i,\ s' \in S_j]$ to a molecule of type $i$ at $(x, s)$ interacting with one of type $j$ at $(x', s')$. Then the second virial coefficient between a molecule of type $i$ and one of type $j$ is

$$B_2^{(ij)} \ =\ \frac{1}{2} \sum_{\substack{s \in S_i \\ s' \in S_j}} \ \sum_{x' \in \mathbb{Z}^d} W_i(s)\, W_j(s') \left[1 - e^{-\mathcal{V}_{ij}((0,s),(x',s'))}\right] . \tag{2.10}$$

In the SAW case, the types are the different lengths $N$, the internal states are the conformations $\omega \in \mathcal{S}_N$ starting at the origin, the Boltzmann weights are $W_N(\omega) =$

---

[10] For a general discussion of universal amplitude ratios in the theory of critical phenomena, see [52].

[11] Very recently, Hara and Slade [48, 49] have proven that the SAW in dimension $d \geq 5$ converges weakly to Brownian motion when $N \to \infty$ with lengths rescaled by $CN^{1/2}$ for a suitable (nonuniversal) constant $C$. It follows from this that (2.6) holds with $\nu = \frac{1}{2}$, and also that (2.7)/(2.8) have the limiting values $A_\infty = \frac{1}{6}$, $B_\infty = \frac{1}{2}$. Earlier, Slade [53, 54, 55] had proven these results for sufficiently high dimension $d$. See also [11].

$1/c_N$ for each $\omega \in \mathcal{S}_N$, and the interaction energies are hard-core repulsions

$$\mathcal{V}_{NN'}\Big((x,\omega),(x',\omega')\Big) \;=\; \begin{cases} +\infty & \text{if } (\omega + x) \cap (\omega' + x') \neq \varnothing \\ 0 & \text{otherwise} \end{cases} \tag{2.11}$$

It follows immediately that

$$B_2^{(N_1,N_2)} \;=\; \frac{c_{N_1,N_2}}{2c_{N_1}c_{N_2}} \;. \tag{2.12}$$

The second virial coefficient $B_2^{(N_1,N_2)}$ is a measure of the "excluded volume" between a pair of SAWs. It is useful to define a *dimensionless* quantity by normalizing $B_2^{(N_1,N_2)}$ by some measure of the "size" of these SAWs. Theorists prefer $\langle R_e^2 \rangle$ as the measure of size, while experimentalists prefer $\langle R_g^2 \rangle$ since it can be measured by light scattering. We follow the experimentalists and define the **interpenetration ratio**

$$\Psi_N \;\equiv\; 2(d/12\pi)^{d/2} \frac{B_2^{(N,N)}}{\langle R_g^2 \rangle_N^{d/2}} \tag{2.13a}$$

$$\;=\; (d/12\pi)^{d/2} \frac{c_{N,N}}{c_N^2 \langle R_g^2 \rangle_N^{d/2}} \tag{2.13b}$$

(for simplicity we consider only $N_1 = N_2 = N$). The numerical prefactor is a convention that arose historically for reasons not worth explaining here. Crudely speaking, $\Psi$ measures the degree of "hardness" of a SAW in its interactions with other SAWs.[12]

$\Psi_N$ is expected to approach a universal value $\Psi^*$ in the limit $N \to \infty$. A deep question is whether $\Psi^*$ is nonzero (this is called **hyperscaling**). It is now known that hyperscaling fails for SAWs in dimension $d > 4$ [11, 48, 49]. It is believed that hyperscaling holds for SAWs in dimension $d < 4$, but the theoretical justification of this fact is a key unsolved problem in the theory of critical phenomena (see e.g. [39]).[13]

Higher virial coefficients can be defined analogously, but the details will not be needed here.

---

[12] A useful standard of comparison is the hard sphere of constant density:

$$\Psi_{hard-sphere} \;=\; \frac{2}{d\,\Gamma(d/2)} \left(\frac{d+2}{3}\right)^{d/2} \;=\; \begin{cases} \approx 1.12838 & \text{in } d=1 \\ 4/3 & \text{in } d=2 \\ \approx 1.61859 & \text{in } d=3 \\ 2 & \text{in } d=4 \end{cases}$$

[13] A very beautiful heuristic argument concerning hyperscaling for SAWs was given by des Cloizeaux [56]. Note first from (2.13b) that $\Psi$ measures, roughly speaking, the probability of intersection of two independent SAWs that start a distance of order $\langle R_g^2 \rangle^{1/2} \sim N^\nu$ apart. Now, by (2.6), we can interpret a long SAW as an object with "fractal dimension" $1/\nu$. Two independent such objects will "generically" intersect if and only if the sum of their fractal dimensions is at least as large as the dimension of the ambient space. So we expect $\Psi^*$ to be nonzero if and only if $1/\nu + 1/\nu \geq d$, i.e. $d\nu \leq 2$. This occurs for $d < 4$. (For $d = 4$ we believe that $d\nu =$ "2 + logs", and thus expect a logarithmic violation of hyperscaling.)

*Remark.* The critical exponents defined here for the SAW are precise analogues of the critical exponents as conventionally defined for ferromagnetic spin systems [57, 58]. Indeed, the generating functions of the SAW are *equal* to the correlation functions of the $n$-vector spin model analytically continued to $n = 0$ [59, 60, 61, 62, 11]. This "polymer-magnet correspondence"[14] is very useful in polymer theory; but we shall not need it in this article.

## 2.2 The Ensembles

Different aspects of the SAW can be probed in four different ensembles[15]:

- Fixed-length, fixed-endpoint ensemble (fixed $N$, fixed $x$)
- Fixed-length, free-endpoint ensemble (fixed $N$, variable $x$)
- Variable-length, fixed-endpoint ensemble (variable $N$, fixed $x$)
- Variable-length, free-endpoint ensemble (variable $N$, variable $x$)

The fixed-length ensembles are best suited for studying the critical exponents $\nu$ and $2\Delta_4 - \gamma$, while the variable-length ensembles are best suited for studying the connective constant $\mu$ and the critical exponents $\alpha_{sing}$ (fixed-endpoint) or $\gamma$ (free-endpoint). Physically, the free-endpoint ensembles correspond to linear polymers, while the fixed-endpoint ensembles with $|x| = 1$ correspond to ring polymers.

All these ensembles give equal weight to all walks of a given length; but the variable-length ensembles have considerable freedom in choosing the relative weights of different chain lengths $N$. The details are as follows:

**Fixed-$N$, fixed-$x$ ensemble.** The state space is $\mathcal{S}_N(x)$, and the probability distribution is $\pi(\omega) = 1/c_N(x)$ for each $\omega \in \mathcal{S}_N(x)$.

**Fixed-$N$, variable-$x$ ensemble.** The state space is $\mathcal{S}_N$, and the probability distribution is $\pi(\omega) = 1/c_N$ for each $\omega \in \mathcal{S}_N$.

**Variable-$N$, fixed-$x$ ensemble.** The state space is $\mathcal{S}(x) \equiv \bigcup_{N=0}^{\infty} \mathcal{S}_N(x)$, and the probability distribution is generally taken to be

$$\pi_{\beta,p}(\omega) \;=\; \beta^{|\omega|}|\omega|^p / Z(\beta, p; x) \qquad \text{for each } \omega \in \mathcal{S}(x) \tag{2.14}$$

[14]It is sometimes called the "polymer-magnet analogy", but this phrase is misleading: at least for SAWs (athermal linear polymers), the correspondence is an *exact mathematical identity* [11, Section 2.3], not merely an "analogy".

[15]The proper terminology for these ensembles is unclear to me. The fixed-length and variable-length ensembles are sometimes called "canonical" and "grand canonical", respectively (based on considering the *monomers* as particles). On the other hand, it might be better to call these ensembles "microcanonical" and "canonical", respectively (considering the *polymers* as particles and the chain length as an "energy") — reserving the term "grand canonical" for ensembles of *many* SAWs. My current preference is to avoid entirely these ambiguous terms, and simply say what one means: "fixed-length", "variable-length", etc.

where

$$Z(\beta, p; x) \;=\; \sum_{N=0}^{\infty} \beta^N N^p c_N(x) \;. \tag{2.15}$$

Here $p \geq 0$ is a fixed number (usually 0 or 1), and $\beta$ is a **monomer fugacity** that can be varied between 0 and $\beta_c \equiv 1/\mu$. By tuning $\beta$ we can control the distribution of walk lengths $N$. Indeed, from (2.2) we have

$$\langle N \rangle \;\approx\; \frac{p + \alpha_{sing} - 1}{1 - \beta\mu} \tag{2.16}$$

as $\beta \uparrow \beta_c$, provided that $p + \alpha_{sing} > 1$.[16] Therefore, to generate a distribution of predominantly long (but not *too* long) walks, it suffices to choose $\beta$ slightly less than (but not *too* close to) $\beta_c$.

**Variable-$N$, variable-$x$ ensemble.** The state space is $\mathcal{S} \equiv \bigcup_{N=0}^{\infty} \mathcal{S}_N$, and the probability distribution is generally taken to be

$$\pi_{\beta,p}(\omega) \;=\; \beta^{|\omega|} |\omega|^p / Z(\beta, p) \qquad \text{for each } \omega \in \mathcal{S} \tag{2.17}$$

where

$$Z(\beta, p) \;=\; \sum_{N=0}^{\infty} \beta^N N^p c_N \;. \tag{2.18}$$

$p$ and $\beta$ are as before, and from (2.1) we have

$$\langle N \rangle \;\approx\; \frac{p + \gamma}{1 - \beta\mu} \tag{2.19}$$

as $\beta \uparrow \beta_c$. (Here the condition $p + \gamma > 0$ is automatically satisfied, as a result of the rigorous theorem $\gamma \geq 1$ [11].)

An unusual two-SAW ensemble is employed in the join-and-cut algorithm, as will be discussed in Section 6.6.2.

# 3 Monte Carlo Methods: A Review

Monte Carlo methods can be classified as static, quasi-static or dynamic. **Static** methods are those that generate a sequence of *statistically independent* samples from the desired probability distribution $\pi$. **Quasi-static** methods are those that generate a sequence of statistically independent *batches* of samples from the desired probability distribution $\pi$; the correlations within a batch are often difficult to describe. **Dynamic** methods are those that generate a sequence of correlated samples from some *stochastic process* (usually a Markov process) having the desired probability distribution $\pi$ as its unique equilibrium distribution.

In this section we review briefly the principles of both static and dynamic Monte Carlo methods, with emphasis on the issues that determine the statistical efficiency of an algorithm.

[16] If $0 < p + \alpha_{sing} < 1$, then $\langle N \rangle \sim (1 - \beta\mu)^{-(p+\alpha_{sing})}$ as $\beta \uparrow \beta_c$, with logarithmic corrections when $p + \alpha_{sing} = 0, 1$. If $p + \alpha_{sing} < 0$, then $\langle N \rangle$ remains bounded as $\beta \uparrow \beta_c$.

## 3.1 Static Monte Carlo Methods

Consider a system with **state space** (**configuration space**) $S$; for notational simplicity, let us assume that $S$ is discrete (i.e. finite or countably infinite). Now let $\pi = \{\pi_x\}_{x\in S}$ be a probability distribution on $S$, and let $A = \{A(x)\}_{x\in S}$ be a real-valued observable. Our goal is to devise a Monte Carlo algorithm for estimating the expectation value

$$\langle A\rangle_\pi \;\equiv\; \sum_{x\in S} \pi_x\, A(x)\,. \tag{3.1}$$

The most straightforward approach (**standard Monte Carlo**) is to generate independent random samples $X_1, \ldots, X_n$ from the distribution $\pi$ (if one can!), and use the **sample mean**

$$\bar{A} \;\equiv\; \frac{1}{n}\sum_{i=1}^{n} A(X_i) \tag{3.2}$$

as an estimate of $\langle A\rangle_\pi$. This estimate is **unbiased**, i.e.

$$\langle \bar{A}\rangle \;=\; \langle A\rangle_\pi\,. \tag{3.3}$$

Its **variance** is

$$\begin{aligned}
\mathrm{var}(\bar{A}) \;&\equiv\; \langle \bar{A}^2\rangle \;-\; \langle \bar{A}\rangle^2 \\
&=\; \frac{1}{n}\mathrm{var}_\pi(A) \\
&\equiv\; \frac{1}{n}\left[\langle A^2\rangle_\pi - \langle A\rangle_\pi^2\right]\,.
\end{aligned} \tag{3.4}$$

However, it is also legitimate to generate samples $X_1, \ldots, X_n$ from *any* probability distribution $\nu$, and then use **weights** $W(x) \equiv \pi_x/\nu_x$. There are two reasons one might want to sample from $\nu$ rather than $\pi$. Firstly, it might be *unfeasible* to generate (efficiently) random samples from $\pi$, so one may be *obliged* to sample instead from some simpler distribution $\nu$. This situation is the typical one in statistical mechanics. Secondly, one might aspire to improve the efficiency (i.e. reduce the variance) by sampling from a cleverly chosen distribution $\nu$.

There are two cases to consider, depending on how well one knows the function $W(x)$:

(a) $W(x)$ is known exactly. [Note that $\sum\limits_{x\in S} \nu_x W(x) = 1$ and $\sum\limits_{x\in S} \pi_x W(x)^{-1} = 1$.]

(b) $W(x)$ is known *except for an unknown multiplicative constant* (normalization factor). This case is common in statistical mechanics: if $\pi_x = Z_\beta^{-1} e^{-\beta H(x)}$ and $\nu_x = Z_{\beta'}^{-1} e^{-\beta' H(x)}$, then $W(x) = (Z_{\beta'}/Z_\beta) e^{-(\beta-\beta')H(x)}$ but we are unlikely to know the ratio of partition functions.

In the first case, we can use as our estimator the **weighted sample mean**

$$\bar{A}^{(W)} \;\equiv\; \frac{1}{n}\sum_{i=1}^{n} W(X_i)\, A(X_i)\,. \tag{3.5}$$

This estimate is unbiased, since

$$\langle \bar{A}^{(W)} \rangle \;=\; \langle WA \rangle_\nu \;=\; \langle A \rangle_\pi \,. \tag{3.6}$$

Its variance is

$$\begin{aligned} \mathrm{var}(\bar{A}^{(W)}) &= \frac{1}{n}\left[\langle (WA)^2\rangle_\nu - \langle WA\rangle_\nu^2\right] \\ &= \frac{1}{n}\left[\langle WA^2\rangle_\pi - \langle A\rangle_\pi^2\right] \end{aligned} \tag{3.7}$$

This estimate can be either better or worse than standard Monte Carlo, depending on the choice of $\nu$. The optimal choice is the one that minimizes $\langle WA^2\rangle_\pi$ subject to the constraint $\langle W^{-1}\rangle_\pi = 1$, namely

$$W(x)^{-1} \;=\; \frac{|A(x)|}{\sum\limits_{x\in S} \pi_x\,|A(x)|} \;, \tag{3.8}$$

or in other words $\nu_x = \text{const} \times |A(x)|\,\pi_x$. In particular, if $A(x) \geq 0$ the resulting estimate has *zero* variance. But it is impractical: in order to know $W(x)$ we must know the denominator in (3.8), which is the quantity we were trying to estimate in the first place! Nevertheless, this result offer some practical guidance: we should choose $W(x)^{-1}$ to mimic $|A(x)|$ as closely as possible, *subject to the constraint that* $\sum\limits_{x\in S} \pi_x W(x)^{-1}$ *be calculable analytically* (and equal to 1).

In the second case, we have to use a **ratio estimator**

$$\bar{A}^{(W,ratio)} \;\equiv\; \frac{\sum\limits_{i=1}^{n} W(X_i)\,A(X_i)}{\sum\limits_{i=1}^{n} W(X_i)} \;; \tag{3.9}$$

here the unknown normalization factor in $W$ cancels out. This estimate is slightly biased: using the small-fluctuations approximation

$$\left\langle \frac{Y}{Z} \right\rangle \;\approx\; \frac{\langle Y\rangle}{\langle Z\rangle}\left[1 - \frac{\mathrm{cov}(Y,Z)}{\langle Y\rangle\langle Z\rangle} + \frac{\mathrm{var}(Z)}{\langle Z\rangle^2}\right] \;, \tag{3.10}$$

we obtain

$$\langle \bar{A}^{(W,ratio)} \rangle \;=\; \langle A\rangle_\pi \;-\; \frac{1}{n}\Big[\langle WA\rangle_\pi - \langle W\rangle_\pi \langle A\rangle_\pi\Big] \;+\; O\!\left(\frac{1}{n^2}\right) \,. \tag{3.11}$$

Since the bias is of order $1/n$, while the standard deviation ($\equiv$ square root of the variance) is of order $1/\sqrt{n}$, the bias is normally negligible compared to the statistical fluctuation.[17] The variance can also be computed by the small-fluctuations approxi-

---

[17]Note that

$$|\langle WA\rangle_\pi - \langle W\rangle_\pi\langle A\rangle_\pi| \;\leq\; \langle W(A - \langle A\rangle_\pi)^2\rangle_\pi^{1/2}\,(\langle W\rangle_\pi - 1)^{1/2}$$

(with equality if and only if $A = c_1 + c_2 W^{-1}$) by the Schwarz inequality with measure $\nu$ applied to the functions $W - 1$ and $W(A - \langle A\rangle_\pi)$. Therefore, from (3.11) and (3.13) we have (to leading order in $1/n$)

$$|\mathrm{bias}(\bar{A}^{(W,ratio)})| \;\leq\; n^{-1/2}\,\mathrm{var}(\bar{A}^{(W,ratio)})^{1/2}\,(\langle W\rangle_\pi - 1)^{1/2} \,.$$

So the bias is $\ll$ the standard deviation unless $\langle W\rangle_\pi$ is enormous.

mation

$$\mathrm{var}\left(\frac{Y}{Z}\right) = \frac{\langle Y\rangle^2}{\langle Z\rangle^2}\,\mathrm{var}\left(\frac{Y}{\langle Y\rangle} - \frac{Z}{\langle Z\rangle}\right) \tag{3.12a}$$

$$= \frac{\mathrm{var}(Y)}{\langle Z\rangle^2} - \frac{2\langle Y\rangle}{\langle Z\rangle^3}\mathrm{cov}(Y,Z) + \frac{\langle Y\rangle^2}{\langle Z\rangle^4}\mathrm{var}(Z)\;; \tag{3.12b}$$

it is

$$\mathrm{var}(\bar{A}^{(W,ratio)}) = \frac{1}{n}\left\langle W(A-\langle A\rangle_\pi)^2\right\rangle_\pi + O\left(\frac{1}{n^2}\right)\,. \tag{3.13}$$

The optimal choice of $\nu$ is the one that minimizes $\langle W(A-\langle A\rangle_\pi)^2\rangle_\pi$ subject to the constraint $\langle W^{-1}\rangle_\pi = 1$, namely

$$W(x)^{-1} = \frac{|A(x)-\langle A\rangle_\pi|}{\sum\limits_{x\in S}\pi_x\,|A(x)-\langle A\rangle_\pi|}\;. \tag{3.14}$$

Let us now try to interpret these formulae. First note that

$$\langle W\rangle_\pi - 1 = \langle W^2\rangle_\nu - \langle W\rangle_\nu^2 \equiv \mathrm{var}_\nu(W) \ge 0\;, \tag{3.15}$$

with equality only if $\nu = \pi$. So $\langle W\rangle_\pi - 1$ measures, in a rough sense, the "mismatch" (or "distance") between $\nu$ and $\pi$. Now assume for simplicity that $A$ is a bounded observable, i.e. $|A(x)| \le M$ for all $x \in S$. Then it is immediate from (3.7) and (3.13) that

$$\mathrm{var}(\bar{A}^{(W)}) \le \frac{M^2}{n}\langle W\rangle_\pi \tag{3.16}$$

$$\mathrm{var}(\bar{A}^{(W,ratio)}) \le \frac{4M^2}{n}\langle W\rangle_\pi + O\left(\frac{1}{n^2}\right) \tag{3.17}$$

So the variances cannot get large unless $\langle W\rangle_\pi \gg 1$, i.e. $\nu$ is very distant from $\pi$; and in this case it is easy to see that the variances *can* get large. The moral is this: when the probability distribution actually simulated ($\nu$) differs considerably from the distribution of interest ($\pi$), the variance of the Monte Carlo estimates can be *vastly* higher than one might expect naively for the given sample size $n$. Heuristically, this is because the states (configurations) that are "typical" for $\pi$ are "rare" for $\nu$, so the *useful* sample size is much smaller than the *total* sample size.

Here is a concrete example: Let $S$ be the set of all $N$-step walks (not necessarily self-avoiding) starting at the origin. Let $\pi$ be uniform measure on *self-avoiding* walks, i.e.

$$\pi_\omega = \begin{cases} 1/c_N & \text{if } \omega \text{ is self-avoiding} \\ 0 & \text{otherwise} \end{cases} \tag{3.18}$$

Unfortunately, it is not easy to generate (efficiently) random samples from $\pi$ (that is the subject of this article!). So let us instead generate *ordinary* random walks, i.e. random samples from

$$\nu_\omega = (2d)^{-N} \qquad \text{for all } \omega \in S\;, \tag{3.19}$$

and then apply the weights $W(\omega) = \pi_\omega/\nu_\omega$. Clearly we have

$$\langle W \rangle_\pi \;=\; \frac{(2d)^N}{c_N} \;=\; \left(\frac{2d}{\mu}\right)^N \;, \tag{3.20}$$

which grows *exponentially* for large $N$. Therefore, the efficiency of this algorithm deteriorates exponentially as $N$ grows.

See [63, Chapter 5] for some more sophisticated static Monte Carlo techniques. It would be interesting to know whether any of them can be applied usefully to the self-avoiding walk.

## 3.2 Dynamic Monte Carlo Methods

In this subsection we review briefly the principles of dynamic Monte Carlo methods, and define some quantities (autocorrelation times) that will play an important role in the remainder of this article.

The idea of *dynamic* Monte Carlo methods is to invent a *stochastic process* with state space $S$ having $\pi$ as its unique equilibrium distribution. We then simulate this stochastic process, starting from an arbitrary initial configuration; once the system has reached equilibrium, we measure time averages, which converge (as the run time tends to infinity) to $\pi$-averages. In physical terms, we are inventing a *stochastic time evolution* for the given system. It must be emphasized, however, that this time evolution *need not correspond to any real "physical" dynamics*: rather, the dynamics is simply a numerical algorithm, and it is to be chosen, like all numerical algorithms, on the basis of its computational efficiency.

In practice, the stochastic process is always taken to be a **Markov process**. We assume that the reader is familiar with the elementary theory of discrete-time Markov chains.[18]

For simplicity let us assume that the state space $S$ is discrete (i.e. finite or countably infinite); this is the case in nearly all the applications considered in this article. Consider a Markov chain with state space $S$ and transition probability matrix $P = \{p(x \to y)\} = \{p_{xy}\}$ satisfying the following two conditions:

(A) For each pair $x, y \in S$, there exists an $n \geq 0$ for which $p^{(n)}_{xy} > 0$. Here $p^{(n)}_{xy} \equiv (P^n)_{xy}$ is the $n$-step transition probability from $x$ to $y$. [This condition is called **irreducibility** (or **ergodicity**); it asserts that each state can eventually be reached from each other state.]

(B) For each $y \in S$,

$$\sum_{x \in S} \pi_x \, p_{xy} = \pi_y . \tag{3.21}$$

[This condition asserts that $\pi$ is a **stationary distribution** (or **equilibrium distribution**) for the Markov chain $P = \{p_{xy}\}$.]

---

[18] The books of Kemeny and Snell [64] and Iosifescu [65] are excellent references on the theory of Markov chains with *finite* state space. At a somewhat higher mathematical level, the books of Chung [66] and Nummelin [67] deal with the cases of *countable* and *general* state space, respectively.

In this case it can be shown [66] that $\pi$ is the *unique* stationary distribution for the Markov chain $P = \{p_{xy}\}$, and that the occupation-time distribution over long time intervals converges (with probability 1) to $\pi$, irrespective of the initial state of the system. If, in addition, $P$ is **aperiodic** [this means that for each pair $x, y \in S$, $p_{xy}^{(n)} > 0$ for *all* sufficiently large $n$], then the probability distribution at any single time in the far future also converges to $\pi$, irrespective of the initial state — that is, $\lim_{n\to\infty} p_{xy}^{(n)} = \pi_y$ for all $x$.

Thus, simulation of the Markov chain $P$ provides a legitimate Monte Carlo method for estimating averages with respect to $\pi$. However, since the successive states $X_0, X_1, \ldots$ of the Markov chain are in general highly correlated, the variance of estimates produced in this way may be much higher than in independent sampling. To make this precise, let $A = \{A(x)\}_{x\in S}$ be a real-valued function defined on the state space $S$ (i.e. a real-valued observable) that is square-integrable with respect to $\pi$. Now consider the *stationary* Markov chain (i.e. start the system in the stationary distribution $\pi$, or equivalently, "thermalize" it for a very long time prior to observing the system). Then $\{A_t\} \equiv \{A(X_t)\}$ is a stationary stochastic process with mean

$$\mu_A \equiv \langle A_t \rangle = \sum_{x\in S} \pi_x \, A(x) \tag{3.22}$$

and **unnormalized autocorrelation function**[19]

$$\begin{aligned} C_{AA}(t) &\equiv \langle A_s A_{s+t} \rangle - \mu_A^2 \\ &= \sum_{x,y\in S} A(x) \, [\pi_x p_{xy}^{(|t|)} - \pi_x \pi_y] \, A(y) \; . \end{aligned} \tag{3.23}$$

The **normalized autocorrelation function** is then

$$\rho_{AA}(t) \equiv C_{AA}(t)/C_{AA}(0) \; . \tag{3.24}$$

Typically $\rho_{AA}(t)$ decays exponentially ($\sim e^{-|t|/\tau}$) for large $t$; we define the **exponential autocorrelation time**

$$\tau_{exp,A} = \limsup_{t\to\infty} \frac{t}{-\log |\rho_{AA}(t)|} \tag{3.25}$$

and

$$\tau_{exp} = \sup_A \; \tau_{exp,A} \; . \tag{3.26}$$

Thus, $\tau_{exp}$ is the relaxation time of the slowest mode in the system. (If the state space is infinite, $\tau_{exp}$ might be $+\infty$!)[20]

---

[19]In the statistics literature, this is called the **autocovariance function**.

[20]An equivalent definition, which is useful for rigorous analysis, involves considering the spectrum of the transition probability matrix $P$ considered as an operator on the Hilbert space $l^2(\pi)$. [$l^2(\pi)$ is the space of complex-valued functions on $S$ that are square-integrable with respect to $\pi$: $\|A\| \equiv (\sum_{x\in S} \pi_x |A(x)|^2)^{1/2} < \infty$. The inner product is given by $(A, B) \equiv \sum_{x\in S} \pi_x A(x)^* B(x)$.] It is not

On the other hand, for a given observable $A$ we define the **integrated autocorrelation time**

$$\tau_{int,A} = \frac{1}{2}\sum_{t=-\infty}^{\infty} \rho_{AA}(t) \qquad (3.27)$$
$$= \frac{1}{2} + \sum_{t=1}^{\infty} \rho_{AA}(t)$$

[The factor of $\frac{1}{2}$ is purely a matter of convention; it is inserted so that $\tau_{int,A} \approx \tau_{exp,A}$ if $\rho_{AA}(t) \sim e^{-|t|/\tau}$ with $\tau \gg 1$.] The integrated autocorrelation time controls the statistical error in Monte Carlo estimates of $\langle A \rangle$. More precisely, the sample mean

$$\bar{A} \equiv \frac{1}{n}\sum_{t=1}^{n} A_t \qquad (3.28)$$

has variance

$$\mathrm{var}(\bar{A}) = \frac{1}{n^2}\sum_{r,s=1}^{n} C_{AA}(r-s) \qquad (3.29)$$
$$= \frac{1}{n}\sum_{t=-(n-1)}^{n-1} \left(1 - \frac{|t|}{n}\right) C_{AA}(t) \qquad (3.30)$$
$$\approx \frac{1}{n}(2\tau_{int,A})\, C_{AA}(0) \quad \text{for } n \gg \tau \qquad (3.31)$$

Thus, the variance of $\bar{A}$ is a factor $2\tau_{int,A}$ larger than it would be if the $\{A_t\}$ were statistically independent. Stated differently, the number of "effectively independent samples" in a run of length $n$ is roughly $n/2\tau_{int,A}$.

In summary, the autocorrelation times $\tau_{exp}$ and $\tau_{int,A}$ play different roles in Monte Carlo simulations. $\tau_{exp}$ controls the relaxation of the *slowest* mode in the system; in particular, it places an upper bound on the number of iterations $n_{disc}$ which should be discarded at the beginning of the run, before the system has attained equilibrium (e.g.

---

hard to prove the following facts about $P$:

(a) The operator $P$ is a contraction. (In particular, its spectrum lies in the closed unit disk.)

(b) 1 is a simple eigenvalue of $P$, as well as of its adjoint $P^*$, with eigenvector equal to the constant function $\mathbf{1}$.

(c) If the Markov chain is aperiodic, then 1 is the only eigenvalue of $P$ (and of $P^*$) on the unit circle.

(d) Let $R$ be the spectral radius of $P$ acting on the orthogonal complement of the constant functions:
$$R \equiv \inf\Big\{r\colon\ \mathrm{spec}\,(P \restriction \mathbf{1}^\perp) \subset \{\lambda\colon |\lambda| \le r\}\Big\}.$$
Then $R = e^{-1/\tau_{exp}}$.

Facts (a)–(c) are a generalized Perron-Frobenius theorem [68]; fact (d) is a consequence of a generalized spectral radius formula [69]. Note that the worst-case rate of convergence to equilibrium from an initial nonequilibrium distribution is controlled by $R$, and hence by $\tau_{exp}$.

$n_{disc} \approx 20\tau_{exp}$ is usually more than adequate). On the other hand, $\tau_{int,A}$ determines the statistical errors in Monte Carlo estimates of $\langle A \rangle$, once equilibrium has been attained.

Most commonly it is assumed that $\tau_{exp}$ and $\tau_{int,A}$ are of the same order of magnitude, at least for "reasonable" observables $A$. But this is *not* true in general. In fact, one usually expects the autocorrelation function $\rho_{AA}(t)$ to obey a dynamic scaling law [70] of the form

$$\rho_{AA}(t;\beta) \;\approx\; |t|^{-a} F\left((\beta-\beta_c)\,|t|^b\right) \tag{3.32}$$

valid in the limit

$$\beta-\beta_c \to 0, \qquad |t|\to\infty, \qquad x \;\equiv\; (\beta-\beta_c)\,|t|^b \text{ fixed.} \tag{3.33}$$

Here $a, b > 0$ are dynamic critical exponents and $F$ is a suitable scaling function; $\beta$ is some "temperature-like" parameter, and $\beta_c$ is the critical point. Now suppose that $F$ is continuous and strictly positive, with $F(x)$ decaying rapidly (e.g. exponentially) as $|x| \to \infty$. Then it is not hard to see that

$$\tau_{exp,A} \;\sim\; |\beta-\beta_c|^{-1/b} \tag{3.34}$$

$$\tau_{int,A} \;\sim\; |\beta-\beta_c|^{-(1-a)/b} \qquad (\text{if } a<1) \tag{3.35}$$

$$\rho_{AA}(t;\,\beta=\beta_c) \;\sim\; |t|^{-a} \tag{3.36}$$

so that $\tau_{exp,A}$ and $\tau_{int,A}$ have *different* critical exponents unless $a = 0$.[21] Actually, this should not be surprising: replacing "time" by "space", we see that $\tau_{exp,A}$ is the analogue of a correlation length, while $\tau_{int,A}$ is the analogue of a susceptibility; and (3.34)–(3.36) are the analogue of the well-known scaling law $\gamma = (2-\eta)\nu$ — clearly $\gamma \neq \nu$ in general! So it is crucial to distinguish between the two types of autocorrelation time.

Returning to the general theory, we note that one convenient way of satisfying the stationarity condition (B) is to satisfy the following *stronger* condition:

$$(\mathrm{B}') \;\; \text{For each pair } x,y \in S,\; \pi_x p_{xy} = \pi_y p_{yx}\,. \tag{3.37}$$

[Summing (B′) over $x$, we recover (B).] (B′) is called the **detailed-balance condition**; a Markov chain satisfying (B′) is called **reversible**.[22] (B′) is equivalent to the *self-adjointness* of $P$ as on operator on the space $l^2(\pi)$. In this case, it follows from the spectral theorem that the autocorrelation function $C_{AA}(t)$ has a spectral representation

$$C_{AA}(t) \;=\; \int_{-1}^{1} \lambda^{|t|}\, d\sigma_{AA}(\lambda) \tag{3.38}$$

[21]Our discussion of this topic in [71] is incorrect. A correct discussion can be found in [72].

[22]For the physical significance of this term, see Kemeny and Snell [64, section 5.3] or Iosifescu [65, section 4.5].

with a *nonnegative* spectral weight $d\sigma_{AA}(\lambda)$ supported on the interval $[-e^{-1/\tau_{exp,A}},\ e^{-1/\tau_{exp,A}}]$. It follows that

$$\tau_{int,A} \;\le\; \frac{1}{2}\left(\frac{1+e^{-1/\tau_{exp,A}}}{1-e^{-1/\tau_{exp,A}}}\right) \;\le\; \frac{1}{2}\left(\frac{1+e^{-1/\tau_{exp}}}{1-e^{-1/\tau_{exp}}}\right) \;\approx\; \tau_{exp}\,. \tag{3.39}$$

There is no particular *advantage* to algorithms satisfying detailed balance (rather than merely satisfying stationarity), but they are easier to analyze mathematically.

Finally, let us make a remark about transition probabilities $P$ that are "built up out of" other transition probabilities $P_1, P_2, \ldots, P_n$:

a) If $P_1, P_2, \ldots, P_n$ satisfy the stationarity condition (resp. the detailed-balance condition) for $\pi$, then so does any convex combination $P = \sum_{i=1}^{n} \lambda_i P_i$. Here $\lambda_i \ge 0$ and $\sum_{i=1}^{n} \lambda_i = 1$.

b) If $P_1, P_2, \ldots, P_n$ satisfy the stationarity condition for $\pi$, then so does the product $P = P_1 P_2 \cdots P_n$. (Note, however, that $P$ does *not* in general satisfy the detailed-balance condition, even if the individual $P_i$ do.[23])

Algorithmically, the convex combination amounts to choosing *randomly*, with probabilities $\{\lambda_i\}$, from among the "elementary operations" $P_i$. (It is crucial here that the $\lambda_i$ are *constants*, independent of the current configuration of the system; only in this case does $P$ leave $\pi$ stationary in general.) Similarly, the product corresponds to performing *sequentially* the operations $P_1, P_2, \ldots, P_n$.

# 4 Static Monte Carlo Methods for the SAW

## 4.1 Simple Sampling and Its Variants

The most obvious static technique for generating a random $N$-step SAW is **simple sampling**: just generate a random $N$-step *ordinary* random walk (ORW), and reject it if it is not self-avoiding; keep trying until success. It is easy to see that this algorithm produces each $N$-step SAW with equal probability. Of course, to save time we should check the self-avoidance as we go along, and reject the walk as soon as a self-intersection is detected. (Methods for testing self-avoidance are discussed in Section 7.1.2.) The algorithm is thus:

[23] Recall that if $A$ and $B$ are self-adjoint operators, then $AB$ is self-adjoint *if and only if* $A$ and $B$ commute.

```
        title Simple sampling.
        function ssamp(N)
        comment This routine returns a random N-step SAW.

        ω_0 ← 0
start:  for i = 1 to N do
            ω_i ← a random nearest neighbor of ω_{i-1}
            if ω_i ∈ {ω_0, ..., ω_{i-1}} goto start
        enddo
        return ω
```

(Here and in what follows, we will express algorithms in "pseudocode". Translation to your favorite language — Fortran, C or whatever — is almost always trivial.)

The trouble with this algorithm is, of course, the exponentially rapid sample attrition for long walks. Clearly, the probability of an $N$-step walk being self-avoiding is $c_N/(2d)^N$, which behaves for large $N$ as

$$\frac{c_N}{(2d)^N} \sim (\mu/2d)^N N^{\gamma-1} \tag{4.1a}$$

$$\sim e^{-\lambda N} N^{\gamma-1} \tag{4.1b}$$

where

$$\lambda = \log(2d/\mu) \tag{4.2}$$

is called the **attrition constant**. Therefore, the mean number of attempts required to generate an $N$-step SAW is $(2d)^N/c_N$, which grows roughly as $e^{\lambda N}$. And the mean CPU time per attempt is of order $\min(1/\lambda, N)$. So this method is extremely inefficient in generating SAWs of length $N \gtrsim 10/\lambda$. For the simple (hyper-)cubic lattices in dimensions 2, 3 and 4, the values of $\lambda$ are approximately 0.42, 0.25 and 0.17, respectively (see Table 1). So it is unfeasible to generate SAWs of length more than $\approx$ 20–60 steps by simple sampling. All alternative SAW Monte Carlo techniques are aimed essentially at alleviating this attrition problem — hopefully without introducing other problems of equal or greater severity!

Some improvement can be obtained by modifying the walk-generation process so as to produce only walks without immediate reversals (such walks are called **non-reversal random walks** (NRRWs) or **memory-2 walks**). The algorithm is thus:

```
        title Non-reversal simple sampling.
        function nrssamp(N)
        comment This routine returns a random N-step SAW.

        ω_0 ← 0
        ω_1 ← a random nearest neighbor of 0
start:  for i = 2 to N do
            ω_i ← a random nearest neighbor of ω_{i-1}, not equal to ω_{i-2}
            if ω_i ∈ {ω_0, ..., ω_{i-1}} goto start
        enddo
        return ω
```

| $d$ | $\mu$ | $\lambda$ | $\lambda'$ |
|---|---|---|---|
| 2 | 2.638 158 5 (10) [73, 27] | 0.416 | 0.129 |
| 3 | 4.683 907 (22) [74] | 0.248 | 0.065 |
| 4 | 6.772 0 (5) [75] | 0.167 | 0.033 |
| 5 | 8.838 6 (8) [76] | 0.123 | 0.018 |
| 6 | 10.878 8 (9) [76] | 0.098 | 0.011 |
| $d \to \infty$ | $2d - 1 - (2d)^{-1} - \ldots$ [77, 78, 79, 50, 51] | $(2d)^{-1} + \ldots$ | $(2d)^{-2} + \ldots$ |

Table 1: Connective constant $\mu$ and attrition constants $\lambda$ and $\lambda'$ for simple (hyper-)cubic lattices in dimensions $2 \le d \le 6$ and $d \to \infty$. Estimated errors in the last digit(s) are shown in parentheses.

Then $(2d)^N$ is replaced by $2d(2d-1)^{N-1}$, and the attrition rate is

$$\lambda' \;=\; \log \frac{2d-1}{\mu} \,. \tag{4.3}$$

For comparison, $\lambda'$ is approximately 0.13, 0.07 and 0.03, respectively, for $d = 2, 3, 4$. This is much smaller than in the unmodified scheme, but the exponential attrition is still prohibitive for walks of length more than $\approx$ 80–300 steps.

The logical next step is to modify the walk-generation process so that walks with loops of length $\le r$ are automatically absent. Let us start by building the walk out of **strides** of $r$ steps [80].[24] That is, let us enumerate in advance all the $r$-step SAWs — call them $\omega^{(1)}, \ldots, \omega^{(c_r)}$. (Obviously this takes a memory of order $rc_r$, and so is feasible only if $r$ is not too large.) We then build up the walk by repeated concatenation of strides. For simplicity let us assume that $N$ is a multiple of $r$:

**title** Simple $r$-step stride sampling.
**function** `simstride`$(r, k)$
**comment** This routine returns a random $kr$-step SAW.

start: $\omega \leftarrow \{0\}$ (zero-step SAW at the origin)
**for** $i = 1$ **to** $k$ **do**
  $\alpha \leftarrow$ a random integer from the set $\{1, \ldots, c_r\}$
  $\omega \leftarrow \omega \circ \omega^{(\alpha)}$ (concatenation)
  **if** $\omega$ is not self-avoiding **goto** start
**enddo**
**return** $\omega$

[24]The treatment in the remainder of this section relies heavily on [11, Section 9.3.1], which is in turn an explication of [63, p. 129].

| $d$ | $r$ | $c_r$ | $rc_r$ | $\lambda^{(r)} \equiv \log(c_r^{1/r}/\mu)$ | $\lambda'^{(r)} \equiv \log\Big(\big(\frac{2d-1}{2d}c_r\big)^{1/r}/\mu\Big)$ |
|---|---|---|---|---|---|
| 2 | 10 | 44100 | 441000 | 0.099 | 0.071 |
| 3 | 7 | 81390 | 569730 | 0.071 | 0.045 |
| 4 | 6 | 127160 | 762960 | 0.046 | 0.024 |
| 5 | 5 | 64250 | 321250 | 0.035 | 0.014 |
| 6 | 5 | 173172 | 865860 | 0.026 | 0.008 |

Table 2: Attrition constants $\lambda^{(r)}$ and $\lambda'^{(r)}$ for simple and non-reversal $r$-stride sampling. For each $d$, we have taken the largest $r$ such that $rc_r \le 10^6$.

The probability of surviving to length $N = kr$ is

$$\frac{c_N}{(c_r)^k} \;\sim\; \left(\frac{\mu}{c_r^{1/r}}\right)^N \;. \tag{4.4}$$

There is still exponential attrition (since $c_r > \mu^r$), but this attrition can in principle be made arbitrarily small by taking $r$ large (since $\lim_{r\to\infty} c_r^{1/r} = \mu$). In practice we can probably handle $rc_r$ of order $10^6$ on a modern-day workstation.[25] The resulting attrition rates $\lambda^{(r)}$ are shown in Table 2. They are far from spectacular; the trouble is that $c_r^{1/r}$ converges rather slowly to $\mu$.[26]

Of course, we can do better by choosing $\omega^{(\alpha)}$ from among only those $r$-step walks whose first step is not opposite to the last step of the current $\omega$. In this **non-reversal $r$-step stride sampling**, the probability of surviving to length $N = kr$ is

$$\frac{c_N}{c_r\left(\frac{2d-1}{2d}c_r\right)^{k-1}} \;\sim\; \left(\frac{\mu}{\left(\frac{2d-1}{2d}c_r\right)^{1/r}}\right)^N \;. \tag{4.5}$$

The resulting attrition rates $\lambda'^{(r)}$ are shown in the last column of Table 2.

Neither of these algorithms in fact eliminates all loops of length $\le r$, because such loops can be formed by the concatenation of two $r$-step strides. But we can eliminate such loops if we are willing to pre-compute the list of legal pairs of strides. That is, for each index $\alpha$ ($1 \le \alpha \le c_r$), we make a list $L_\alpha$ containing those indices $\beta$ such that $\omega^{(\alpha)} \circ \omega^{(\beta)}$ is self-avoiding. (This takes a memory of order $c_r^2$.) Now, it would *not* be correct to choose at each stage of the algorithm a random SAW from the appropriate list $L_\alpha$; the trouble is that the lists do not all have the same number of

[25] Of course, we can reduce the memory requirements by at least a factor $2d$ by exploiting symmetry, e.g. storing only those $r$-step SAWs whose first step is in some particular direction. But this only increases the feasible $r$ by about 1.

[26] From (2.1) we have $c_r^{1/r} \approx \mu\left[1 + \frac{(\gamma-1)\log r}{r} + O(\frac{1}{r})\right]$.

elements, and as a result the walks would not be generated with uniform probability (see also Section 4.2). Instead, we must allow the possibility of "rejections". Let $c_r^* \equiv \max_{1\le\alpha\le c_r} |L_\alpha|$ be the number of elements in the *largest* list. We can then perform:

**title** Super-duper $r$-step stride sampling.
**function** `supstride`$(r, k)$
**comment** This routine returns a random $kr$-step SAW.

$\alpha_1 \leftarrow$ a random integer from the set $\{1, \ldots, c_r\}$
start: $\omega \leftarrow \omega^{(\alpha_1)}$
**for** $i = 2$ **to** $k$ **do**
  $m \leftarrow$ a random integer from the set $\{1, \ldots, c_r^*\}$
  **if** $m > |L_{\alpha_{i-1}}|$ **goto** start  (this is the "rejection")
  $\alpha_i \leftarrow$ the $m^{th}$ element from $L_{\alpha_{i-1}}$
  $\omega \leftarrow \omega \circ \omega^{(\alpha_i)}$
  **if** $\omega$ is not self-avoiding **goto** start
**enddo**
**return** $\omega$

Clearly the probability of surviving to length $N = kr$ is

$$\frac{c_N}{c_r (c_r^*)^{k-1}} \sim \left(\frac{\mu}{(c_r^*)^{1/r}}\right)^N . \tag{4.6}$$

Little is known about $c_r^*$, but it is certainly $\le \frac{2d-1}{2d} c_r$, and probably not much less. In practice, the extravagant memory requirements of this method limit $r$ to very small values; for a given amount of memory, non-reversal stride sampling probably works better.[27]

## 4.2 Inversely Restricted Sampling (Rosenbluth-Rosenbluth Algorithm)

The exponential attrition of simple sampling and its variants arises from the fact that each new step of the walk might lead to a self-intersection. So it is tempting to envisage an algorithm in which one chooses randomly (with equal probability) from among only those next steps which do *not* lead to a self-intersection (assuming such steps exist). Unfortunately, this means that SAWs are not generated with uniform probability; rather, the probability that this algorithm generates a given $N$-step SAW $\omega$ is

$$P(\omega) = \text{const} \times \prod_{i=1}^{N} \frac{1}{k_i} , \tag{4.7}$$

[27] Actually, it suffices to store the list *lengths* $|L_\alpha|$, and not the lists themselves. One can then choose $\alpha_i$ by repeatedly trying to find a stride compatible with $\omega^{(\alpha_{i-1})}$, together with a probability $1 - |L_{\alpha_{i-1}}|/c_r^*$ of giving up ("rejection") at each try.

where $k_i \equiv k_i(\omega_0, \ldots, \omega_{i-1})$ is the number of choices available at step $i$.[28] Therefore, each walk must be assigned a *weight* $W(\omega) \sim 1/P(\omega)$, and the mean value of an observable $\mathcal{O}(\omega)$ must be estimated from a sample of walks $\omega^{(1)}, \ldots, \omega^{(n)}$ by a ratio of weighted averages:

$$\langle \mathcal{O}(\omega) \rangle \;\approx\; \frac{\sum\limits_{i=1}^{n} W(\omega^{(j)})\, \mathcal{O}(\omega^{(j)})}{\sum\limits_{i=1}^{n} W(\omega^{(j)})}\;. \qquad (4.8)$$

This method is known as **inversely restricted sampling** [81, 82]:

**title** Inversely restricted sampling.
**function** `irsamp`($N$)
**comment** This routine returns an $N$-step SAW and its weight factor.

$\omega_0 \leftarrow 0$
start: *weight* $\leftarrow 1/[2d(2d-1)^{N-1}]$ (this is merely a convenient normalization)
**for** $i = 1$ **to** $N$ **do**
  $S_i \leftarrow$ set of all nearest neighbors of $\omega_{i-1}$ not contained in $\{\omega_0, \ldots, \omega_{i-1}\}$
  **if** $S_i = \varnothing$ **goto** start (the walk is "trapped")
  $\omega_i \leftarrow$ a random element of $S_i$
  *weight* $\leftarrow$ *weight* $\times |S_i|$
**enddo**
**return** ($\omega$, *weight*)

This method has several difficulties: Firstly, there is still exponential sample attrition for long walks (although at a much slower rate than in simple sampling): it is caused now not by mere self-intersection, but by "trapping". This is most serious in $d = 2$ [83, 84, 85, 86]. Secondly, a ratio estimator (4.8) is slightly biased; however, as discussed in Section 3.1, this difficulty is negligible for large sample size $n$ [87]. The most serious difficulty is that "the weights are almost certain to get out of hand, a few of them being very much larger than all the rest. This means that the greater part of the data, corresponding to the negligible weights, gets ignored" [63, p. 131]. Thus, the variance of the estimates will be *vastly* higher than one might expect naively for the given sample size $n$. This is in fact a general problem in any Monte Carlo work in which the probability distribution actually simulated differs considerably from the distribution of interest (see Section 3.1). In the case at hand, one expects that the discrepancy between the two distributions will grow *exponentially* as the chain length $N$ gets large. This has been verified numerically by Batoulis and Kremer [86], who conclude that for large $N$ inversely restricted sampling is inferior to non-reversal simple sampling.

Fraser and Winnik [88] have proposed a generalization of inversely restricted sampling, based on strides with cleverly chosen probabilities. Meirovitch [89, 90, 91, 92] has introduced a slightly different generalization (which he calls the "scanning method"), in which the algorithm looks ahead more than one step. But neither of these methods appears to avoid the exponential explosion of weights, although they may reduce it.

[28] Here "const" is in fact the reciprocal of the probability of surviving in this algorithm to $N$ steps without getting "trapped".

## 4.3 Dimerization

The **dimerization** algorithm [93] is an implementation of the computer scientists' principle of "divide and conquer" [94, Section 1.3] [95]. To generate an $N$-step self-avoiding walk, we generate two independent $(N/2)$-step SAWs ("dimers") and attempt to concatenate them. If the result is self-avoiding, we are done; otherwise, we discard the two walks and start again from scratch. This procedure can now be repeated recursively: to generate each of the $(N/2)$-step SAWs, we generate a pair of $(N/4)$-step SAWs and attempt to concatenate them, and so on. For $N \leq$ some cutoff $N_0$, we stop the recursion and generate the SAWs by some primitive method, such as non-reversal simple sampling. The dimerization algorithm can thus be written recursively as follows:

**title** Dimerization (recursive version).
**function** `dim`$(N)$
**comment** This routine returns a random $N$-step SAW.

**if** $N \leq N_0$ **then**
    $\omega \leftarrow$ `nrssamp`$(N)$
    **return** $\omega$
**else**
    $N_1 \leftarrow \lfloor N/2 \rfloor$ (integer part)
    $N_2 \leftarrow N - N_1$
    start:
      $\omega^{(1)} \leftarrow$ `dim`$(N_1)$
      $\omega^{(2)} \leftarrow$ `dim`$(N_2)$
      $\omega \leftarrow \omega^{(1)} \circ \omega^{(2)}$ (concatenation)
      **if** $\omega$ is not self-avoiding **goto** start
      **return** $\omega$
**endif**

It is easy to prove inductively that algorithm `dim` produces each $N$-step SAW with equal probability, using the fact that the subroutine `nrssamp` does so. It is crucial here that after a failure we discard *both* walks and start again *from scratch*.

A non-recursive description of this same algorithm is given by Suzuki [93].

Let us analyze [96] the efficiency of the dimerization algorithm under the scaling hypothesis

$$c_N \;\approx\; A\mu^N N^{\gamma-1} \tag{4.9}$$

[cf. (2.1)]. Let $T_N$ be the mean CPU time needed to generate an $N$-step SAW by algorithm `dim`. Now, the probability that the concatenation of two random $(N/2)$-step

SAWs yields an $N$-step SAW is

$$p_N \;=\; \frac{c_N}{(c_{N/2})^2} \;\approx\; B^{-1} N^{-(\gamma-1)} \;, \tag{4.10}$$

where $B = A/4^{\gamma-1}$. We will need to generate, on average, $1/p_N$ pairs of $(N/2)$-step SAWs in order to get a single $N$-step SAW; hence

$$T_N \;\approx\; B N^{\gamma-1} 2 T_{N/2} \,. \tag{4.11}$$

(We have neglected here the time needed for checking the intersections of the two dimers; this time is linear in $N$, which, as will be seen shortly, is negligible compared to the time $2T_{N/2}$ for generating the two dimers.) Iterating this $k$ times, where $k = \log_2(N/N_0)$ is the number of levels, we obtain

$$\begin{aligned} T_N \;&\approx\; \frac{(2BN^{\gamma-1})^k}{2^{(\gamma-1)k(k-1)/2}} T_{N_0} \\ &=\; C_0 N^{C_1 \log_2 N \,+\, C_2} \;, \end{aligned} \tag{4.12a}$$

where

$$C_1 \;=\; \frac{\gamma-1}{2} \tag{4.13a}$$

$$C_2 \;=\; \frac{\gamma-1}{2} + \log_2(2B) \;=\; \frac{5-3\gamma}{2} + \log_2 A \tag{4.13b}$$

and $C_0$ depends on $N_0$. Thus, the growth of $T_N$ is slower than exponential in $N$; but if $\gamma > 1$ (which occurs for $d < 4$) it is faster than any polynomial in $N$. Fortunately, however, the constants $C_1$ and $C_2$ are very small: see Table 3. For $d = 2$ (resp. $d = 3$) this means that in practice $T_N$ behaves like $N^{\approx 2-3}$ up to $N$ of order several thousand (resp. several million). In $d = 4$ the behavior is presumably $\sim N^{\widetilde{C}_1 \log\log N + \widetilde{C}_2}$. For $d \geq 5$ we have $C_1 = 0$ and $C_2$ only very slightly larger than 1, so dimerization is extraordinarily efficient: a random $N$-step SAW can be produced in a CPU time that grows only slightly faster than linearly in $N$.

The efficiency can be improved further by choosing the first step of $\omega^{(2)}$ to avoid reversing the last step of $\omega^{(1)}$. This effectively replaces $A$ by $[(2d-1)/2d]A$ in (4.13b), and hence $C_2$ by $C_2' \equiv C_2 - \log_2[2d/(2d-1)]$. See Table 3 for the effect of this change.

It is an open question whether for $d \leq 4$ there exists any static Monte Carlo algorithm for generating a random $N$-step SAW (with exactly uniform distribution) in a mean CPU time that is bounded by a polynomial in $N$.

For discussion of some statistical issues related to dimerization, see [11, Section 9.3.2].

**Remark.** A slightly different version of the dimerization algorithm was invented independently by Alexandrowicz [97, 98] and used subsequently by many others. In this version, one begins by producing a large initial batch of $M$-step SAWs, where $M$ is some small number ($\approx N_0$); one then joins randomly chosen pairs to form walks of length $2M$, and so forth. The trouble with this method is that the same subchains of length $M$, $2M$, $4M$, etc. tend to occur repeatedly in the SAWs produced. Thus, the sample of SAWs produced is both slightly biased (SAWs with two or more identical subchains of length $M$ are favored) and somewhat correlated, but it is difficult to assess these effects quantitatively. Both effects can be reduced by using an extremely large initial batch of SAWs, and by making not too many dimerization attempts per batch, but this is likely to be inefficient or unreliable or both. In my opinion Alexandrowicz' version of dimerization should *not* be used; algorithm `dim` is simpler, more efficient, and — above all — is correct.

| $d$ | $\mu$ (est.) | $\gamma$ (est.) | $A$ (est.) | $C_1$ | $C_2$ | $C_2'$ |
|---|---|---|---|---|---|---|
| 2 | 2.6381585 | 43/32 | 1.1775 | $11/64 \approx 0.172$ | 0.72 | 0.31 |
| 3 | 4.6839066 | 1.162 | 1.1845 | 0.081 | 1.00 | 0.74 |
| 4 | 6.7720 | $1\ (\times \log^{1/4})$ | — | — | — | — |
| 5 | 8.83861 | 1 | 1.25 | 0 | 1.32 | 1.17 |
| 6 | 10.87879 | 1 | 1.16 | 0 | 1.21 | 1.09 |
| all $d \geq 5$ | — | 1 | $\leq 1.493$ | 0 | $\leq 1.58$ | $\leq 1.58$ |

Table 3: Efficiency of dimerization algorithm as a function of lattice dimension $d$. Estimates are obtained by extrapolation of the available counts $c_N$ [11, Tables C.1 and C.4]. The last line is a rigorous bound valid for all $d \geq 5$ [11, p. 172]. The case $d = 4$ is somewhat anomalous, as it is believed [14] that $c_N \sim \mu^N (\log N)^{1/4}$ in contrast to the usual power-law behavior (2.1).

# 5 Quasi-Static Monte Carlo Methods for the SAW

## 5.1 Quasi-Static Simple Sampling

In each of the foregoing static methods, a slight improvement in efficiency can be obtained by working with several different values of $N$ at once, and noticing that "a walk that intersects itself for the first time at the $M^{th}$ step provides [unbiased] instances of $N$-step self-avoiding walks for all $N < M$" [63, p. 129]. The resulting method is quasi-static in our classification: each pass through the algorithm produces a batch of SAWs (of various lengths) which are highly correlated among themselves, although successive batches are independent of each other. Unfortunately — and this seems to be characteristic of quasi-static methods — it appears difficult to estimate quantitatively the degree of correlation between the various SAWs in a given batch, and therefore difficult to estimate the actual statistical efficiency of the method.

## 5.2 Enrichment

One method of generating long SAWs with much less attrition than (non-reversal) simple sampling is the **enrichment technique** [99]: if a walk survives to $s$ steps, then several ($t$) copies are made of it and each copy is used independently as a starting point for further attempts to add steps; and likewise at $2s$, $3s$, etc. (The same idea was used earlier in Monte Carlo work on neutron-transport problems [63, Section 8.2].) The free parameters $s$ and $t$ must be chosen judiciously (see below). This method is quasi-static: the SAWs produced in a single pass through the algorithm (i.e. the progeny of a single $s$-step SAW) are manifestly correlated, since they all have the same initial $s$ steps and many of them have the same initial $2s, 3s, \ldots$ steps as well. However, as before, it is difficult to assess this correlation quantitatively. (Indeed, the quasi-static simple sampling method can be considered to be the special case of the enrichment method with $s = t = 1$.)

The enrichment algorithm can be analyzed semi-quantitatively as follows [100] [11, Section 9.3.3]: Let $M_{ns}$ (a random variable) be the number of $ns$-step walks that are produced in a single pass through the algorithm; by definition of "single pass" we have $M_s = 1$. (The data produced by successive passes through the algorithm are obviously independent.) A SAW of length $ns$ gives rise to $t$ copies, each of which survives to length $(n+1)s$ with (average) probability[29]

$$\alpha_n \equiv \frac{c_{(n+1)s}}{(2d-1)^s c_{ns}} \approx \left(\frac{\mu}{2d-1}\right)^s . \tag{5.1}$$

So we can regard $M_s, M_{2s}, \ldots$ as a **branching process** [102] in which $M_{ns}$ is the number of "individuals" alive in the $n^{th}$ generation. We assume that each individual reproduces independently, producing a number of "children" which is a binomial random variable with parameters $t$ and $\alpha = [\mu/(2d-1)]^s$.[30]

It is easy to see that $\langle M_{ns} \rangle = (t\alpha)^n$. Thus, if $t\alpha < 1$ there is exponential sample attrition, just as in simple sampling. If $t\alpha > 1$, some (although not all) starts lead to an exponential explosion of progeny; this is undesirable, as great computational effort will be expended in producing these samples, but the information contained in them is less than proportional to that effort, because they are highly correlated. The most interesting case is $t\alpha = 1$: every start dies out eventually, but the mean lifetime of a start is infinite. More precisely, it can be shown [102] that

$$\text{Prob}(M_{ns} = 0) \approx 1 - \frac{2}{\sigma^2 n} \tag{5.2}$$

$$\text{Prob}(M_{ns} > \beta n) \approx \frac{2}{\sigma^2 n} e^{-2\beta/\sigma^2} \qquad (\beta \geq 0) \tag{5.3}$$

[29]For large $n$, one has [from (2.1)] $\alpha_n \approx \left(\frac{\mu}{2d-1}\right)^s \left(1 + \frac{1}{n}\right)^{\gamma - 1} \to \left(\frac{\mu}{2d-1}\right)^s$ as $n \to \infty$, irrespective of the value of $s$ or of $\gamma$. For *even* $s$ this convergence is actually a rigorous theorem [101].

[30]This assumption is not really correct: the trouble is that some SAWs will have higher or lower "fertility" than others (i.e. be harder or easier to intersect with); and these fertilities are somewhat *correlated* between different walks in the process, as all these walks share some common segments (the degree of correlation obviously depends on the relative location of the two SAWs in the "family tree"). Nevertheless, this assumption seems to be a reasonable approximation.

where $\sigma^2 \equiv (1-\alpha)t = t-1$. Thus, $M_{ns}$ is nonzero only with probability of order $1/n$; but when it is nonzero it is typically of order $n$.[31] On the other hand, it is not hard to see that the mean CPU time per start is (when $t\alpha = 1$) of order $n$ times the CPU time needed to produce a single $s$-step segment.[32] So even if we make the over-pessimistic assumption that all the walks in the $n^{th}$ generation of a given start are *perfectly* correlated (and thus carry only as much information as *one* walk), it follows that we can obtain one independent $ns$-step walk in a CPU time of order $n^2$ (i.e. $\sim n$ starts each taking a CPU time $\sim n$). This $\sim N^2$ behavior is exceptionally good: it is better than all known static algorithms (e.g. dimerization) in dimension $d \leq 4$, and it is comparable to most of the new dynamic algorithms (see Section 6). The enrichment algorithm definitely deserves a systematic theoretical and empirical study.

*Remarks.* 1. The integer $t$ can also be a random variable; in this case the role of $t\alpha$ is played by $\langle t \rangle \alpha$. This generalization is useful in permitting fine-tuning of $\langle t \rangle \alpha$.

2. The $\sim N^2$ obtained here may really be $\sim N^{\gamma+1}$ if one takes account of the corrections to the $\approx$ sign in (5.1). See also Sections 5.3 and 6.6.1.

## 5.3 Incomplete Enumeration (Redner-Reynolds Algorithm)

**Incomplete enumeration** [104, 105] is a quasi-static algorithm that generates a batch of SAWs from the variable-$N$, variable-$x$ ensemble (2.17) with $p = 0$. The idea is to take a standard algorithm for systematically enumerating *all* self-avoiding walks up to some length $N_{max}$, and modify it so that it pursues each branch of the "SAW tree" only with some probability $\beta < 1$:

[31] This prediction is strikingly confirmed by Grishman's [103] empirical observations for $t\alpha \approx 1$ and large but fixed $n$: "if the enrichment parameters are adjusted to make the total number of generated walks approximately equal to the number of starts, one will find that most of the starts 'die out' before generating any walks, while a few starts each lead to a large number of walks."

[32] Here I assume that one tests self-avoidance using a method that takes a CPU time of order 1 per added step (see Section 7.1.2).

```
title Incomplete enumeration (recursive version).
subroutine incenum(ω, β, N_max)
comment This routine performs an incomplete enumeration of the SAW tree
    beginning at ω, with probability parameter β, up to maximum length N_max.

append ω to output list
if |ω| ≥ N_max return
for i = 1 to 2d do
    U ← random number uniformly distributed on [0,1]
    if U ≤ β then
        ω' ← one-step extension of ω in direction i
        if ω' is self-avoiding incenum(ω', β, N_max)
    endif
enddo
return
```

To perform an incomplete enumeration of the SAW tree beginning at the "root", one simply invokes incenum($\{0\}, \beta, N_{max}$), where $\{0\}$ is the zero-step walk at the origin. Typically one chooses $\beta < \beta_c \equiv 1/\mu$; in this case it is safe to set $N_{max} = \infty$. (If one sets $N_{max} = \infty$ when $\beta > \beta_c$, then with nonzero probability the algorithm will run forever!)

A non-recursive implementation of incomplete enumeration can be found in [11, Section 9.3.3].

Incomplete enumeration is very closely related to enrichment: indeed, it is nearly identical to the enrichment algorithm in which $s = 1$ and $t$ is a binomial random variable with parameters $2d$ and $\beta$. The only difference is that the enrichment algorithm performs "sampling with replacement" among the various one-step extensions of $\omega$ (i.e. the same $\omega'$ could occur more than once), while incomplete enumeration performs "sampling without replacement" (each $\omega'$ occurs at most once). So one expects incomplete enumeration to be slightly more efficient, in that the resulting batch of SAWs will be less correlated. (Indeed, there may even be some *anti*correlation arising from the sampling without replacement: this certainly occurs if, for example, $\beta = 1$. On the other hand, for $\beta \ll 1$ one still has strong positive correlation for the same reason as in the enrichment algorithm: many walks share the same initial segments.)

The CPU time for one invocation of incomplete enumeration is obviously proportional to the total number of walks encountered during the enumeration.[33] On the average this is

$$Z(\beta) \equiv \sum_{N=0}^{\infty} \beta^N c_N \sim (1-\beta\mu)^{-\gamma} \sim \langle N \rangle^{\gamma} . \tag{5.4}$$

On the other hand, it is reasonable to guess that, as in enrichment,

$$\text{Prob(at least one } N\text{-step SAW is produced)} \sim 1/N \qquad \text{for } N \sim \langle N \rangle . \tag{5.5}$$

[33] Here I assume that one tests self-avoidance using an $O(1)$ algorithm (see Section 7.1.2).

So if we make the over-pessimistic assumption that all the walks of length $N$ in a given batch are *perfectly* correlated (and thus carry only as much information as *one* walk), it follows that we can obtain one independent $N$-step walk in a CPU time of order $N^{\gamma+1}$ (i.e. choose $\beta$ so that $\langle N \rangle \sim N$, then make $\sim N$ starts each taking a CPU time $\sim N^{\gamma}$).

This differs from the estimate made previously for enrichment, because $\gamma$ is in general slightly larger than 1 (namely 43/32 in dimension $d = 2$, $\approx 1.16$ in $d = 3$, and 1 in $d \geq 4$). However, I suspect that the correct answer in both cases is $N^{\gamma+1}$, and that this would be obtained also in the branching-process analysis if one were to take account of the $n$-dependence of $\alpha_n$ [i.e. $\alpha_n \approx (\frac{\mu}{2d-1})^s (1 + \frac{1}{n})^{\gamma-1}$]. The incomplete-enumeration algorithm is also closely related to the slithering-tortoise algorithm discussed in Section 6.6.1, and the same $\langle N \rangle^2$ vs. $\langle N \rangle^{\gamma+1}$ issue arises there.

In any case, the incomplete-enumeration algorithm is potentially quite efficient, and deserves a systematic theoretical and empirical study.

# 6 Dynamic Monte Carlo Methods for the SAW

In this section I discuss dynamic Monte Carlo methods for generating SAWs in various ensembles. I emphasize that the stochastic dynamics is merely a numerical algorithm, whose goal is to provide statistically independent samples from the desired distribution $\pi$ in the smallest CPU time possible. *It need not correspond to any real "physical" dynamics!* Indeed, the most efficient algorithms typically make non-local moves which would be impossible for real polymer molecules.

## 6.1 General Considerations

This subsection contains some exceedingly pedantic — but I hope useful — general considerations on dynamic Monte Carlo algorithms.

The study of a Monte Carlo algorithm can be divided into three stages:

- *specification* of the algorithm;
- verification of the algorithm's *validity*; and
- study (by mathematical, heuristic and/or numerical means) of the algorithm's *efficiency*.

Let us consider these aspects in turn:

To *specify* a dynamic Monte Carlo algorithm, we must specify three things:

1) The state space $S$.
2) The desired equilibrium measure $\pi$.
3) The transition probability matrix $P = \{p(\omega \to \omega')\}$.

Here $S$ and $\pi$ specify the *model* that we wish to study, while $P$ specifies the *numerical method* by which we propose to study it.[34] With regard to $P$, it is useful to subdivide the issue further, by specifying successively:

3a) The set of allowed **elementary moves**, i.e. the transitions $\omega \to \omega'$ for which $p(\omega \to \omega') > 0$.

3b) The **probabilities** $p(\omega \to \omega')$ for the allowed elementary moves.

After specifying the algorithm, we must prove the algorithm's *validity*. As discussed in Section 3.2, this has two parts:

(A) We must prove **ergodicity** (**irreducibility**), i.e. we must prove that we can get from any state $\omega \in S$ to any other state $\omega' \in S$ by some finite sequence of allowed elementary moves.

(B) We must prove the **stationarity** of $\pi$ with respect to $P$.

Ergodicity is a combinatorial problem; it depends only on the set of allowed elementary moves, not on their precise probabilities. [This is the motivation for distinguishing sub-questions (3a) and (3b).] For many SAW algorithms, the proof of ergodicity is highly non-trivial. Indeed, in several embarrassing cases, an algorithm was used for many years before it was realized to be *non*-ergodic. As for stationarity, this is usually (though not always) verified by showing that $P$ satisfies detailed balance for $\pi$ [cf. (3.37)] or is built up out of constituents $P_1, \ldots, P_n$ which do so [as discussed at the end of Section 3.2].

Once the algorithm is known to be valid, we can investigate its *computational efficiency*, as measured by the amount of CPU time it takes to generate one "statistically independent" sample from the distribution $\pi$. This study also has two parts:

i) Investigate the **autocorrelation times** $\tau_{int,A}$ (for observables $A$ of interest) and $\tau_{exp}$.

ii) Investigate the **computational complexity** of the algorithm, i.e. the mean CPU time per iteration (henceforth denoted $T_{CPU}$).

The CPU time per "statistically independent" sample is then $2\tau_{int,A}T_{CPU}$, and this provides a criterion for comparing algorithms: for given $S$ and $\pi$, the best algorithm is the one with the smallest product $\tau_{int,A}T_{CPU}$. (This criterion may of course depend on the observable $A$.) Both aspects (i) and (ii) should be studied by all the methods at our disposal: rigorous mathematical analysis, heuristic physical reasoning, and numerical experimentation. Since we are primarily interested in *long* self-avoiding walks, we will emphasize the asymptotic behavior of $\tau_{int,A}$, $\tau_{exp}$ and $T_{CPU}$ as $N \to \infty$ (or $\langle N \rangle \to \infty$ for the variable-$N$ algorithms). Usually these quantities behave as some power of $N$ or $\langle N \rangle$, in which case we can identify a **dynamic critical exponent**.

[34] Of course, the choice of $P$ trivially determines $S$; and if $P$ is ergodic (as it should be if the algorithm is to be valid), then it also determines $\pi$. But it is nevertheless useful conceptually to distinguish the three ingredients, which typically correspond to the chronological stages in inventing a new Monte Carlo algorithm.

*Remarks.* 1. We are here always considering the relaxation time of the autocorrelation functions *in equilibrium*, as it is this relaxation time which is of primary importance for determining the statistical accuracy of the Monte Carlo method (see Section 3.2). Some studies [106, 107, 108, 109] have focussed instead on the relaxation *to equilibrium* from a highly non-equilibrium initial state; but as Kranbuehl and Verdier [110] have pointed out, this quantity may well have a strictly smaller dynamic critical exponent. An analogous situation arises in the Glauber dynamics for the Ising model [111].[35]

2. In this article we always measure time in units of attempted elementary moves. Much of the literature on dynamic SAW models uses a time scale of attempted elementary moves *per bead*; autocorrelation times expressed in this way should be multiplied by $N$ (actually $N + 1$) before comparing them to the present paper.

## 6.2 Classification of Moves

The elementary moves in a SAW Monte Carlo algorithm can be classified according to whether they are

- $N$-conserving or $N$-changing
- endpoint-conserving or endpoint-changing
- local or non-local

Obviously, fixed-$N$ algorithms must use only $N$-conserving moves, while variable-$N$ algorithms are free to use both $N$-conserving and/or $N$-changing moves (and indeed *must* use some of the latter in order to satisfy ergodicity). An analogous statement holds for fixed-$x$ and variable-$x$ algorithms with regard to endpoint-conserving and endpoint-changing moves. The distinction between local and non-local moves will be explained later.

Within these limitations, the elementary moves to be discussed below can be mixed more or less freely to make "hybrid" algorithms. The art is, of course, to find a *useful* combination. (A cocktail can be made by mixing any set of liqueurs in any proportions, but there is no guarantee that the resulting concoction will taste good! In particular, it may or may not taste better than the individual ingredients taken separately.) Thus, a hybrid algorithm is useful when its performance is superior to that of either of its "pure" components, in the sense of having a smaller product $\tau_{int,A} T_{CPU}$. Roughly speaking, this occurs when the "slow modes" under one type of move are speeded up by the other type of move. (An extreme case of this is when the combined algorithm is ergodic while both pure algorithms are non-ergodic.)

[35] In principle, the *asymptotic* behavior as $t \to \infty$ of the relaxation from nonequilibrium is controlled by the same $\tau_{exp}$ as the autocorrelation in equilibrium. But to observe this asymptotic relaxation in practice would require enormous statistics (e.g. tens of thousands of repeat runs). Most of these studies have focussed instead on the initial relaxation — e.g. the time to relax to within $1/e$ [106, 109], the time(s) in a phenomenological fit [107, 108], or the area under the relaxation curve [112, 113, 111] — and all of these are likely to exhibit a smaller dynamic critical exponent than either $\tau_{exp}$ or $\tau_{int,A}$.

Note that a hybrid algorithm always contains one or more free parameters — namely the relative probabilities of the different types of moves — which can be tuned to optimize the computational efficiency. (Often this requires a tradeoff between autocorrelation time and computational complexity, particularly if one move is local while the other is non-local.) This optimization is a non-trivial problem, and it must be approached systematically if the results are to be of any use. (The behavior of the algorithm at some *arbitrarily chosen* set of parameter values may be completely misleading.) *A priori* there are three possibilities:

a) One of the pure algorithms is superior to any non-trivial hybrid.

b) The optimal hybrid is better than either of the pure algorithms, but only by a bounded factor (e.g. 2 or 10 or whatever).

c) The optimal hybrid has a better dynamic critical exponent than either of the pure algorithms, so that its superiority factor grows without bound as $N \to \infty$.

The most interesting hybrid algorithms are of course those of type (c). But those of type (b) should not be sniffed at, if the gain is large enough (and the human effort required to find an optimal or nearly-optimal mixture is not too large).

Most of the elementary moves to be discussed below have the property that the resulting walk $\omega'$ is not *guaranteed* to be self-avoiding. Therefore, it is necessary to *test* whether $\omega'$ is self-avoiding (see Section 7.1.2 for methods for doing this). If $\omega'$ is self-avoiding, then the proposed move is accepted; otherwise, it is rejected, and the old walk is counted once again in the sample. This procedure can be understood as the Metropolis criterion for the energy function

$$E(\omega) \;=\; \begin{cases} 0 & \text{if } \omega \text{ is self-avoiding} \\ +\infty & \text{otherwise} \end{cases} \tag{6.1}$$

One final remark: Up to now we have adopted the convention that all walks start at the origin ($\omega_0 = 0$) unless specified otherwise. However, some of the moves to be discussed below may alter the initial site of the walk so that it is no longer at the origin. One could, of course, imagine that the resulting walk is then translated back to the origin. But a more convenient approach is to consider all translates of a given SAW to be *equivalent*; then we need not worry where the initial point is.[36] This is how the algorithms are most efficiently implemented in practice. (Every once in a while one should translate the walk back to the origin, just as a precaution against integer overflow.)

## 6.3 Examples of Moves

### 6.3.1 Local $N$-Conserving Moves

A **local** move is one that alters only a few consecutive sites ("beads") of the SAW, leaving the other sites unchanged. Otherwise put, a local move excises a small piece from the original SAW and splices in a new local conformation in its place. (Of course, it is always necessary to verify that the proposed new walk is indeed self-avoiding.) In this subsection we concentrate on **$N$-conserving** local moves, i.e. those in which the excised and spliced-in subwalks have the same number of beads.

[36] In mathematical language, we redefine $\mathcal{S}_N$ to be the set of *equivalence classes* of $N$-step SAWs modulo translation. (And likewise for the other state spaces.) Note also that the observables (2.3)–(2.5) are unaffected by translation of the walk, hence they are well-defined on equivalence classes.

More precisely, let $\omega$ and $\omega'$ be $N$-step SAWs, and let $k$ be an integer $\geq 1$. We say that **$\omega$ and $\omega'$ are interconvertible by a $k$-bead move** if there exists an index $i_{min}$ $(0 \leq i_{min} \leq N-k+1)$ such that $\omega_i = \omega'_i$ for all $i$ *except possibly* $i = i_{min}, i_{min}+1, \ldots, i_{min}+k-1$. (We shall also assume that $\omega_{i_{min}} \neq \omega'_{i_{min}}$ and $\omega_{i_{min}+k-1} \neq \omega'_{i_{min}+k-1}$, since otherwise $\omega$ and $\omega'$ would be interconvertible also by some move of *less than* $k$ beads.) If $i_{min} = 0$ or $N-k+1$, we call this an **end-group** $k$-bead move; otherwise we call it an **internal** $k$-bead move. Clearly, internal moves are endpoint-conserving, while end-group moves are not.

In Figure 1 we show all the possible one-bead moves (on a hypercubic lattice). Move A is a "one-bead flip" (also called "kink-jump"); it is the only one-bead internal move. Moves B and C are end-bond rotations.

In Figure 2 we show all the possible *internal* two-bead moves. Move D is a "180° crankshaft". Move E is a "90° crankshaft"; of course it is possible only in dimension $d \geq 3$. Move F is a "two-bead L-flip". Move G permutes three successive mutually perpendicular steps (which lie along the edges of a cube); again this is possible only in dimension $d \geq 3$.

We leave it to the reader to construct the list of two-bead end-group moves. There are numerous three-bead moves; a few interesting ones are shown in Figure 3.

### 6.3.2 Bilocal $N$-Conserving Moves

A **bilocal** move is one that alters *two* disjoint small groups of consecutive sites (or steps) of the walk; these two groups may in general be very far from each other. One *trivial* way of making an $N$-conserving bilocal move is to make two independent (nonoverlapping) $N$-conserving local moves. Here we are interested in the *nontrivial* $N$-conserving bilocal moves, i.e. those in which one subwalk loses sites and the other subwalk gains them. Instead of formalizing the concept, let us simply give some examples:

- The **slithering-snake** (or **reptation**) move, which deletes a bond from one end of the walk and appends a new bond (in an arbitrary direction) at the other end [Figure 4].

- The **kink transport** move, which deletes a kink at one location along the walk and inserts a kink (in an arbitrary orientation) at another location [Figure 5].

- The **kink-end reptation** move, which deletes a kink at one location along the walk and appends two new bonds (in arbitrary directions) at one of the ends of the walk [Figure 6⟶].

- The **end-kink reptation** move, which deletes two bonds from one of the ends of the walk and inserts a kink (in an arbitrary orientation) at some location along the walk [Figure 6⟵].

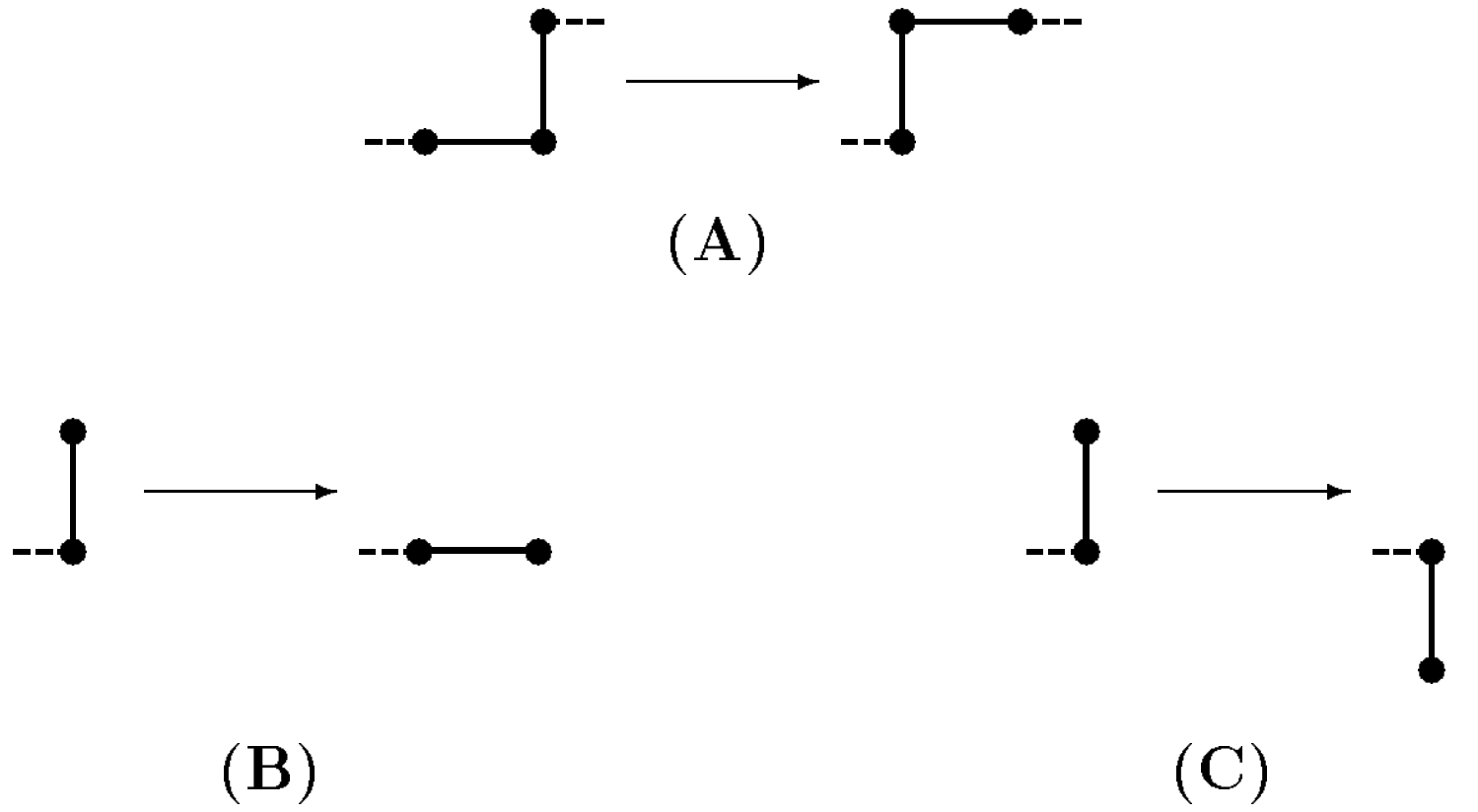

Figure 1: All one-bead local moves. (A) One-bead flip. (B) 90° end-bond rotation. (C) 180° end-bond rotation.

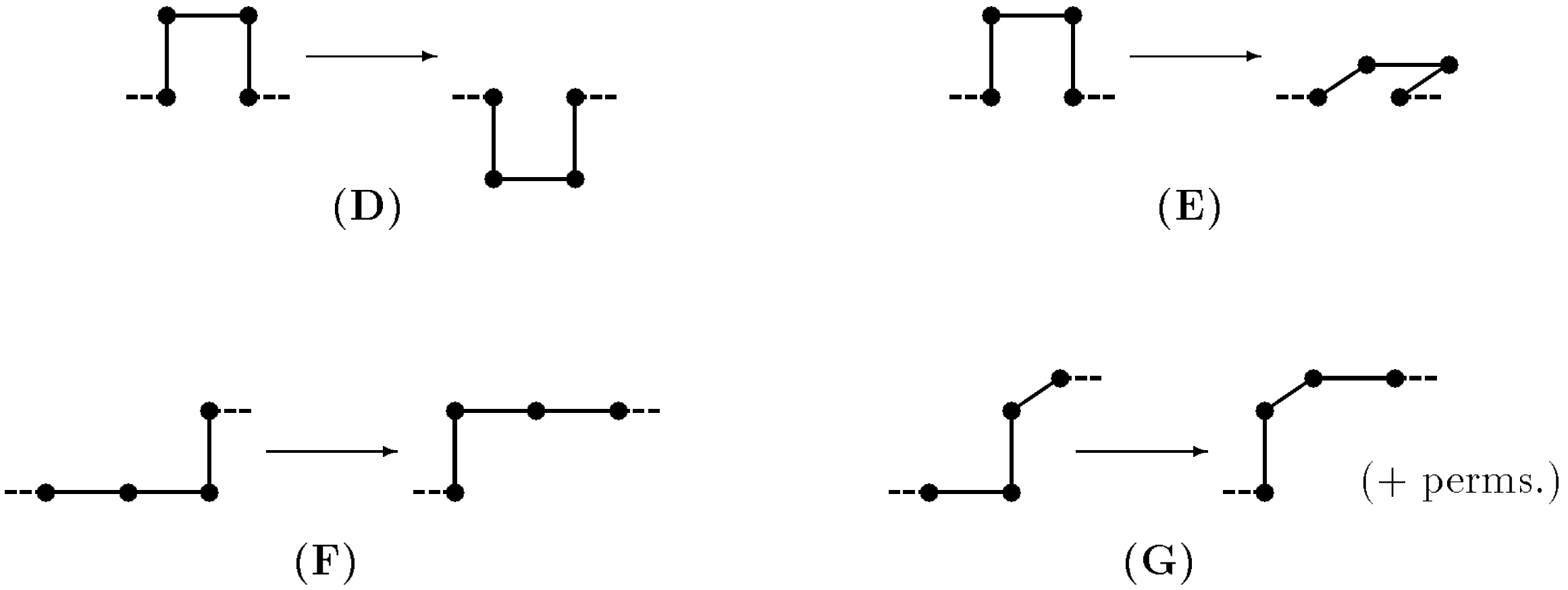

Figure 2: All internal two-bead local moves. (D) 180° crankshaft. (E) 90° crankshaft ($d \geq 3$ only). (F) Two-bead L-flip. (G) Cube permutation ($d \geq 3$ only).

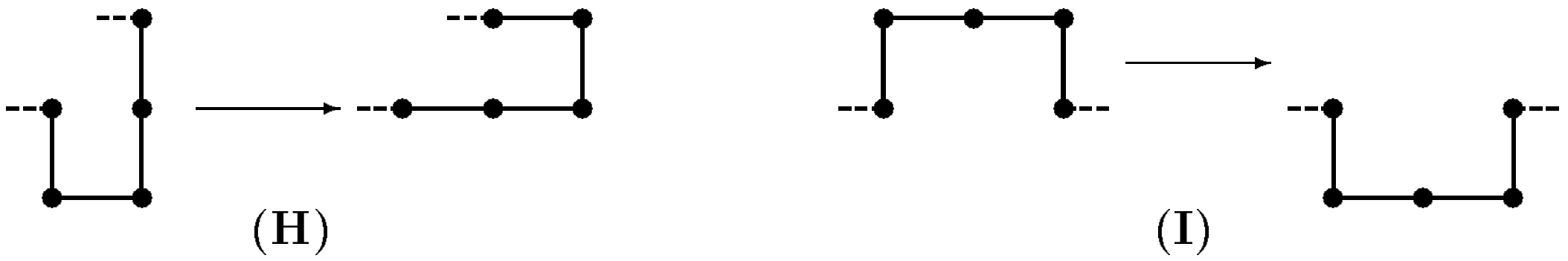

Figure 3: Some selected internal three-bead local moves. (H) Three-bead J-flip. (I) Three-bead 180° crankshaft.

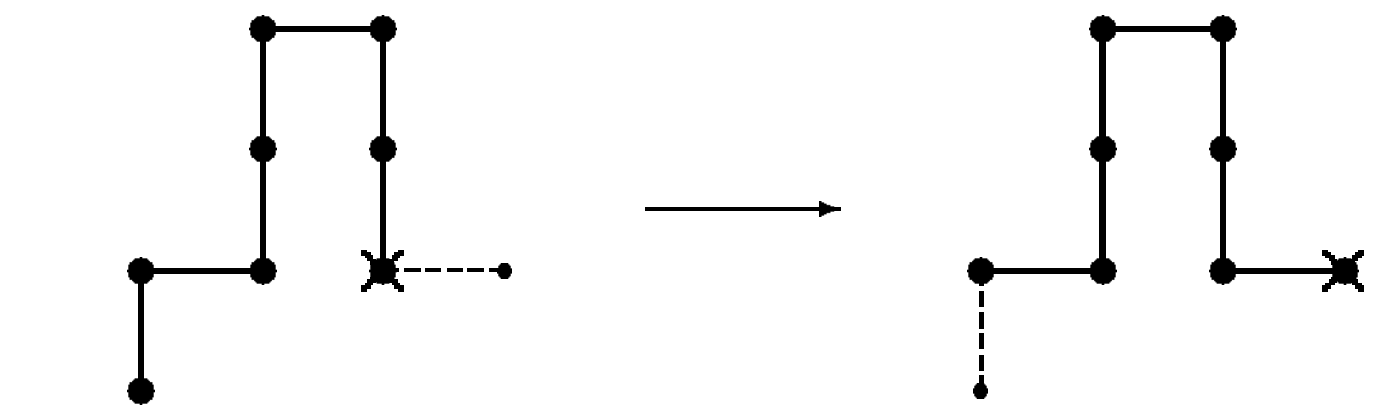

Figure 4: The slithering-snake (reptation) move. Head of the walk is indicated by $\times$. Dashed lines indicate the proposed new step (resp. the just-abandoned old step).

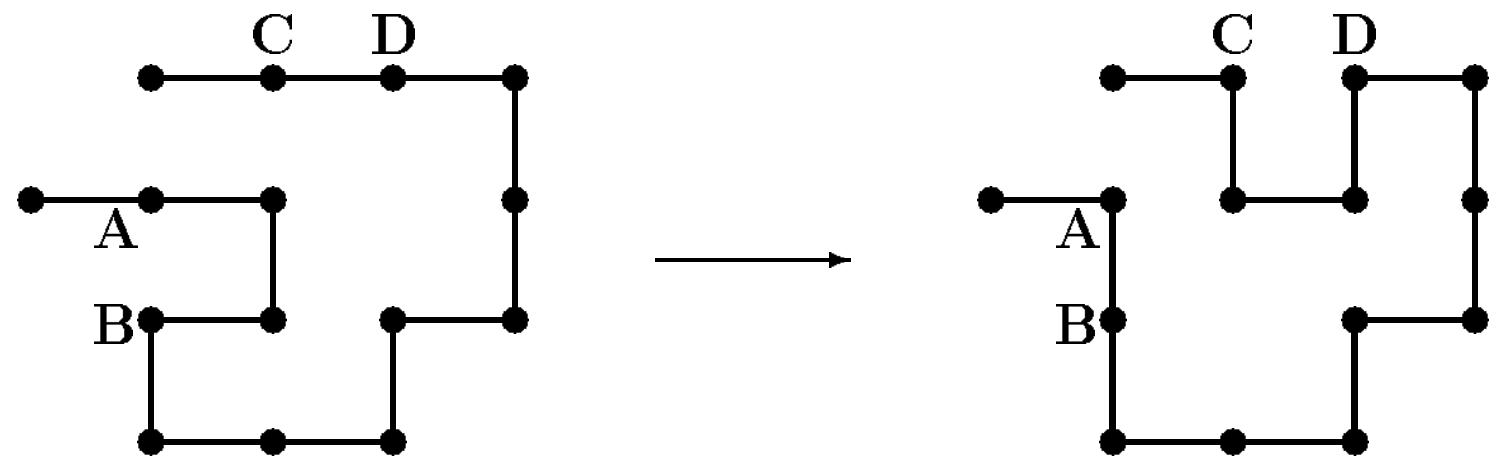

Figure 5: The kink-transport move. A kink has been cleaved from AB and attached at CD. Note that the new kink is permitted to occupy one or both of the sites abandoned by the old kink.

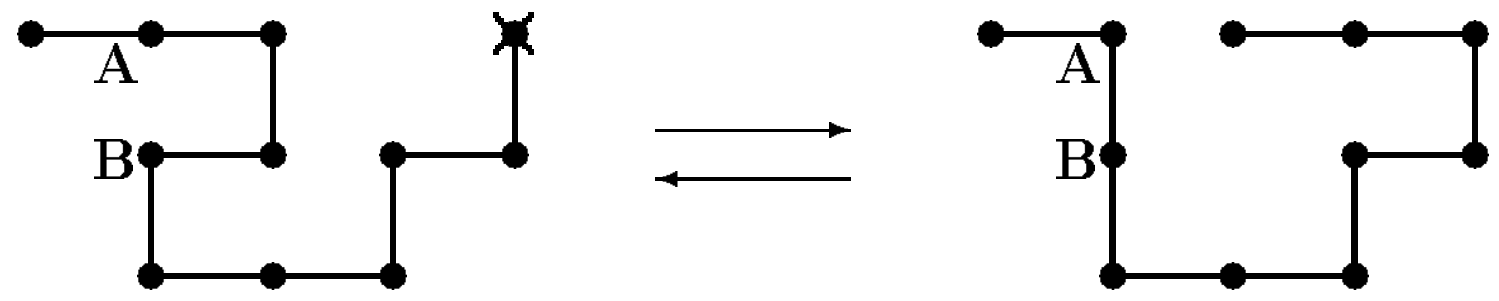

Figure 6: The kink-end reptation ($\longrightarrow$) and end-kink reptation ($\longleftarrow$) moves. In $\longrightarrow$, a kink has been cleaved from AB and two new steps have been attached at the end marked $\times$. Note that the new end steps are permitted to occupy one or both of the sites abandoned by the kink.

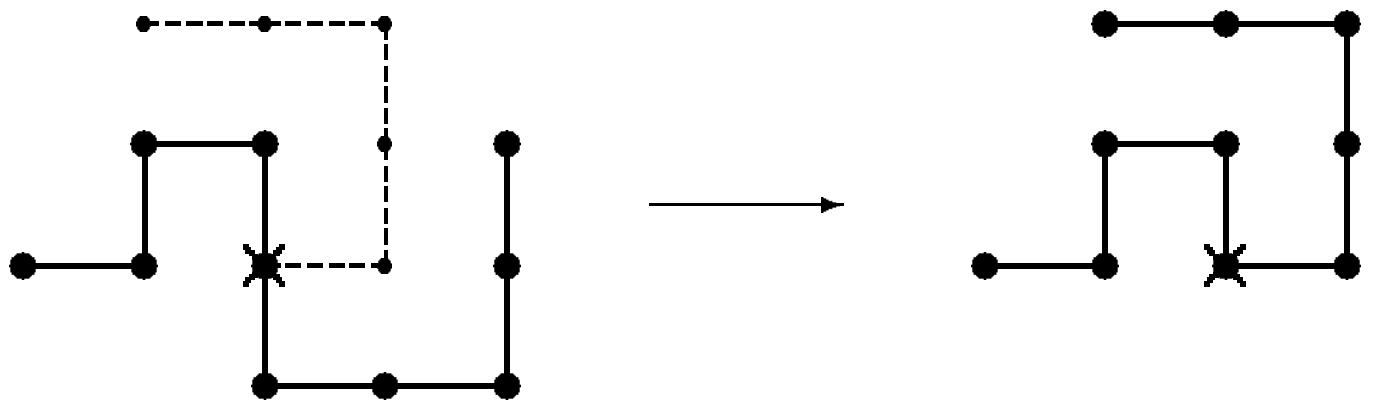

Figure 7: The pivot move (here a $+90°$ rotation). The pivot site is indicated with an $\times$. Dashed lines indicate the proposed new segment.

#### 6.3.3 Non-Local $N$-Conserving Moves

Here we move definitively out of the realm of systematic classification and into the realm of ingenuity. The possibilities for non-local moves are almost endless, but it is very difficult to find one which is *useful* in a Monte Carlo algorithm. There are two reasons for this: Firstly, since a non-local move is very radical, the proposed new walk usually violates the self-avoidance constraint. (If you move a large number of beads around, it becomes very likely that *somewhere* along the walk two beads will collide.) It is therefore a non-trivial problem to invent a non-local move whose acceptance probability does not go to zero too rapidly as $N \to \infty$. Secondly, a non-local move usually costs a CPU time of order $N$ (or in any case $N^p$ with $p > 0$), in contrast to order 1 for a local or bilocal move. It is non-trivial to find moves whose effects justify this expenditure (by reducing $\tau_{int,A}$ more than they increase $T_{CPU}$).

The following paragraphs are, therefore, nothing more than a brief listing of those non-local moves which, *as of 1993*, have been demonstrated to be useful for Monte Carlo algorithms in at least one of the SAW ensembles. Tomorrow someone could invent a new and even better non-local move; I hope that some reader of this essay will be moved (pardon the pun) to do so. This is a wide-open area of research.

Only two broad types of useful non-local moves are known at present: *pivot* moves, and *cut-and-paste* moves.

In a **pivot** move, we choose some site $\omega_k$ along the walk as a pivot point, and apply some symmetry operation of the lattice (e.g. rotation or reflection) to the part of the walk subsequent to the pivot point, using the pivot point as the origin [Figure 7]. That is, the proposed new walk is

$$\omega'_i = \begin{cases} \omega_i & \text{for } 0 \le i \le k \\ \omega_k + g(\omega_i - \omega_k) & \text{for } k+1 \le i \le N \end{cases} \tag{6.2}$$

where $g$ is the chosen symmetry operation.

In a **cut-and-paste** move, we cut the walk into two or more pieces, invert and/or reflect and/or permute the pieces, and finally reassemble the pieces. For example, one may invert the subwalk $\omega^{[k,l]}$ (Figure 8):

$$\omega'_i = \begin{cases} \omega_k + \omega_l - \omega_{k+l-i} & \text{for } k \le i \le l \\ \omega_i & \text{otherwise} \end{cases} \tag{6.3}$$

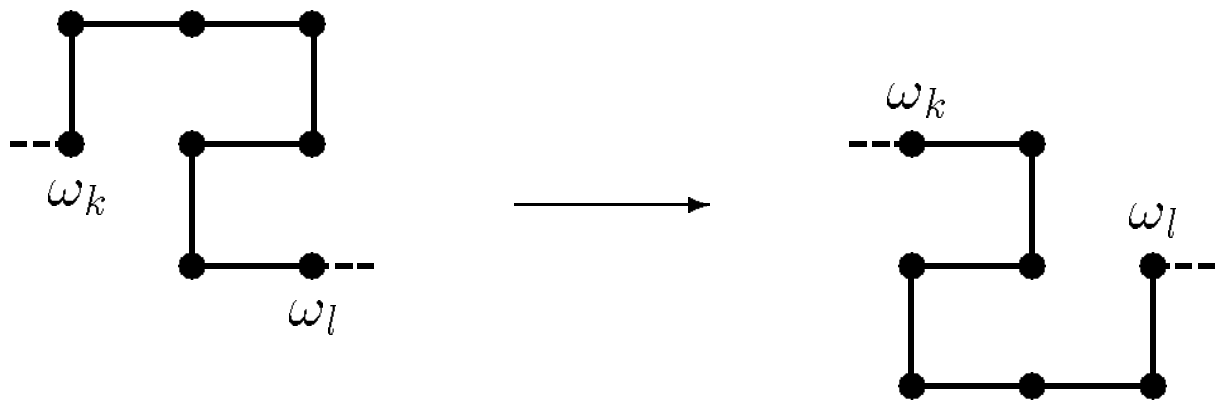

Figure 8: Inversion of the subwalk $\omega^{[k,l]}$.

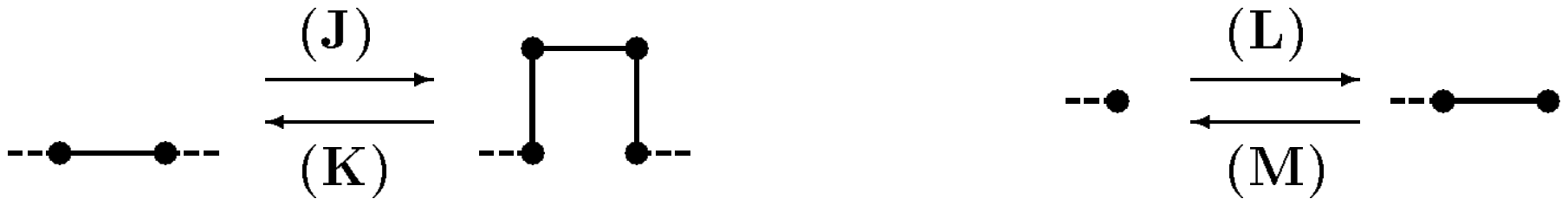

Figure 9: Some local $N$-changing moves. (J) Kink insertion ($\Delta N = +2$). (K) Kink deletion ($\Delta N = -2$). (L) End-bond addition ($\Delta N = +1$). (M) End-bond deletion ($\Delta N = -1$).

More generally, one may apply to $\omega^{[k,l]}$ any lattice symmetry operation that leaves $\omega_k$ and $\omega_l$ invariant; or any lattice symmetry operation that interchanges $\omega_k$ and $\omega_l$, followed by the inversion (6.3). Such symmetry operations exist if $\omega_l - \omega_k$ lies on certain axis or diagonal hyperplanes.

Empirically, the pivot and cut-and-paste moves have a reasonable acceptance probability even at large $N$, of order $\sim N^{-p}$ with $p$ typically of order 0.1–0.4. The heuristic reason is that these moves conserve most of the chain structure (and hence its self-avoidance), except near the pivot point or cut point(s). As a result, the acceptance probability for a move involving a *single* pivot or cut point is roughly similar to that encountered when concatenating two independent $(N/2)$-step walks, namely $c_N/c_{N/2}^2 \sim N^{-(\gamma-1)}$. By the same reasoning, a move with $n$ cut points might be expected to have an acceptance probability roughly like $\sim N^{-n(\gamma-1)}$.

### 6.3.4 Local $N$-Changing Moves

Recall that a local move is one that excises a small piece from the original SAW and splices in a new local conformation in its place. An **$N$-changing** local move has the freedom to splice in a piece with fewer or more sites than the original piece.

The simplest **internal** local $N$-changing moves are kink insertion and kink deletion (J and K in Figure 9); these have $\Delta N = +2$ and $\Delta N = -2$, respectively. The simplest **end-group** local $N$-changing moves are end-bond addition and end-bond deletion (L and M in Figure 9); these have $\Delta N = +1$ and $\Delta N = -1$, respectively.

Note that each local $N$-changing move is simply "one half" of some bilocal move.

| Scheme | References | Elementary Moves (see Figs. 1–2) | Autocorrelation Time $\tau$ (in elementary moves) |
|---|---|---|---|
| Verdier-Stockmayer | [106, 115, 116] | A,B | $\sim N^{\approx 3+2\nu}$ (?) |
| Modified Verdier-Stockmayer | [116, 109] | A,B,C | $\sim N^{\approx 3+2\nu}$ (?) |
| Heilmann II | [107, 108, 117] | A,B,E | $\sim N^{\approx 2+2\nu}$ (?) |
| Birshtein *et al.*/ Heilmann-Rotne 3 | [118, 108] | A,B,D | $\sim N^{\approx 2+2\nu}$ (?) |
| Taran-Stroganov | [119] | A,B,D,E | $\sim N^{\approx 2+2\nu}$ (?) |
| Verdier-Kranbuehl | [120, 121] | B,D,F | $\sim N^{\approx 2+2\nu}$ (?) |
| Kranbuehl-Verdier | [110, 122, 123] | A,B,D,F | $\sim N^{\approx 2+2\nu}$ (?) |
| Delbrück | [114] | most one- and two-bead | $\sim N^{\approx 2+2\nu}$ (?) |
| Meirovitch | [124] | most three-bead | $\sim N^{\approx 2+2\nu}$ (?) |
| Lal *et al.* | [126, 127] | all three-bead | $\sim N^{\approx 2+2\nu}$ (?) |
| Gény-Monnerie/ Kremer *et al.* | [128, 129, 130, 131] | some two- and three-bead (tetrahedral lattice) | $\sim N^{\approx 2+2\nu}$ (?) |

Table 4: Some local $N$-conserving algorithms.

## 6.4 Fixed-$N$, Variable-$x$ Algorithms

### 6.4.1 Local Algorithms

Historically the earliest dynamic Monte Carlo algorithms for the SAW were local $N$-conserving algorithms: they date back to the work of Delbrück [114] and Verdier and Stockmayer [106], both published in 1962. During the subsequent two decades, numerous variants on this theme were proposed (see Table 4). Most of these algorithms employ some subset of moves A–F from Figures 1 and 2.

Unfortunately, all local $N$-conserving algorithms have a fatal flaw: they are *nonergodic*. The nonergodicity of the Verdier-Stockmayer algorithm was noticed as early as 1968 [107, 115], but this fact did not seem to deter usage of the algorithm, or to provoke serious discussion about the ergodicity or nonergodicity of related algorithms (one exception is [108]). Finally, in 1987 Madras and Sokal [125] proved that *all* local $N$-conserving algorithms for SAWs in dimensions $d = 2, 3$ are nonergodic for sufficiently large $N$.[37] Furthermore, they proved that for large $N$, each ergodic class contains only an *exponentially small fraction* of the SAW configuration space. (These results are probably true also in dimension $d \geq 4$, but have not yet been proven.)

This nonergodicity is in fact quite easy to see: Consider the **double cul-de-sac** configuration shown in Figure 10(a). This SAW is completely "frozen" under elementary moves A, B, D and F: it cannot transform itself into any other state, nor can it be reached from any other state. It follows that the original Verdier-Stockmayer algorithm [106, 115] and most of its generalizations [107, 108, 117, 118, 119, 120, 110, 122, 123] are nonergodic (in $d = 2$) already for $N = 11$. If move C is allowed, then the configuration of Figure 10(a) is no longer frozen, but that of Figure 10(b) still is. Thus, any algorithm based on moves A–D and F is nonergodic (in $d = 2$) for $N = 15$. If two-bead end-group moves are allowed, then the configuration of Figure 10(b) is no longer frozen, but that of Figure 10(c) still is. Thus, any algorithm based on one-bead and two-bead moves is nonergodic (in $d = 2$) for $N = 19$.

[37]For algorithms based on moves of $k$ or fewer beads, nonergodicity arises in dimension $d = 2$ for all $N \geq 16k + 63$, and for quite a few smaller $N$ as well [125, Theorem 1].

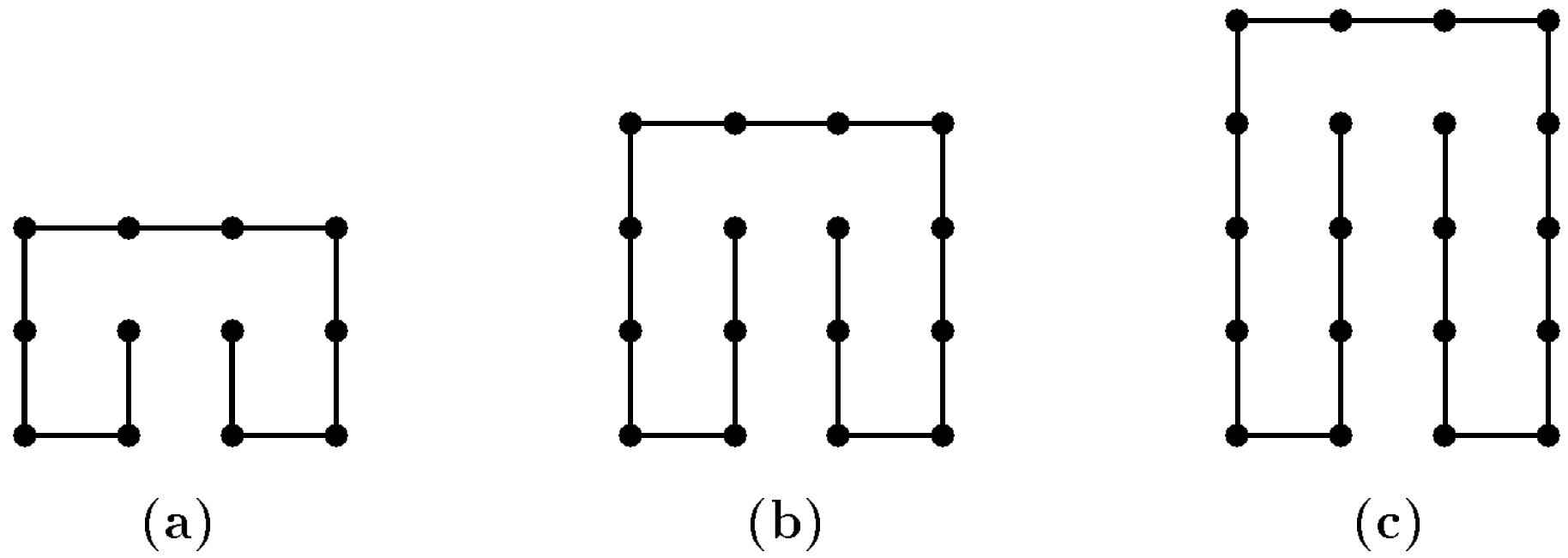

Figure 10: Some "double cul-de-sac" configurations, frozen in the Verdier-Stockmayer algorithm and its generalizations.

When three-bead moves are allowed, it is not sufficient simply to make the double cul-de-sac taller. Indeed, *any* double cul-de-sac of the kind shown in Figure 10, no matter how tall, can be unfolded into a straight rod by repeated use of the moves A, B and H. (The reader might find it amusing to work out the required sequence of moves.) But only one additional trick is needed: by folding the double cul-de-sac once more, as in Figure 11, a frozen configuration can be obtained for the $k$-bead algorithm for arbitrary $k$. This is the idea behind the Madras-Sokal proof.

The nonergodicity of the Verdier-Stockmayer algorithm due to double culs-de-sac was noticed already by Verdier [115] in 1969.

An entirely different type of nonergodicity arises in dimension $d = 3$ (and only there) because of the possibility of **knots**, as was first pointed out by Heilmann [107] in 1968. The simplest knotted configuration[38] is shown in Figure 12: it has $N = 18$, and although it is not completely "frozen", it nevertheless cannot be deformed to a straight rod using moves A–F. It is likely that analogous knots can be constructed for the $k$-bead algorithm for arbitrary $k$.

For additional historical discussion, along with extensive discussion of the practical implications of nonergodicity, see [125, Sections 3 and 4].

Since the local $N$-conserving algorithms are nonergodic, their autocorrelation times $\tau_{exp}$ and $\tau_{int,A}$ are by definition $+\infty$: the simulation *never* converges to the correct ensemble average. Nevertheless, we can imagine starting the simulation in some particular configuration (e.g. a straight rod), and ask how long it takes to ex-

[38]This is not a knot in the true topological sense, since true knots can occur only for *closed* walks (ring polymers). We are therefore here using the word "knot" in a loose sense to describe the general shape of the chain.

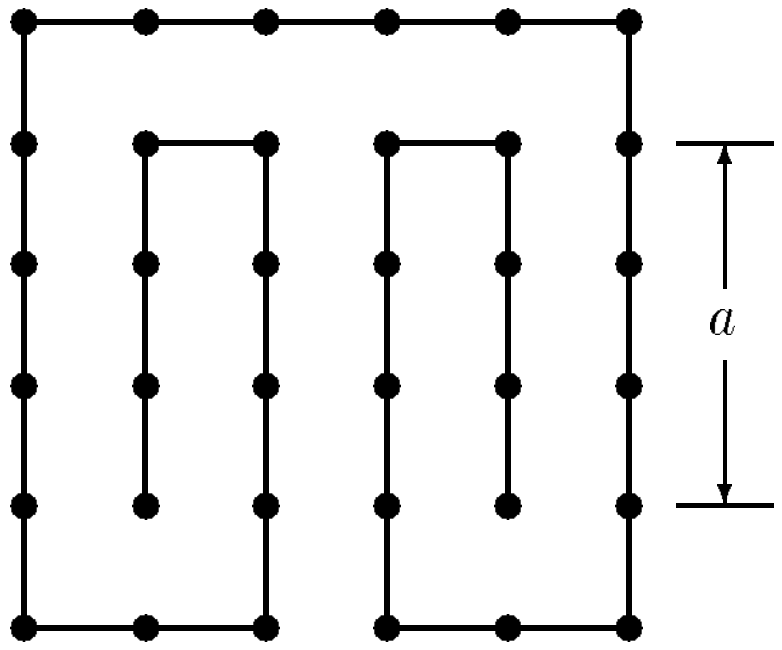

Figure 11: A "folded double cul-de-sac" configuration, drawn here for $a = 3$. Such configurations are frozen under under all $k$-bead local $N$-conserving moves with $k \leq a$.

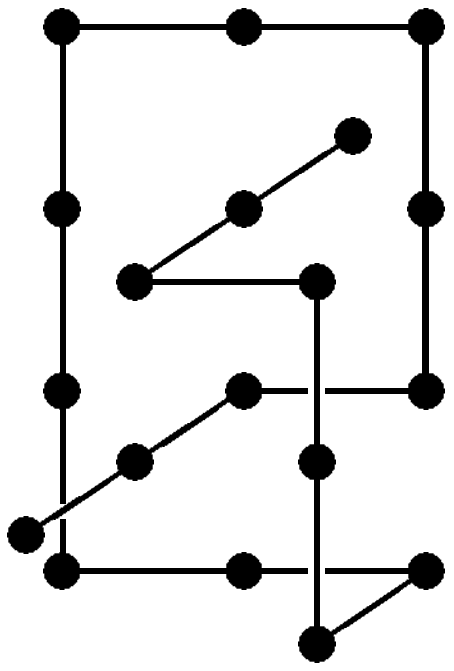

Figure 12: A "knot" that cannot be deformed to a straight rod using moves A–F.

plore adequately the ergodic class of that particular configuration. Here is a heuristic argument [71] giving a crude estimate of $\tau$ for such a restricted simulation:

Let us consider, as a typical quantity, the evolution of the squared radius of gyration of the chain. At each elementary move, a few beads of the chain move a distance of order 1; it follows from (2.4c) that the change in $R_g^2$ is of order $N^{\nu-1}$. In order to traverse its equilibrium distribution, $R_g^2$ must change by something of order its standard deviation, which is $\sim N^{2\nu}$. Assuming that $R_g^2$ executes a random walk, it takes about $(N^{2\nu}/N^{\nu-1})^2 \sim N^{2+2\nu}$ elementary moves for this to occur. So we predict $\tau \sim N^{2+2\nu}$.[39] This basic estimate ought to be applicable to the dynamics of ordinary random walks (ORWs, free-flight chains) and non-reversal random walks (NRRWs) as well as SAWs.

This reasoning can easily be converted into a more-or-less rigorous proof of the *lower bound* $\tau_{exp} \gtrsim \tau_{int,R_g^2} \geq \text{const} \times N^{2+2\nu}$ (see [132, p. 51] for a slight variant). However, the true $\tau$ could be significantly *larger* than this estimate if there are modes which relax essentially more slowly (i.e. with a larger dynamic critical exponent) than the radius of gyration. It also could be wrong if there are, in a particular model, special conservation laws or quasi-conservation laws which inhibit the relaxation of the radius of gyration; this indeed can occur (see the next paragraph). Finally, it is not clear whether the result $\tau \sim N^{2+2\nu}$ is exact — and if so, for *which* $\tau$ — or is merely a reasonable first approximation. Usually dynamic critical exponents are *not* expressible in terms of static ones, except for trivial (Gaussian-like) models. It is thus probable that $\tau \sim N^{2+2\nu}$ is not exact for the SAW, although it may be exact for the NRRW; it is known to be exact for the ORW, at least in the Verdier-Stockmayer dynamics [133, 134, 135, 136, 137, 138, 139, 140]. The numerical evidence is inconclusive (see references in [71]).

The Verdier-Stockmayer (pure one-bead) and Kranbuehl-Verdier (pure two-bead) algorithms for the SAW or NRRW (but not the ORW) have peculiar conservation laws which inhibit the relaxation of the chain, thereby increasing the relaxation time above the usual $N^{2+2\nu}$ by at least an extra factor of $N$ [116, 121].[40] These conservation laws can be broken by mixing one-bead and two-bead moves.

---

[39] The same order of magnitude is obtained if one considers the evolution of the end-to-end distance vector, the vector from the center-of-mass to an endpoint, or any similar global quantity.

[40] Hilhorst, Deutch and Boots [116, 121] have given a very beautiful analysis of these dynamics, using a mapping onto a gas of one-dimensional random walkers interacting via exclusion. It would be interesting to reexamine these questions using mathematical results on this latter model [141, 142, 143]; it might be possible to compute the exact asymptotic behavior, at least in the NRRW case.

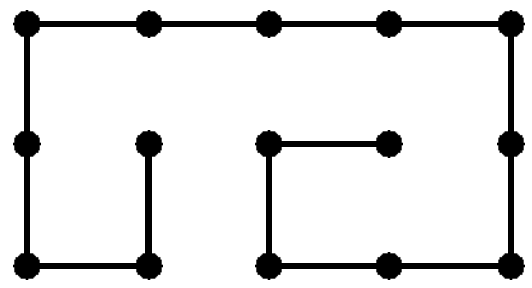

Figure 13: A "double cul-de-sac" configuration which is frozen in the reptation algorithm.

### 6.4.2 Bilocal Algorithms

The **slithering-snake (reptation) algorithm**[41] was invented independently by Kron [144, 145] and Wall and Mandel [146, 147].[42] The elementary moves of this algorithm are "slithering" motions: one step is appended at one end of the walk, and one step is simultaneously deleted from the other end (Figure 4). Such a move is $N$-conserving and *bi*local (but not local).

Unfortunately, the reptation algorithm is *nonergodic*, as was observed already by the original inventors [145, 146]. In this algorithm, frozen configurations occur when both ends of the walk are trapped in culs-de-sac. The simplest such configuration is shown (for $d = 2$) in Figure 13, and has $N = 14$.[43]

As with the local $N$-conserving algorithms, we can inquire about the autocorrelation times of a simulation *within* a particular ergodic class. A plausible heuristic argument [147] suggests that $\tau \sim N^2$. This is because the SAW transforms itself by random back-and-forth slithering along the chain; after $\sim N^2$ moves, this slithering will have carried it $N$ steps, and thus all the original bonds of the chain will have disappeared and been replaced by random new ones. (Alternatively, $R_g^2$ changes by an amount of order $N^{2\nu-1}$ per elementary move; thus, it takes $\sim N^2$ elementary moves for $R_g^2$ to traverse a distribution of width $\sim N^{2\nu}$.) It is easy to see that this argument is exact for ORWs and NRRWs; it is not clear whether it is exact or merely approximate for SAWs.

[41]The "slithering-snake (reptation) *algorithm*" should not be confused with the "reptation *conjecture*" of de Gennes [5, Section VIII.2], who argues that the *most rapid* modes of polymer diffusion in the true *physical* dynamics (in a melt consisting of many entangled polymer chains) are the slithering modes.

[42]There are at least three variants of the slithering-snake scheme. In the first version of Wall and Mandel [146, pp. 4592–4593], the "head" and "tail" of the chain are interchanged only when an attempted move is rejected; this satisfies the stationarity condition but not the detailed-balance condition. (In fact, one step of the algorithm *followed by a head-tail interchange* satisfies detailed balance [148]. This is analogous to the situation in the Hamilton/Langevin hybrid algorithm for spin models [37, p. 237, footnote 18].) In the second version of Wall and Mandel [146, top of p. 4594], adopted also by Webman *et al.* [149], the "head" and "tail" are chosen anew at each cycle, randomly with 50%–50% probability; this satisfies detailed balance. The original algorithm proposed by Kron [144] is much more complicated than either of the preceding versions; it appears to satisfy stationarity but not detailed balance.

[43]The superficial resemblance between Figures 10–11 and 13 is, however, very misleading: the nonergodicities in the two types of algorithm are of radically different natures. In the local $N$-conserving algorithms, nonergodicity is caused by the occurrence of a frozen conformation *anywhere* along the walk. In the reptation algorithm, by contrast, nonergodicity occurs only if both of the *ends* of the walk are trapped in culs-de-sac. It turns out that the ergodic class of the straight rod in the reptation algorithm contains at least a fraction $N^{-(\gamma-1)/2}$ of the SAW configuration space, whereas in the local algorithms it contains only an exponentially small fraction. See [125, Section 3] for further discussion.

The nonergodicity of the reptation algorithm can be cured by adjoining additional bilocal moves to the algorithm [150]. Indeed, the following bilocal (or mixed local/bilocal) algorithms are known to be ergodic:

- reptation + kink-end and end-kink reptation [150]
- kink-end and end-kink reptation + 90° end-bond rotation [148]
- reptation + kink transport + one-bead flip, at least in dimension $d = 2$ [151]
- kink transport + one-bead flip + 90° and 180° end-bond rotations, at least in $d = 2$ [151]

All of these algorithms deserve systematic study, in particular of their dynamic critical behavior. One would like to know whether they all lie in the same dynamic universality class, and whether the conjecture $\tau \sim N^2$ is exact, approximate or wrong.

An alternative way of making the reptation algorithm ergodic is to switch to the variable-$N$ ensemble and introduce $\Delta N = \pm 1$ moves (L and M in Figure 9), as proposed by Kron *et al.* [145]. But once one does this, there is no great reason to retain the "slithering-snake" moves; one can just as well use *only* the $\Delta N = \pm 1$ moves. This leads to the slithering-tortoise (Berretti-Sokal) algorithm (see Section 6.6.1).

### 6.4.3 Pivot Algorithm

The **pivot algorithm** has a curious history. It was invented by Lal [152] in 1969, and then promptly forgotten by almost everyone.[44] It was reinvented in 1985 by MacDonald *et al.* [161, 162, 163], and reinvented a short time later by Madras. This third reinvention led to a comprehensive analytical and numerical study by Madras and Sokal [96], which showed that the pivot algorithm is *by far the most efficient algorithm yet invented* for simulating the fixed-$N$, variable-$x$ SAW ensemble. After that the applications took off.[45]

The elementary move of the pivot algorithm is as follows: Choose at random a pivot point $k$ along the walk ($0 \leq k \leq N-1$); choose at random an element $g$ of the symmetry group of the lattice (rotation or reflection or a combination thereof); then apply $g$ to the part of the walk subsequent to the pivot point (namely $\omega_{k+1}, \ldots, \omega_N$), using $\omega_k$ as the temporary "origin" [cf. (6.2)]. The resulting walk is accepted if it is self-avoiding; otherwise, it is rejected, and the old walk $\omega$ is counted once more in the sample. It is easy to see that this algorithm satisfies detailed balance for the standard equal-weight SAW distribution. Ergodicity is less obvious, but it can be proven [96, 179].

[44] The only exceptions I know of are Olaj and Pelinka [153] and Clark and Lal [154]. Continuum analogues of the pivot algorithm were used by Stellman and Gans [155, 156], Freire and Horta [157], Curro [158, 159], Scott [160] and possibly others. For additional history, see [96, footnote 3].

[45] Here is a very incomplete list: Caracciolo, Madras, Pelissetto, Sokal and collaborators [132, 164, 165, 166, 39], Zifferer [167, 168, 169, 170, 171, 172], Bishop and collaborators [173, 174, 175, 176], Chorin [177], Eizenberg and Klafter [178].

At first thought this seems to be a terrible algorithm: for $N$ large, nearly all the proposed moves will get rejected. In fact, this latter statement is true, but the hasty conclusion drawn from it is radically false! The acceptance fraction $f$ does indeed go to zero as $N \to \infty$, roughly like $N^{-p}$; empirically, it is found that the exponent $p$ is $\approx 0.19$ in $d = 2$ [96] and $\approx 0.11$ in $d = 3$ [96, 167, 178]. But this means that roughly once every $N^p$ moves one gets an acceptance. And the pivot moves are very radical: one might surmise that after very few accepted moves (say, 5 or 10) the SAW will have reached an "essentially new" configuration. One conjectures, therefore, that the autocorrelation time $\tau$ of the pivot algorithm behaves as $\sim N^p$. Things are in fact somewhat more subtle (see below), but roughly speaking (and modulo a possible logarithm) this conjecture appears to be true. On the other hand, a careful analysis of the computational complexity of the pivot algorithm (see also below) shows that one accepted move can be produced in a computer time of order $N$. Combining these two facts, we conclude that one "effectively independent" sample (at least as regards *global* observables) can be produced in a computer time of order $N$ (or perhaps $N \log N$). This is vastly better than the $N^{2+2\nu} \sim N^{\approx 3.2}$ of the local $N$-conserving algorithms and the $N^{\approx 2}$ of the bilocal algorithms. Indeed, this order of efficiency cannot be surpassed by any algorithm which computes each site on successive SAWs, for it takes a time of order $N$ simply to *write down* an $N$-step walk!

Let us mention briefly some relevant issues; the reader is referred to [96] for a fuller discussion.

*Variants of the pivot algorithm.* Different variants of the pivot algorithm are obtained by specifying different distributions when we "choose at random":

1) The pivot point $k$ can be chosen according to any pre-set family of strictly positive probabilities $p_0, \ldots, p_{N-1}$. The strict positivity ($p_k > 0$ for all $k$) is necessary to ensure the ergodicity of the algorithm. Most work has used a uniform distribution, but there could be some (probably minor) advantage in using a non-uniform distribution.

2) Let $G$ be the group of orthogonal transformations (about the origin) which leave invariant the lattice $\mathbb{Z}^d$. Then the symmetry operation $g \in G$ can be chosen according to any pre-set probability distribution $\{p_g\}_{g \in G}$ which satisfies $p_g = p_{g^{-1}}$ for all $g$, and which has enough nonzero entries to ensure ergodicity (see below). The condition $p_g = p_{g^{-1}}$ is easily seen to be both necessary and sufficient to ensure detailed balance with respect to the equal-weight distribution $\pi$.

In dimension $d = 2$, $G$ is the dihedral group $D_4$, which has eight elements: the identity, two 90° rotations, the 180° rotation, two axis reflections, and two diagonal reflections. Detailed balance holds provided that $p_{+90^\circ\, rot} = p_{-90^\circ\, rot}$. Ergodicity holds if the probabilities $p_g$ are nonzero for either

- both diagonal reflections [179]; or

- $\pm 90°$ rotations and the $180°$ rotation [96]; or
- $\pm 90°$ rotations and both axis reflections [96].

Most work has used a uniform distribution over the seven non-identity operations, but some gain could probably be achieved by using a non-uniform distribution.

In general dimension $d$, the symmetry group of $\mathbb{Z}^d$ is the group $O(d, \mathbb{Z})$ of orthogonal matrices with integer entries. In fact every such matrix is of the form

$$g_{ij} \;=\; \sigma_i \delta_{i,\pi(j)} \;, \tag{6.4}$$

where $\sigma_1, \ldots, \sigma_d = \pm 1$ and $\pi$ is a permutation of $\{1, \ldots, d\}$. Using this description, the pivot algorithm can be programmed very easily in any dimension.

*Acceptance fraction and autocorrelation time.* Suppose we know that the acceptance fraction $f$ in the pivot algorithm behaves as $f \sim N^{-p}$ as $N \to \infty$. Then, as argued above, after a few successful pivots — i.e. a time of order $1/f \sim N^p$ — the *global* conformation of the walk should have reached an "essentially new" state. Thus, we expect that for observables $A$ which measure the *global* properties of the walk — such as the squared end-to-end distance $R_e^2$ or the squared radius of gyration $R_g^2$ — the autocorrelation time $\tau_{int,A}$ should be a few times $1/f$. This is confirmed numerically [96, Section 4.3]. On the other hand, it is important to recognize that *local* observables — such as the angle between the $17^{th}$ and $18^{th}$ steps of the walk — may evolve a factor of $N$ more slowly than global observables. For example, the observable mentioned in the preceding sentence changes only when $\omega_{17}$ serves as a successful pivot point; and this happens, on average, only once every $N/f$ attempted moves. Thus, for *local* observables $A$ we expect $\tau_{int,A}$ to be of order $N/f$. By (3.39), $\tau_{exp}$ must be of at least this order; and if we have not overlooked any slow modes in the system, then $\tau_{exp}$ should be of exactly this order. Finally, even the global observables are unlikely to be precisely orthogonal [in $l^2(\pi)$] to the slowest mode; so it is reasonable to expect that $\tau_{exp,A}$ be of order $N/f$ for these observables too. In other words, for global observables $A$ we expect the autocorrelation function $\rho_{AA}(t)$ to have an extremely-slowly-decaying tail which, however, contributes little to the area under the curve. This behavior is illustrated by the exact solution of the pivot dynamics for the case of ordinary random walk [96, Section 3.3], and by numerical calculations for the SAW.

The foregoing heuristic argument is, of course, far from a rigorous proof. It is not in general possible to find upper bounds on the autocorrelation time in terms of the acceptance fraction; the problem is that the state space could contain "bottlenecks" through which passage is unusually difficult. No one has suggested any reason why such bottlenecks should occur in the pivot algorithm, but neither does there exist any proof of their nonexistence.

A heuristic argument [96, Section 3.2] suggests that $f \sim N^{-p}$ with $p = \gamma - 1$ ($=$ $11/32$ in $d = 2$, $\approx 0.16$ in $d = 3$, and 0 in $d \geq 4$). Quantitatively this prediction is incorrect; numerical experiments yield $p \approx 0.19$ in $d = 2$ [96] and $p \approx 0.11$ in $d = 3$ [96, 167, 178]. However, the argument does correctly predict that $p$ is small and that it gets smaller in higher dimension. It is an open question to find a better theoretical prediction for $p$, and in particular to express it in terms of static exponents for the SAW.

*Computational complexity.* A very important issue in any algorithm — but especially in a non-local one — is the CPU time per iteration. By using a hash table (see Section 7.1.2), the self-avoidance of a proposed new walk can be checked in a time of order $N$. But one can do even better: by starting the checking at the pivot point and working outwards, *failures* can be detected in a mean time of order $N^{1-p}$ [96, Sections 3.4 and 4.4]. The mean CPU time *per successful pivot* is therefore $\sim N^{1-p}$ for each of $\sim N^p$ failures, plus $\sim N$ for one success, or $\sim N$ in all. Combining this with the observations made previously, we conclude that one "effectively independent" sample — as regards *global* observables — can be produced in a computer time of order $N$.

*Initialization.* There are two main approaches:

1) *Equilibrium start.* Generate the initial configuration by dimerization (Section 4.3); then the Markov chain is in equilibrium from the beginning, and no data need be discarded. This approach is feasible (and recommended) at least up to $N$ of order a few thousand. There is no harm in spending even days of CPU time on this step, provided that this time is small compared to the rest of the run; after all, the algorithm need only be initialized once.

2) *"Thermalization".* Start in an arbitrary initial configuration, and then discard the first $n_{disc} \gg \tau_{exp} \sim N/f$ iterations. This is painful, because $\tau_{exp}$ is a factor $\sim N$ larger than $\tau_{int,A}$ for global observables $A$; thus, for very large $N$ ($\gtrsim 10^5$), the CPU time of the algorithm could end up being dominated by the thermalization. Nevertheless, one must resist the temptation to cut corners here, as even a small initialization bias can lead to systematically erroneous results, especially if the statistical error is small [39]. Some modest gain can probably be obtained by using closer-to-equilibrium initial configurations (e.g. [168]), but it is still prudent to take $n_{disc}$ at least several times $N/f$.

Initialization will become a more important issue in the future, as faster computers permit simulations at ever-larger chain lengths.

## 6.5 Fixed-$N$, Fixed-$x$ Algorithms

This ensemble has not been considered until very recently; it seems difficult to devise algorithms which are ergodic under the severe constraints of fixed chain length and fixed endpoints.

### 6.5.1 Bilocal Algorithms

I do not know whether there are any bilocal (or mixed local/bilocal) algorithms that are ergodic for this ensemble.

### 6.5.2 Cut-and-Paste Algorithms

These algorithms were devised by Dubins, Madras, Orlitsky, Reeds and Shepp [180, 179]. In dimension $d = 2$ the simplest cut-and-paste algorithm uses two moves:

- Inversion of the subwalk $\omega^{[k,l]}$ (Figure 8).
- If $\omega_l - \omega_k$ lies on a $\pm 45°$ diagonal line, then one can reflect the subwalk $\omega^{[k,l]}$ through the perpendicular bisector of the segment $\omega_k \omega_l$, followed by an inversion.

This algorithm is ergodic [179]. In dimension $d \geq 3$ one must adjoin coordinate-interchange moves, and the algorithm is then again ergodic [179].

The dynamic critical behavior of this algorithm is unclear. One's first guess is a behavior similar to that of the pivot algorithm, i.e. an acceptance fraction $f \sim N^{-q}$ for some small power $q$, and an autocorrelation time $\tau_{int,A} \sim N^q$ for *global* observables. However, one should remember that the diagonal-reflection moves — which are *necessary* for ergodicity[46] — are possible only for a fraction $\sim N^{-\nu}$ of pairs $k, l$. So perhaps one should expect $\tau_{int,A} \sim N^{q+\nu}$. (Numerically, this might be seen most easily by looking at an observable which is sensitive to the diagonal-reflection moves, e.g. the numbers $N_1, \ldots, N_d$ of steps along each axis.) On the other hand, failures of diagonal reflection due to disallowed $\omega_l - \omega_k$ can be detected in a CPU time of order 1; so such failures may not affect the scaling of the product $\tau_{int,A} T_{CPU}$.

See [181] for a first study of the cut-and-paste algorithm (on the face-centered-cubic lattice).

## 6.6 Variable-$N$, Variable-$x$ Algorithms

### 6.6.1 Slithering-Tortoise (Berretti-Sokal) Algorithm

The **slithering-tortoise algorithm**[47] [182] is a Markov chain with state space $\mathcal{S} \equiv \bigcup_{N=0}^{\infty} \mathcal{S}_N$ and invariant probability distribution (2.17) with $p = 0$, i.e.

$$\pi_\beta(\omega) \;=\; \text{const} \times \beta^{|\omega|} \,. \tag{6.5}$$

The algorithm's elementary moves are as follows: either one attempts to append a new step to the walk, with equal probability in each of the $2d$ possible directions; or else one deletes the last step from the walk. In the former case, one must check that the proposed new walk is self-avoiding; if it isn't, then the attempted move is rejected and the old configuration is counted again in the sample ("null transition"). If an attempt is made to delete a step from an already-empty walk, then a null transition

[46] Without them, the numbers $N_1, \ldots, N_d$ of steps along each axis would be *separately* conserved.

[47] Because the walk sticks out and retracts its "head", like a tortoise. This algorithm is also known as the Berretti-Sokal algorithm, or BS for short.

is also made. The relative probabilities of $\Delta N = +1$ and $\Delta N = -1$ attempts are chosen to be

$$P(\Delta N = +1 \text{ attempt}) = \frac{2d\beta}{1+2d\beta} \tag{6.6}$$

$$P(\Delta N = -1 \text{ attempt}) = \frac{1}{1+2d\beta} \tag{6.7}$$

It is easily verified that this transition matrix satisfies detailed balance for the distribution $\pi_\beta$. It is also easy to see that the algorithm is ergodic: to get from a walk $\omega$ to another walk $\omega'$, it suffices to use $\Delta N = -1$ moves to transform $\omega$ into the empty walk, and then use $\Delta N = +1$ moves to build up the walk $\omega'$.

Regarding the critical slowing-down of the slithering-tortoise algorithm, we can argue heuristically that

$$\tau \sim \langle N \rangle^2 . \tag{6.8}$$

To see this, consider the quantity $N(t) = |\omega|(t)$, the number of steps in the walk at time $t$. This quantity executes, crudely speaking, a random walk (with drift) on the nonnegative integers; the average time to go from some point $N$ to the point 0 (i.e. the empty walk) is of order $N^2$. Now, each time the empty walk is reached, all memory of the past is erased; future walks are then independent of past ones. Thus, the autocorrelation time ought to be of order $\langle N^2 \rangle$, or equivalently $\langle N \rangle^2$.

This heuristic argument can be turned into a rigorous proof of a *lower bound* $\tau_{exp} \gtrsim \tau_{int,N} \geq \text{const} \times \langle N \rangle^2$ [69]. However, as an argument for an *upper bound* of the same form, it is not entirely convincing, as it assumes without proof that the *slowest* mode is the one represented by $N(t)$. With considerably more work, it is possible to prove an upper bound on $\tau$ that is only slightly weaker than the heuristic prediction: $\tau_{exp} \leq \text{const} \times \langle N \rangle^{1+\gamma}$ [69, 183, 184]. (Note that the critical exponent $\gamma$ is believed to equal 43/32 in dimension $d = 2$, $\approx$1.16 in $d = 3$, and 1 in $d \geq 4$.) In fact, there is reason to believe [185] that the true behavior is $\tau \sim \langle N \rangle^p$ for some exponent $p$ *strictly between* 2 and $1+\gamma$. A deeper understanding of the dynamic critical behavior of the slithering-tortoise algorithm would be of definite value.

### 6.6.2 Join-and-Cut Algorithm

The behavior $\tau \gtrsim \langle N \rangle^2$ of the slithering-tortoise algorithm is in fact characteristic of *any* variable-$N$ algorithm whose elementary moves make *bounded* changes in $N$: roughly speaking, $N$ must perform a random walk on the nonnegative integers, and the autocorrelation time of such a random walk satisfies

$$\tau \gtrsim \text{var}(N) \equiv \langle N^2 \rangle - \langle N \rangle^2 \sim \langle N \rangle^2 \tag{6.9}$$

[132, Theorems A.6 and A.7]. Therefore, if one wants to do better than $\tau \gtrsim \langle N \rangle^2$, it is necessary to make *unbounded* changes in $N$.

An amazingly simple way of doing this was proposed recently by Caracciolo, Pelissetto and Sokal [165]. Their algorithm works in the unorthodox ensemble $\mathcal{T}_{N_{tot}}$ consisting of all *pairs* of SAWs $(\omega^{(1)}, \omega^{(2)})$ [each walk starts at the origin and ends anywhere] such that the *total* number of steps in the two walks is some fixed number $N_{tot}$:

$$\begin{aligned} \mathcal{T}_{N_{tot}} &\equiv \{(\omega^{(1)}, \omega^{(2)})\colon\ \omega^{(1)}, \omega^{(2)} \text{ are self-avoiding, with } |\omega^{(1)}| + |\omega^{(2)}| = N_{tot}\} \\ &\equiv \bigcup_{N_1=0}^{N_{tot}} (\mathcal{S}_{N_1} \times \mathcal{S}_{N_{tot}-N_1}) \ . \end{aligned} \tag{6.10}$$

Each pair $(\omega^{(1)}, \omega^{(2)})$ in this ensemble is given equal weight; therefore, the two walks are non-interacting except for the constraint on the sum of their lengths. One sweep of the algorithm consists of two steps:

(a) We update independently each of the two walks, using some $N$-conserving ergodic algorithm (e.g. the pivot algorithm).

(b) We perform a **join-and-cut** move: we concatenate the two walks $\omega^{(1)}$ and $\omega^{(2)}$, forming a new (not necessarily self-avoiding) walk $\omega^{(1)} \circ \omega^{(2)}$; then we cut $\omega^{(1)} \circ \omega^{(2)}$ at a random position, creating two new walks $\omega'^{(1)}$ and $\omega'^{(2)}$. If $\omega'^{(1)}$ and $\omega'^{(2)}$ are both self-avoiding, we keep them; otherwise the move is rejected and we stay with $\omega^{(1)}$ and $\omega^{(2)}$.

The join-and-cut move is illustrated in Figure 14.

Since the algorithm used in step (a) is ergodic in the ensemble of fixed-length walks, it is easy to see that the full algorithm is ergodic. (If $\omega^{(1)}$ and $\omega^{(2)}$ are perpendicular rods, then the join-and-cut move will always succeed.) It is also easy to see that the algorithm satisfies the detailed-balance condition with respect to the equal-weight measure

$$\pi(\omega^{(1)}, \omega^{(2)}) \;=\; \frac{1}{Z(N_{tot})} \qquad \text{for each } (\omega^{(1)}, \omega^{(2)}) \in \mathcal{T}_{N_{tot}} \ , \tag{6.11}$$

where

$$Z(N_{tot}) \;=\; \sum_{N_1=0}^{N_{tot}} c_{N_1} c_{N_{tot}-N_1} \ . \tag{6.12}$$

Therefore, from the probability distribution of the random variable $N_1 \equiv |\omega^{(1)}|$ (the length of the first walk) in the measure (6.11), we can obtain estimates of the critical exponent $\gamma$ by the maximum-likelihood method (see Section 7.3). Since the join-and-cut move can make large jumps in $N_1$ in a single step, this evades the bound (6.9).

The dynamic critical behavior of the pivot + join-and-cut algorithm was studied in [165] by both analytical and numerical methods. For the relevant observable $X \equiv \log[N_1(N_{tot} - N_1)]$, the autocorrelation time in units of elementary moves is found to grow as $\tau_{int,X} \sim N_{tot}^{\approx 0.70}$ in $d = 2$. On the other hand, the mean CPU time per elementary move behaves as $T_{CPU} \sim N^{\approx 0.81}$, just as in the pivot algorithm. Hence the autocorrelation time in CPU units behaves as $\tau_{int,X} T_{CPU} \sim N^{\approx 1.51}$, which is a significant improvement over the $\tau \sim N^{\approx 2}$ of the slithering-tortoise algorithm. Moreover, the behavior is expected to be even better in higher dimensions.

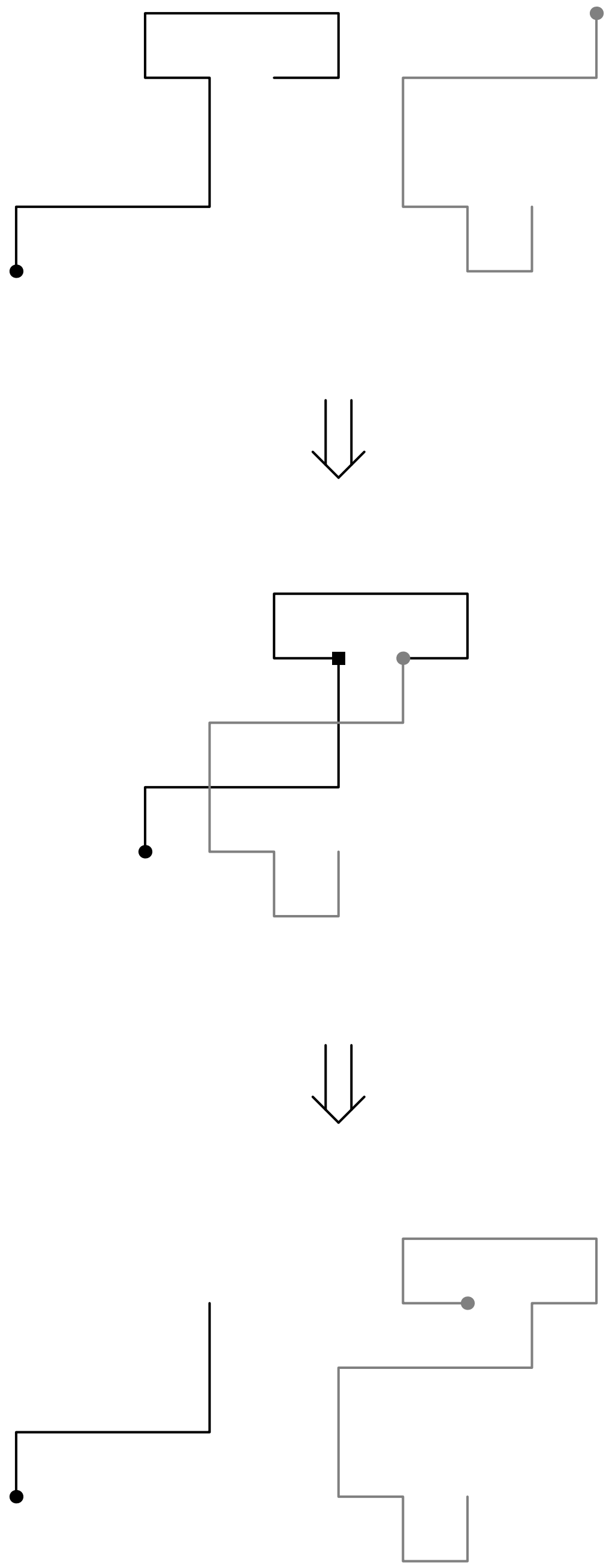

Figure 14: Join-and-cut move: the two SAWs (upper figure) are concatenated (middle figure) and then cut at the point marked with a square (lower figure). Note that the concatenated walk *need not* be self-avoiding.

## 6.7 Variable-$N$, Fixed-$x$ Algorithms

### 6.7.1 BFACF Algorithm

The BFACF algorithm was invented by Berg and Foerster [186] and Aragão de Carvalho, Caracciolo and Fröhlich [61, 187].[48] It is a Markov chain with state space $\mathcal{S}(x) \equiv \bigcup\limits_{N=0}^{\infty} \mathcal{S}_N(x)$ and invariant probability distribution (2.14) with $p = 1$, i.e.

$$\pi_\beta(\omega) \;=\; \text{const} \,\times\, |\omega| \beta^{|\omega|} \,. \tag{6.13}$$

The elementary moves of the BFACF algorithm are the following local deformations:

- The one-bead flip (move A of Figure 1), which has $\Delta N = 0$.
- Kink insertion (move J of Figure 9), which has $\Delta N = +2$.
- Kink deletion (move K of Figure 9), which has $\Delta N = -2$.

Note that all these moves can be generated by displacing the middle bond of a three-bond group by one lattice unit perpendicular to itself in one of the $2d - 2$ possible directions. Therefore, one iteration of the BFACF algorithm consists of the following operations:

1) Choose at random a bond of the current walk $\omega$ (with equal probability for each bond).

2) Enumerate the $2d - 2$ possible deformations of that bond; choose randomly among these deformations, giving each deformation a probability $p(\Delta N)$ depending only on $\Delta N \equiv |\omega'| - |\omega|$. (If the sum of these probabilities is $s < 1$, then make a "null transition" $\omega \to \omega$ with probability $1 - s$.) The probabilities $p(\Delta N)$ will be specified below.

3) Check whether the proposed new walk $\omega'$ is self-avoiding. If it is, keep it; otherwise, make a null transition.

This algorithm satisfies detailed balance for $\pi_\beta$ provided that

$$p(+2) \;=\; \beta^2 p(-2) \,. \tag{6.14}$$

On the other hand, the sum of the probabilities must in all cases be $\leq 1$; this imposes the inequalities [132]

$$[1 + (2d-3)\beta^2]\, p(+2) \;\leq\; \beta^2 \tag{6.15a}$$

$$2\,p(0) + (2d-4)p(+2) \;\leq\; 1 \tag{6.15b}$$

[48]The exposition in the original paper [187] suffers from an unfortunate confusion regarding the meaning of $p(\Delta N)$. Here I follow the corrected exposition in [132, Section 3.1].

A nearly optimal choice is

$$p(-2) \;=\; \frac{1}{1+(2d-3)\beta^2} \tag{6.16a}$$

$$p(0) \;=\; \frac{1+\beta^2}{2\,[1+(2d-3)\beta^2]} \tag{6.16b}$$

$$p(+2) \;=\; \frac{\beta^2}{1+(2d-3)\beta^2} \tag{6.16c}$$

(see [132] for further discussion).

The *ergodicity* of the BFACF algorithm is an extremely subtle problem. The known results are the following:

- In dimension $d = 2$, the BFACF algorithm is *ergodic*, for all choices of $x$ [11].
- In dimension $d = 3$, the algorithm is *non-ergodic* if $\|x\|_\infty \equiv \max(|x_1|, |x_2|, |x_3|) = 1$, due to knotted configurations which cannot be untied [188].[49] However, it can be made ergodic[50] by adjoining the three-bead crankshaft move (I of Figure 3) [188].
- In dimension $d = 3$, the algorithm is *ergodic* if $\|x\|_\infty \geq 2$ [189]. (It is then possible to disentangle any knot, no matter how large and complicated, by a motion which passes one strand at a time between the endpoints.)
- In dimension $d \geq 4$, the algorithm is presumably ergodic for all $x$, but a rigorous proof appears to be lacking.

The *dynamical behavior* of the BFACF algorithm is also an extremely subtle problem. The known results are the following:

- $\tau_{exp} = +\infty$ for *all* $\beta > 0$ [190]. This surprising result arises from the existence of arbitrarily slowly-relaxing modes associated with sequences of transitions $\omega \to \cdots \to \omega'$ with $\mathcal{A}(\omega, \omega') \gg \max(|\omega|, |\omega'|)$, where $\mathcal{A}(\omega, \omega')$ is the minimum surface area spanned by the union of $\omega$ and $\omega'$.
- $\tau_{int,\mathcal{A}} \geq \text{const} \times [\langle \mathcal{A}^2 \rangle - \langle \mathcal{A} \rangle^2]$, where $\mathcal{A} \equiv \mathcal{A}(\omega, \omega^\circ)$ for some fixed walk $\omega^\circ$ from 0 to $x$ [132]. Assuming the usual scaling behavior [191], this implies $\tau_{int,\mathcal{A}} \gtrsim \langle N \rangle^{4\nu}$.
- Numerically, it appears that $\tau_{int,N} \sim \langle N \rangle^p$ with $p$ equal or very nearly equal to $4\nu$ [191].

[49] For the BFACF algorithm applied to *unrooted polygons* (ring polymers), the ergodic classes are precisely the knot classes [188]. This is probably true also for the BFACF algorithm applied to *walks* with $\|x\|_\infty = 1$, but it has apparently not yet been proven.

[50] At least in the case of unrooted polygons. It is probably true also for the BFACF algorithm applied to *walks* with $\|x\|_\infty = 1$, but it has apparently not yet been proven.

However, in dimension $d = 3$ (for $\|x\|_\infty \geq 2$) the relaxation of long chains may (possibly) be further slowed down by the appearance of "quasi-knots", i.e. configurations whose decay is extremely slow due to the complicated sequence of moves needed to untie them. Since untying a quasi-knot requires passing the end of the chain through the knotted region (which may be near the middle of the chain), this process becomes more time-consuming the larger $\langle N \rangle$ is. This effect might possibly increase the dynamic critical exponent above $4\nu$, at least for observables that are sensitive to (quasi-)knots.[51]

### 6.7.2 BFACF + Cut-and-Paste Algorithm

Caracciolo, Pelissetto and Sokal [132] have recently proposed to speed up the BFACF algorithm by adjoining some non-local (cut-and-paste) moves. The hope is that these non-local moves will destabilize the long-lived (metastable) configurations that are responsible for the slow equilibration of the BFACF algorithm. Thus, the algorithm is a hybrid in which the non-local moves hopefully assure the rapid equilibration *within* subspaces of fixed $N$, while the local (BFACF) moves assure equilibration *between* different $N$ (and in particular make the algorithm ergodic). The algorithm has a free parameter $p_{nl}$ — the percentage of non-local moves — which can be tuned as a function of $\langle N \rangle$ to optimize the computational efficiency.

The best one can hope for is an autocorrelation time $\tau \sim \langle N \rangle^2$: for even if the non-local moves were to cause instant equilibration at fixed $N$, the local moves would still have to carry out a random walk in $N$. Such a behavior, if achieved, would be a significant improvement over the pure BFACF algorithm. This estimate refers, however, to *physical* time units; since the non-local moves require a mean CPU time per move that grows as a fractional power of $\langle N \rangle$, it is a subtle matter to choose $p_{nl}$ so as to minimize the autocorrelation time as measured in *computer* (CPU) time units.

Numerical experiments in $d = 2$ [132] confirm that the autocorrelation time (in units of elementary moves) at *fixed* $p_{nl} > 0$ scales as $\tau_{int,N} \sim \langle N \rangle^{\approx 2}$. Taking into account the CPU time, it is found that the optimal $p_{nl}$ scales as $\sim 1/\langle N \rangle^{\approx 0.8}$, and the autocorrelation time in CPU units then scales as $\tau_{int,N} T_{CPU} \sim \langle N \rangle^{\approx 2.3}$. In practice, at $\langle N \rangle \approx 100$ it is found that the physical (resp. CPU) autocorrelation time of the hybrid algorithm with $p_{nl} = 0.05$ is a factor 6 (resp. 4) smaller than that of the pure BFACF algorithm. The BFACF + cut-and-paste algorithm provides, therefore, a significant (though not exactly earth-shaking) improvement over previous algorithms for variable-$N$, fixed-endpoint SAWs. But more research is needed to establish conclusively the dynamic critical behavior of this algorithm.

[51] For example, any one of the standard knot invariants [192, 193] applied to $\omega \circ \omega^\circ$, where $\omega^\circ$ is some fixed path from $x$ to 0 in $\mathbb{R}^d \setminus \mathbb{Z}^d$.

# 7 Miscellaneous Issues

## 7.1 Data Structures

The new algorithms described in Section 6 are potentially very efficient; but in order to realize this potential, it is necessary to choose carefully the data structures and the low-level "bookkeeping" algorithms. For example, a local or bilocal update can be carried out, with *suitable* data structures, in a CPU time of order 1 (as one would expect); but a *naive* choice of data structures might force us into "garbage collection" (mass copying of data from one storage location to another) costing a time of order $N$. Likewise, checking the global self-avoidance of a walk $\omega$ (as arises e.g. in the pivot algorithm) can be carried out, with suitable data structures, in a CPU time of order $N$; but the naive method (comparing $\omega_i$ to $\omega_j$ for each pair $i, j$) takes a time of order $N^2$ — which would nullify the advantages of the pivot algorithm.

We divide the discussion of data structures into two parts: those needed in *updating the walk*, and those needed in *testing self-avoidance*. As will be seen, it is usually necessary to maintain *redundant* data structures, in order that both operations can be carried out efficiently. For the first task, one typically uses **linear (or circular) lists** of various kinds; for the second task, one uses **bit tables** or **hash tables**. A lucid discussion of these structures can be found in [94, Chapters 11 and 12].

### 7.1.1 Updating the Walk

Each of the algorithms described in Section 6 requires one or more of the following operations in order to update the walk:

1. Choose a random site (or a random step) of the walk.
2. Update the coordinates of a given site of the walk.
3. Find the successor (or predecessor) of a given site on the walk.
4. Insert one or more new sites
   (a) at the beginning of the walk
   (b) at the end of the walk
   (c) in the interior of the walk
5. Delete one or more sites from (a), (b) or (c).

Depending on the particular operations needed, we will choose to represent the walk as one or another type of list.

The simplest type of list is a **sequentially allocated linear list**: the walk coordinates $\omega_0, \ldots, \omega_N$ are stored in order in successive memory locations (e.g. a Fortran array). Clearly this structure permits the first three operations to be carried out in a time of order 1. In particular, this suffices for local $N$-conserving moves, pivot moves, and cut-and-paste moves.[52] Insertion and deletion at the beginning and end of the walk can also be carried out in a time of order 1, *provided that* an upper bound is known in advance on how far the walk can grow in each direction (so that sufficient storage can be allocated). This suffices for the slithering-tortoise algorithm, because one can take e.g. $N_{max} \approx 70\langle N \rangle$ and be virtually certain that this walk length will never be exceeded (at least not before the sun runs out of hydrogen and engulfs the earth).

For the slithering-snake (reptation) algorithm one should use a **sequentially allocated circular list**: the walk coordinates are stored in sequential memory locations, but in general in a cyclically permuted order, i.e. $\omega_j, \omega_{j+1}, \ldots, \omega_N, \omega_0, \omega_1, \ldots, \omega_{j-1}$. A pointer then indicates which element is $\omega_0$. (A flag may also indicate which direction is "forward": this is useful in implementing the first version of the algorithm [146, pp. 4592–4593], in which "head" and "tail" are sometimes interchanged.)

None of the foregoing structures are adequate for those algorithms which insert or delete sites in the *interior* of the walk (BFACF, most bilocal algorithms, etc.): insertion or deletion of even one site would necessitate shifting a large part of the list ("garbage collection"), which would take a time of order $N$. A more flexible data structure is the *doubly linked list*: here the walk coordinates $\omega_0, \ldots, \omega_N$ may be stored anywhere in memory (neither contiguously nor in order); but together with each coordinate $\omega_i$ we maintain link fields $\ell^-(i)$ and $\ell^+(i)$, which indicate the storage locations where the $\omega_{i-1}$ and $\omega_{i+1}$, respectively, are to be found. Unused but available memory is also stored in a linked list, the so-called *free list* (here the *two-way* linking is unnecessary). It is easy to insert or delete into the interior of a doubly linked list, by using the link fields $\ell^-(i)$ and $\ell^+(i)$ to locate the predecessor and successor sites (which take part in the relinking). On the other hand, since the $\{\omega_i\}$ may be scattered throughout memory, it is not easy to choose a random site or step of the walk. (One would have to "thread through" the list sequentially, taking a time of order $N$.)

To get the virtues of both sequential allocation and linked list, one can use a **contiguously allocated doubly linked linear list** (see e.g. [132, Section 3.3]). That is, the walk coordinates are stored in a contiguous array $\{s(i)\}_{i=0}^N$, but *not in any particular order*; that is, $s(i)$ is *some* site of the walk, but *not* necessarily $\omega_i$. To keep track of the sequence of steps along the walk, we use forward pointers $\{\ell^+(i)\}_{i=0}^N$ and backward pointers $\{\ell^-(i)\}_{i=0}^N$; here $\ell^+(i)$ [resp. $\ell^-(i)$] is the index corresponding to the site following [resp. preceding] the site whose index is $i$, or $-1$ if no such site exists. It is often convenient that the initial and final points of the walk be assigned to indices 0 and 1, respectively. Therefore,

[52] In the case of pivot or cut-and-paste moves, one maintains *two* linear lists — one "active" and one "scratch" — together with a flag saying which is currently which. The coordinates of a *proposed* new walk are successively calculated and placed in the "scratch" list, and self-avoidance is simultaneously tested (see Section 7.1.2); if the self-avoidance test is passed, the flag is flipped and the "scratch" list now becomes "active".

$$\begin{aligned} s(0) &= \omega_0 \, (= 0) \\ s(\ell^+(0)) &= \omega_1 \\ s(\ell^+(\ell^+(0))) &= \omega_2 \\ &\vdots \end{aligned}$$

and so on. Likewise,

$$\begin{aligned} s(1) &= \omega_N \, (= x) \\ s(\ell^-(1)) &= \omega_{N-1} \\ s(\ell^-(\ell^-(1))) &= \omega_{N-2} \\ &\vdots \end{aligned}$$

and so on. It is then trivial to choose a random site of the walk, as this is equivalent to choosing a random index $i$.[53] It is also trivial to insert new sites: just put them in locations $N+1$, $N+2$, etc. and then relink. Deletion requires "garbage collection" to maintain the contiguous allocation, but this also can be performed in a time of order 1: move the entries from locations $N$, $N-1$, etc. into the just-vacated locations, with appropriate updates to the link fields. So the contiguously allocated doubly linked linear list is the appropriate data structure for the BFACF and bilocal algorithms.

### 7.1.2 Testing Self-Avoidance

If the walk configuration were stored only as a linear list (whether sequentially allocated or linked), then this entire list would have to be searched each time we want to add *one* new site to the walk, in order to check the self-avoidance constraint. This would take a time of order $N$, which is a disaster.

The solution is to maintain *two* (redundant) data structures to store the current walk configuration: a linear or circular list as described in the preceding subsection, and a **bit table** or **hash table**.[54] The latter can be defined abstractly as data structures that perform the following functions: Given a finite (but typically enormous) set $K$ of "possible keywords", we wish to store a subset $H \subset K$ (of cardinality $\leq$ some maximum $M$) in such a way that, for any $x \in K$, the following operations can be carried out rapidly:

1. *Query.* Is $x \in H$?
2. *Insertion.* Insert $x$ into $H$ (if it is not in $H$ already).
3. *Deletion.* Delete $x$ from $H$ (if it is in $H$ currently).

[53] Note that this works only for choice with *uniform* probability (or uniform excluding one or both endpoints). It would *not* work if one wanted to choose sites with a nonuniform probability depending on their location along the walk.

[54] To my knowledge, the first use of a bit table for self-avoidance checking was by McCrackin [194], and the first such uses of a hash table were by Alexandrowicz [97] and Jurs and Reissner [195].

Here "rapidly" means "in a time of order 1, on the average". In our application, the set $K$ of possible keywords will be the set of all points in some box $B \subset \mathbb{Z}^d$ which is large enough to contain all possible points in the walk $\omega$: e.g. a cube of side $\geq 2N$ centered at the origin for a fixed-$N$ simulation, or a cube of side $\gtrsim 140\langle N \rangle$ for a variable-$N$ simulation.

A **bit table** is simple to describe: it is a large block of memory in which each possible keyword $x \in K$ (i.e. each site of the large box $B$) is assigned one distinct bit. That bit is set to 1 if $x \in H$ (i.e. if the site in question currently belongs to the walk), and 0 otherwise. Clearly, the three operations of query, insertion and deletion can each be carried out in a time of order 1. The trouble is that for large $N$, the memory requirements become prohibitive, especially in dimension $d > 2$: e.g. at $N = 1000$ the memory needed is 0.5 megabyte for $d = 2$, and 1 gigabyte for $d = 3$!

The memory requirements can be reduced by a clever trick called the **sliding bit table** [196, 197]. This approach is adequate in algorithms in which the new sites are all being added (or old sites being deleted) in the same small vicinity. In practice this means the slithering-tortoise algorithm.

A more general approach is provided by the **hash table** [94, Chapter 12] [198, Section 6.4]: An array of $M$ words is assigned, and each keyword $x \in K$ is assigned a *primary address* $h(x)$ in this array. Since in general $M \ll |K|$, the "hash function" $h$ is necessarily many-to-one, i.e. many distinct keywords may share the same primary address, leading to the possibility of *collisions*. The various hash-coding algorithms are distinguished by the method they use to resolve collisions, i.e. to decide where to store a keyword if its primary address happens to be occupied by some other keyword. One of the simplest collision-resolution schemes is **linear probing**: if the primary address $h(x)$ is occupied, the algorithm searches successively in addresses $h(x) + 1$, $h(x) + 2$, ... (modulo $M$) until it finds either the keyword $x$ or an empty slot. Other approaches involve **chaining** via linked lists.

In the worst possible case, a single query or insertion into a hash table containing $N$ entries could take a time of order $N$. However, it can be shown [198, Section 6.4] that as long as the hash table does not get close to full (i.e. $N$ does not get near $M$), then the *average* time for a single query or insertion (assuming random distribution of the points) is of order 1. So the hash-table method is nearly as fast as the bit-table method, and far more space-effective.

We remark that *deletion* from a linear-probing hash table requires some care: if done naively, entries can get "lost" [198, pp. 526–527]. However, these subtleties are irrelevant if deletions always occur in a last-in-first-out manner (as in the slithering-tortoise algorithm), or occur only when cleaning up the table at the end (as in the pivot and cut-and-paste algorithms). In the latter case it suffices to keep a linear list of the memory locations in which elements have been inserted; these locations can then be cleared at the end.

Depending on the application, it may be desirable to maintain the bit table/hash table either as a *permanent* data structure (containing the current walk $\omega$) or as a *scratch* data structure (for checking self-intersection of proposed new walks $\omega'$); or it may be desirable to maintain one of each, together with a flag saying which is which. See [132, Section 3.3] and [165, Section 3.2] for details.

## 7.2 Measuring Virial Coefficients

The virial coefficients $B_k$ play a central role in the theory of dilute polymer solutions. But to measure them it is *not* necessary (or even desirable) to simulate a many-chain system; rather, it suffices to simulate $k$ *independent* polymer chains and then measure a suitable overlap observable. Consider, for concreteness, the second virial coefficient $B_2^{(N_1,N_2)}$, defined by (2.10)–(2.12). We have

$$B_2^{(N_1,N_2)} \;=\; \frac{1}{2}\langle T(\omega^{(1)},\omega^{(2)})\rangle_{N_1,N_2} \;, \tag{7.1}$$

where $\omega^{(1)}$ and $\omega^{(2)}$ are *independent* SAWs of $N_1$ and $N_2$ steps, respectively, and $T(\omega^{(1)},\omega^{(2)})$ is the number of translates of $\omega^{(2)}$ which somewhere intersect $\omega^{(1)}$:

$$T(\omega^{(1)},\omega^{(2)}) \;=\; \#\{x\in\mathbb{Z}^d\colon\, \omega^{(1)}\cap(\omega^{(2)}+x)\neq\emptyset\}\;. \tag{7.2}$$

So we can run in parallel two independent simulations (using e.g. the pivot algorithm), and then every once in a while measure the observable $T(\omega^{(1)},\omega^{(2)})$.

The straightforward method for determining $T(\omega^{(1)},\omega^{(2)})$ is to compute $x=\omega_i^{(1)}-\omega_j^{(2)}$ for each of the $(N_1+1)(N_2+1)$ pairs $i,j$, write these points into a hash table (see Section 7.1.2), and count how many distinct values of $x$ are obtained. Unfortunately, this takes a CPU time of order $N_1N_2$, i.e. order $N^2$ if $N_1=N_2=N$. By contrast, we expect that one "effectively independent" sample of the pair $(\omega^{(1)},\omega^{(2)})$ can be produced by the pivot algorithm in a CPU time of order $N$. So this approach would spend more time analyzing the data than generating it!

Fortunately, there exist *Monte Carlo* algorithms which can produce an *unbiased estimate* of $T(\omega^{(1)},\omega^{(2)})$, with statistical errors comparable to or smaller than those already intrinsic in the observable $T(\omega^{(1)},\omega^{(2)})$, in a mean CPU time of order $N$. So the idea is to perform a "Monte Carlo within a Monte Carlo". At least two such algorithms are known: the "hit-or-miss" algorithm [39], and the Karp-Luby algorithm [199, 200]. See [96, Section 5.3] for a preliminary discussion, and [39] for a fuller account.

The "hit-or-miss" algorithm can easily be generalized to compute higher virial coefficients. I do not know whether the Karp-Luby algorithm can be generalized in this way.

## 7.3 Statistical Analysis

For the most part, the statistical analysis of SAW Monte Carlo data uses the same methods as are employed in other types of Monte Carlo simulation. In particular, with dynamic Monte Carlo data it is essential to carry out a thorough **autocorrelation analysis**: only in this way can one test the adequacy of the thermalization interval and the run length and properly assess the statistical error bars. For details, see e.g. [36, Section 3], [37, Section 2], [96, Appendix C] and [11, Sections 9.2.2 and 9.2.3].

Typically one carries out fixed-$N$ simulations at a (wide) range of values of $N$, and then uses **weighted least-squares estimation** [201] to extract the critical exponent $\nu$ and the various critical amplitudes. However, for high-precision work it is important to take account of **corrections to scaling**. According to renormalization-group theory [202, 203], the mean value of any global observable $\mathcal{O}$ behaves as $N \to \infty$ as

$$\begin{aligned} \langle \mathcal{O} \rangle_N \;=\; & AN^p \Bigg[ 1 + \frac{a_1}{N} + \frac{a_2}{N^2} + \ldots + \frac{b_0}{N^{\Delta_1}} + \frac{b_1}{N^{\Delta_1+1}} + \frac{b_2}{N^{\Delta_1+2}} + \ldots \\ & + \frac{c_0}{N^{\Delta_2}} + \frac{c_1}{N^{\Delta_2+1}} + \frac{c_2}{N^{\Delta_2+2}} + \ldots \Bigg] \; . \end{aligned} \tag{7.3}$$

Thus, in addition to "analytic" corrections to scaling of the form $a_k/N^k$, there are "non-analytic" corrections to scaling of the form $b_k/N^{\Delta_1+k}$, $c_k/N^{\Delta_2+k}$ and so forth, as well as more complicated terms not shown in (7.3). The leading exponent $p$ and the correction-to-scaling exponents $\Delta_1 < \Delta_2 < \ldots$ are universal; $p$ of course depends on the observable in question, but the $\Delta_i$ do not. [Please note that the exponents $\Delta_1 < \Delta_2 < \ldots$ have no relation whatsoever to the gap exponent $\Delta_4$ defined in (2.9). The notation used here is standard but unfortunate.] The various amplitudes (both leading and subleading) are all nonuniversal. However, *ratios* of the corresponding amplitudes $A$, $b_0$ and $c_0$ (but not $a_k$ or the higher $b_k, c_k$) for different observables are universal [52, 203].

Obviously it is hopeless to try to estimate from noisy Monte Carlo data more than the first one or two terms in (7.3), i.e.

$$\langle \mathcal{O} \rangle_N \;=\; AN^p \left[ 1 + \frac{a}{N^{\Delta}} \right] \tag{7.4}$$

where $\Delta \equiv \min(\Delta_1, 1)$. A reasonable approach is as follows: First truncate the series at zeroth order ($\langle \mathcal{O} \rangle_N = AN^p$) and perform a weighted least-squares fit using only the data at $N \geq N_{min}$, for a sequence of successively larger values $N_{min}$. For each such fit, the $\chi^2$ value indicates whether the hypothesis $\langle \mathcal{O} \rangle_N = AN^p$ for $N \geq N_{min}$ is consistent with the data — or in other words, whether the corrections to scaling for $N \geq N_{min}$ (which surely exist) are statistically significant.[55] If they are not, then one is done; the estimates of $p$ and $A$ ought to be roughly independent of $N_{min}$, within statistical error bars. If the corrections are significant, then one can insert the first correction-to-scaling term and redo the least-squares fit and $\chi^2$ test.[56] However, one must keep in mind that the estimate of $\Delta$ (and the correction amplitude) produced by such a fit is merely an *effective exponent* which may be imitating the combined effect of *several* correction-to-scaling terms over some particular range of $N$. Such an effective exponent is of no intrinsic physical interest; this approach should simply be thought of as a semi-empirical method for extrapolating the data to $N \to \infty$. In principle the true $\Delta$ can be found by taking $N_{min}$ very large — large enough so that the second correction-to-scaling term is negligible compared to the first — but this means that the first correction-to-scaling term will also be negligible compared to the leading term, and the *statistical* errors in $\Delta$ and the corresponding amplitude will therefore be enormous. At present it seems feasible to obtain only *rough* estimates of $\Delta$ by direct Monte Carlo simulation [204, 39]. However, the situation may improve in the future, as more powerful computers become available.

[55] Statistical significance (resp. insignificance) of the corrections for $N \geq N_{min}$ means only that these corrections are larger than (resp. comparable to or smaller than) the statistical errors *in this particular simulation*. By making sufficiently long runs, the statistical error bars can in principle be made arbitrarily small, and so the corrections to scaling will *always* eventually be found to be statistically significant.

[56] One may either make a guess for $\Delta$ and then estimate the amplitude via linear least-squares, or estimate simultaneously both $\Delta$ and the amplitude by nonlinear least-squares.

A novel point arises when estimating $\mu$ and $\gamma$ (or $\mu$ and $\alpha_{sing}$) from a variable-$N$ simulation. Thanks to (2.1)/(2.17) [or (2.2)/(2.14)], the probability distribution of chain lengths $N$ is known *exactly* except for the two unknown parameters $(\mu, \gamma)$ [or $(\mu, \alpha_{sing})$], provided that we temporarily neglect corrections to scaling. This fact allows us to use **maximum-likelihood estimation** [201] to obtain estimates of $\mu$ and $\gamma$ which are not only demonstrably optimal in a rigorous statistical sense — that is, they achieve the minimum possible mean-square error for a given quantity of Monte Carlo data — but which also provide *a priori* error estimates. This means that the statistical errors can be computed reliably, *in advance* of performing the Monte Carlo simulation. Or to put it more strikingly: before performing the simulation, one cannot know what the final central estimates will be, but one *can* know the error bars! See [182, Section 4.2] and [165, Section 4] for details.

# 8 Some Applications of the Algorithms

The development over the past decade of efficient Monte Carlo algorithms for the SAW (Section 6) has combined with recent advances in computer hardware (notably clusters of high-speed RISC workstations) to make possible high-precision studies of SAWs that would have been unimaginable only a few years ago. For example, a recent study [39] of SAWs in $d = 2$ and $d = 3$ has employed chain lengths up to $N = 80000$, obtaining error bars of order 0.1–0.3%. (To be sure, this work took several *years* of CPU time!)

In this section I cannot hope to do justice to all the applications of the new algorithms. Rather, I shall limit myself to giving an informal account of a few illustrative examples drawn from my own work (most of which is joint work with Sergio Caracciolo, Bin Li, Neal Madras and Andrea Pelissetto).

## 8.1 Linear Polymers in Dimension $d = 3$

Probably our most important work is a high-precision study of the critical exponents $\nu$ and $2\Delta_4 - \gamma$ (and in particular the hyperscaling law $d\nu = 2\Delta_4 - \gamma$) and universal amplitude ratios for SAWs in both two and three dimensions, using the pivot algorithm [39]. In dimension $d = 3$, the renormalization-group prediction for the exponent $\nu$ is $0.5880 \pm 0.0015$ [16, 17, 18, 19, 20] or $0.5872 \pm 0.0014$ [21]. However, this method is susceptible to serious (and quite possibly undetectable) systematic errors arising from a confluent singularity at the RG fixed point [25, 26]; this led me to be rather skeptical of the claimed error bars. Indeed, when we began this work five years ago, the series-extrapolation [205] and Monte Carlo [206, 96] estimates of $\nu$ were significantly higher, around $0.592 \pm 0.0015$ (one standard deviation). I was therefore looking forward to improving the statistics on the Monte Carlo studies by a factor of 10 or so (i.e. reducing the error bar by a factor of about 3), so as to definitively rule out the RG prediction (or at least its claimed error bar). But what actually happened is rather different from what I had envisaged!

The original Monte Carlo studies [206, 96] predicting $\nu \approx 0.592$ were based on walks of length $N \lesssim 3000$. But as we studied longer and longer walks, the apparent exponent fell.[57] Eventually things stabilized, but we had to use walks up to $N = 80000$ and (if we fit to a pure power law) to *throw away* all walks with $N \lesssim 5000$! And the value at which things stabilized was — surprise! — almost exactly the RG value: our preliminary estimate is $\nu = 0.5879 \pm 0.0005$ (68% confidence limits), taking account of both statistical errors and corrections to scaling. (This estimate may change when we do the definitive data analysis, but probably not by much.) The moral is that corrections to scaling are an extremely serious effect in high-precision Monte Carlo studies; it is necessary to be very careful, and sometimes to go to enormous chain lengths, to escape from their effects. The extraordinary accuracy of the RG prediction remains a mystery (at least to me).[58]

Another startling conclusion from this study concerns the interpenetration ratio $\Psi$ [cf. (2.13)], and goes to the heart of polymer theory. But this requires a brief historical digression.

For several decades, most work on the behavior of long-chain polymer molecules in dilute solution [208, 12, 5, 209, 15, 210] has been based on the so-called "two-parameter theory" in one or another of its variants: traditional (Flory-type)[59], pseudo-traditional (modified Flory-type)[60] or modern (continuous-chain-type)[61]. All two-parameter theories predict that in the limit of zero concentration, the mean-square end-to-end distance $\langle R_e^2 \rangle$, the mean-square radius of gyration $\langle R_g^2 \rangle$ and the interpenetration ratio $\Psi$ depend on the degree of polymerization $N$ (or equivalently on the molecular weight $M = NM_{monomer}$) according to

$$\langle R_e^2 \rangle \quad = \quad ANF_{R_e}(bN) \qquad (8.1a)$$

[57] For example, a recent study of Eizenberg and Klafter [178] using walks of length $N \lesssim 7200$ found $\nu \approx 0.5909 \pm 0.0003$.

[58] It may be related to the apparent fact [207, 26] that the confluent exponent $\Delta_2/\Delta_1$ is very close to an integer (namely, it is $\approx 2$).

[59] See Yamakawa [12], Sections 11 and 16 (pp. 69–73 and 110–118) and parts of Sections 15, 20b and 21b (pp. 94–110, 153–164 and 167–169). See also DesCloizeaux and Jannink [15], Section 8.1 (pp. 289–313).

[60] See Yamakawa [12], most of Section 15 (pp. 94–110) and parts of Sections 20b and 21b (pp. 153–164 and 167–169). See also Domb and Barrett [211].

[61] These theories take as their starting point the Edwards model of a weakly self-avoiding continuous chain [212, 213, 214, 215, 216, 209, 15]. (The Edwards model is also equivalent to the continuum $\varphi^4$ field theory with $n = 0$ components.) See DesCloizeaux and Jannink [15] for a detailed treatment of the Edwards model.

$$\langle R_g^2 \rangle = A N F_{R_g}(bN) \tag{8.1b}$$
$$\Psi = F_\Psi(bN) \tag{8.1c}$$

where $F_{R_e}, F_{R_g}, F_\Psi$ are claimed to be *universal* functions (which each specific two-parameter theory should predict), and $A$ and $b$ are *non-universal* scale factors depending on the polymer, solvent and temperature but independent of $N$. [The conventional notation is $\alpha_R^2 = F_{R_e}$, $\alpha_S^2 = 6F_{R_g}$, $h = \alpha_S^d F_\Psi / z$ and $z = (bN)^{2-d/2}$ in spatial dimension $d$.] Moreover, virtually all the theories — and in particular the modern continuous-chain-based theories — predict that $F_\Psi$ is a monotone increasing and concave function of its argument $bN$, which approaches a limiting value $\Psi^* \approx 0.2 - 0.3$ as $bN \to \infty$.

But our Monte Carlo data show precisely the opposite behavior: $\Psi$ is a *decreasing* and *convex* function of $N$, which approaches a limiting value $\Psi^* \approx 0.247$ as $N \to \infty$ (Figure 15). The same behavior was found by Nickel [203]. Indeed, there is *experimental* evidence that for real polymers in a sufficiently good solvent, the approach to $\Psi^*$ is also from above, contrary to the two-parameter theory [217, 218, 219, 210]. This behavior was considered to be a perplexing "anomalous effect", and various explanations were advanced [220, 221, 218]. What is going on here?

The correct explanation, in my opinion, was given two years ago by Nickel [203] (see also [222, 223]): theories of two-parameter type are simply wrong. Indeed, they are wrong not merely because they make incorrect predictions, but for a more fundamental reason: they purport to make universal predictions for quantities that are not in fact universal. Two-parameter theories predict, among other things, that $\Psi$ is a universal function of the expansion factor $\alpha_S^2 \equiv \langle R_g^2 \rangle / \langle R_g^2 \rangle_{T_\theta}$; in particular, $\Psi$ is claimed to depend on molecular weight and temperature only through the particular combination $\alpha_S^2(M, T)$. This prediction is quite simply incorrect, both for model systems and for real polymers. Indeed, even the *sign* of the deviation from the limiting value $\Psi^*$ is not universal.

All this has a very simple renormalization-group explanation [203], so it is surprising that it was not noticed earlier. As mentioned already in Section 7.3, standard RG arguments predict, for any real or model polymer chain, the asymptotic behavior

$$\langle R_e^2 \rangle = A_{R_e} N^{2\nu} (1 + b_{R_e} N^{-\Delta_1} + \ldots) \tag{8.2a}$$
$$\langle R_g^2 \rangle = A_{R_g} N^{2\nu} (1 + b_{R_g} N^{-\Delta_1} + \ldots) \tag{8.2b}$$
$$\Psi = \Psi^* (1 + b_\Psi N^{-\Delta_1} + \ldots) \tag{8.2c}$$

as $N \to \infty$ at fixed temperature $T > T_\theta$. The critical exponents $\nu$ and $\Delta_1$ are universal. The amplitudes $A_{R_e}, A_{R_g}, b_{R_e}, b_{R_g}, b_\Psi$ are nonuniversal; in fact, even the *signs* of the correction-to-scaling amplitudes $b_{R_e}, b_{R_g}, b_\Psi$ are nonuniversal. However, the RG theory also predicts that the dimensionless amplitude *ratios* $A_{R_g}/A_{R_e}$, $\Psi^*$, $b_{R_g}/b_{R_e}$ and $b_\Psi / b_{R_e}$ are universal [203, 52].

So there is no reason why the correction-to-scaling amplitudes should have any particular sign. In the continuum Edwards model, the effective exponents $\nu_{\mathit{eff},R_e} \equiv \frac{1}{2} d \log \langle R_e^2 \rangle / d \log N$ and $\nu_{\mathit{eff},R_g} \equiv \frac{1}{2} d \log \langle R_g^2 \rangle / d \log N$ and the interpenetration ratio $\Psi$ all approach their asymptotic values *from below* [209, 15, 22, 23, 24]: that is, $b_{R_e}, b_{R_g} > 0$ and $b_\Psi < 0$. On the other hand, in lattice self-avoiding walks, these quantities approach their asymptotic values *from above* [203, 39]; and the same occurs in the bead-rod model with sufficiently large bead diameter [224]. Indeed, this latter behavior is almost obvious qualitatively: short self-avoiding walks behave roughly like hard spheres; only at larger $N$ does one see the softer excluded volume (smaller $\Psi$) characteristic of a fractal object. All these models agree closely, as they should, for the leading *universal* quantities $\nu$, $A_{R_g}/A_{R_e}$ and $\Psi^*$; and they agree reasonably well for the universal correction-to-scaling quantities $\Delta_1$, $b_{R_g}/b_{R_e}$ and $b_\Psi/b_{R_e}$.

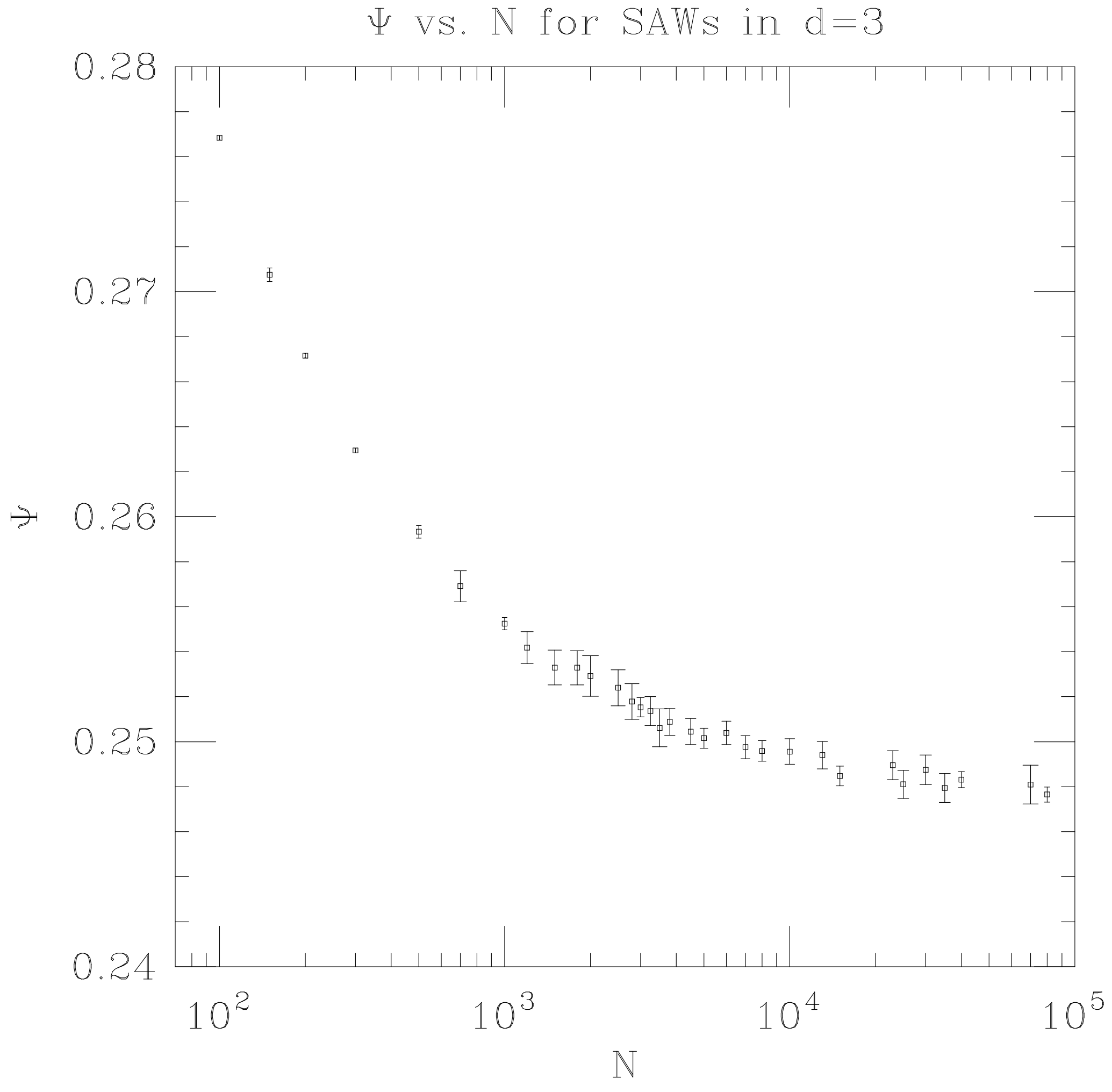

Figure 15: Interpenetration ratio $\Psi$ versus chain length $N$, for SAWs in $d = 3$. Error bar is one standard deviation. Data from [39].

In summary, the error of all two-parameter theories is to fail to distinguish correctly which quantities are universal and which are non-universal. In particular, the modern two-parameter theory begins from one special model — the continuum Edwards model — and assumes (incorrectly) that it can describe certain aspects of polymer behavior (e.g. the sign of approach to $\Psi^*$) which in reality are non-universal.

However, this is not the end of the story: The continuum Edwards model *does* in fact describe universal properties of polymer molecules, albeit not the behavior as $N \to \infty$ at fixed temperature $T > T_\theta$. Rather, this theory describes the *universal crossover scaling behavior in an infinitesimal region just above the theta temperature*, namely the limit $N \to \infty$, $T \to T_\theta$ with $x \equiv N^\phi (T - T_\theta)$ fixed, where $\phi$ is a suitable **crossover exponent**. More precisely, for suitably chosen exponents $\phi$ and $\nu_\theta$, the following limits are expected to exist:

$$f_{R_e}(x) \;\equiv\; \lim_{\substack{N\to\infty \\ T\to T_\theta \\ x\equiv N^\phi(T-T_\theta)\ \text{fixed}}} \frac{\langle R_e^2\rangle_{N,T}}{N^{2\nu_\theta}} \tag{8.3a}$$

$$f_{R_g}(x) \;\equiv\; \lim_{\substack{N\to\infty \\ T\to T_\theta \\ x\equiv N^\phi(T-T_\theta)\ \text{fixed}}} \frac{\langle R_g^2\rangle_{N,T}}{N^{2\nu_\theta}} \tag{8.3b}$$

$$f_{\Psi}(x) \;\equiv\; \lim_{\substack{N\to\infty \\ T\to T_\theta \\ x\equiv N^\phi(T-T_\theta)\ \text{fixed}}} \Psi(N,T) \tag{8.3c}$$

The exponents $\phi$ and $\nu_\theta$ are universal, and the **crossover scaling functions** $f_{R_e}$, $f_{R_g}$ and $f_\Psi$ are universal modulo a rescaling of abscissa and ordinate. The exponents are believed to take the values

$$\phi \;=\; \begin{cases} 2-\frac{d}{2} & \text{for } 3<d<4 \\ \frac{1}{2}\times\log^{3/22} & \text{for } d=3 \text{ [225, 226]} \\ \frac{1}{2}+\frac{3\epsilon}{22}+\ldots & \text{for } d=3-\epsilon \text{ [44, 45, 46, 225, 226]} \\ \frac{3}{7} & \text{for } d=2 \text{ [47]} \end{cases} \tag{8.4}$$

$$\nu_\theta \;=\; \begin{cases} \frac{1}{2} & \text{for } d\geq 3 \\ \frac{1}{2}+\frac{2\epsilon^2}{363}+\ldots & \text{for } d=3-\epsilon \text{ [44, 45, 46, 225, 226]} \\ \frac{4}{7} & \text{for } d=2 \text{ [47]} \end{cases} \tag{8.5}$$

The functions $f_{R_e}$ and $f_{R_g}$ (and $f_\Psi$ at least for $x \geq 0$) are monotonically increasing functions of their argument $x \equiv N^\phi(T - T_\theta)$, with the asymptotic behavior

$$f_{R_e}(x), f_{R_g}(x) \sim \begin{cases} x^{2(\nu-\nu_\theta)/\phi} & \text{as } x \to +\infty \\ (-x)^{2(\nu_{coll}-\nu_\theta)/\phi} & \text{as } x \to -\infty \end{cases} \tag{8.6a}$$

$$f_\Psi(x) \sim \begin{cases} \Psi^* \, (> 0) & \text{as } x \to +\infty \\ \text{unknown} \, (< 0) & \text{as } x \to -\infty \end{cases} \tag{8.6b}$$

where $\nu_{coll} = 1/d$. Then the claim [222] is that, for $3 \leq d < 4$, the functions $f_{R_e}(x)$, $f_{R_g}(x)$ and $f_\Psi(x)$ for $x \geq 0$ are given precisely by the continuum Edwards model, modulo the nonuniversal rescaling of abscissa and ordinate:

$$f_{R_e}(x) = K_1 \, \alpha_R^2(K_2 x) \tag{8.7a}$$

$$f_{R_g}(x) = (K_1/6) \, \alpha_S^2(K_2 x) \tag{8.7b}$$

$$f_\Psi(x) = \widetilde{h}(K_2 x)/\alpha_S^d(K_2 x) \tag{8.7c}$$

Here $\alpha_R^2(z)$, $\alpha_S^2(z)$ and $\widetilde{h}(z) \equiv z h(z)$ are the conventional expansion and second virial factors of the continuum Edwards model [12, 15, 210, 22, 23, 24], and $K_1$ and $K_2$ are nonuniversal scale factors. Thus, the continuum Edwards model *is* a correct theory for a certain limiting regime in the molecular-weight/temperature plane — but this regime is *not* the one previously thought. The explanation of (8.7) relies on a Wilson-deGennes-type renormalization group [13, 227]; see [222] for details, and [223] for further discussion.

It will be very interesting to test the predictions (8.7) numerically. But this will require better algorithms for simulating SAWs near the theta point (see Section 9.2).

## 8.2 Linear Polymers in Dimension $d = 2$

Dimension $d = 2$ is very special: for many statistical-mechanical systems, the critical exponents can be determined *exactly* (though non-rigorously) by Coulomb-gas [228, 229] and/or conformal-invariance [230, 231, 232] arguments. This is the case for the two-dimensional SAW, for which we know the exact exponents $\nu = 3/4$ and $\gamma = 43/32$ [228]. The unknown quantities in this model are the various universal amplitude ratios, which together determine the typical shape of a SAW and the strength of its interactions with other SAWs. The pivot algorithm has been employed [164] to obtain extremely accurate values for the limiting ratios

$$A_\infty = \lim_{N \to \infty} \frac{\langle R_g^2 \rangle_N}{\langle R_e^2 \rangle_N} = 0.14026 \pm 0.00011 \tag{8.8}$$

$$B_\infty = \lim_{N \to \infty} \frac{\langle R_m^2 \rangle_N}{\langle R_e^2 \rangle_N} = 0.43962 \pm 0.00033 \tag{8.9}$$

In particular, this confirms the beautiful conformal-invariance prediction [233, 164]

$$\frac{246}{91} A_\infty - 2B_\infty + \frac{1}{2} = 0 \, . \tag{8.10}$$

Another open question, which has attracted a lot of work, concerns the correction-to-scaling exponents in the two-dimensional SAW. We are currently using the pivot algorithm to investigate this question [204].

# 9 Conclusions

## 9.1 Practical Recommendations

What is the upshot of all this for the practicing polymer scientist, who wants to know which algorithm to use when? Here are a few recommendations:

1) When simulating linear polymers for the purpose of studying *global* observables (e.g. the critical exponent $\nu$, universal amplitude ratios, etc.), use the pivot algorithm. For initialization, use dimerization if this can be done in a CPU time less than half of your total planned run length; otherwise use the method proposed in [168], but be careful to discard at least $\sim N/f$ iterations at the beginning of the run.

2) When simulating linear polymers for the purpose of studying *local* observables (e.g. number of bends, nearest-neighbor contacts, etc.), use the pivot algorithm as described above, but make sure that the run length is $\gtrsim 1000N/f$ iterations. Alternatively, use the slithering-tortoise algorithm or the incomplete-enumeration algorithm.

3) To obtain the critical exponent $\gamma$, use the join-and-cut algorithm (together with the pivot algorithm for the $N$-conserving moves).

4) To obtain the critical exponent $\alpha_{sing}$, use the BFACF + cut-and-paste algorithm.

5) To obtain the connective constant $\mu$, use the slithering-tortoise algorithm or the incomplete-enumeration algorithm.

Of course, these recommendations are not engraved in stone; other algorithms could be useful in certain situations.

## 9.2 Open Problems

There are numerous open problems concerning the behavior of the algorithms discussed in this article. Among the most important ones are:

1) For $d \leq 4$, does there exist any static Monte Carlo algorithm for generating a random $N$-step SAW (with exactly uniform distribution) in a mean CPU time that is bounded by a polynomial in $N$? [Section 4.3]

2) What is the precise behavior of the enrichment algorithm as $N \to \infty$? [Section 5.2]

3) What is the precise behavior of the incomplete-enumeration algorithm as $N \to \infty$? [Section 5.3]

4) Are local $N$-conserving algorithms necessarily nonergodic also in dimension $d \geq 4$? [Section 6.4.1]

5) What is the dynamic critical exponent of the various local $N$-conserving algorithms (restricted to the ergodic class of a straight rod)? Is it exactly $2 + 2\nu$ for algorithms not having special conservation laws? What about for algorithms having special conservation laws, such as Verdier-Stockmayer? [Section 6.4.1]

6) What is the dynamic critical exponent for the various bilocal (or mixed local/bilocal) algorithms? Is the conjecture $\tau \sim N^2$ exact, approximate or wrong? [Section 6.4.2]

7) Can we improve our theoretical understanding of the acceptance fraction and dynamic critical behavior of the pivot algorithm? [Section 6.4.3]

8) Are there any bilocal (or mixed local/bilocal) algorithms that are ergodic for the fixed-$N$, fixed-$x$ ensemble? If so, what is their dynamic critical behavior? [Section 6.5.1]

9) What is the dynamic critical exponent of the fixed-$N$ cut-and-paste algorithm? [Section 6.5.2]

10) What is the precise dynamic critical exponent of the slithering-tortoise algorithm? Is it strictly between 2 and $1 + \gamma$? [Section 6.6.1]

11) What is the precise dynamic critical exponent of the BFACF algorithm? Is it exactly $4\nu$? [Section 6.7.1]

12) What is the precise dynamic critical behavior of the BFACF + cut-and-paste algorithm, as a function of $\langle N \rangle$ and $p_{nl}$? [Section 6.7.2]

13) Can the Karp-Luby algorithm be generalized to compute virial coefficients $B_k$ for $k \geq 3$? [Section 7.2]

In addition, there are many interesting open problems concerning the adaptation of these algorithms — or the invention of new algorithms — for polymeric systems more complicated than a single SAW (athermal linear polymer) in infinite space. For lack of space, I can merely list these problems and give a few (grossly incomplete) bibliographical references.

*SAWs in confined geometries.* In recent years there has been much interest in studying SAWs attached to surfaces or confined to specified regions (wedges, slabs, tubes, etc.) [234, 235]. Most of the algorithms described here do generalize to such situations; but it is no longer guaranteed that they are even ergodic, much less efficient. This requires a case-by-case study. See [40, Section 2.4.2] for some references.

*SAWs with nearest-neighbor attraction (→ theta point).* The transition of polymer conformation from the high-temperature (good-solvent) regime to the theta point to the collapsed regime is well modelled by the self-avoiding walk with nearest-neighbor attraction. One of the most fundamental problems in polymer physics is to understand quantitatively the details of this crossover, both in $d = 3$ [225, 226, 222] and in $d = 2$ [236]. The dynamic algorithms described here can easily be modified to handle a nearest-neighbor interaction, by inserting a Metropolis accept/reject step. But their efficiency may deteriorate markedly in the neighborhood of the theta point and even more drastically in the collapsed regime; this requires a detailed study for each algorithm. See [171, 172, 237, 238] for some recent preliminary work on the pivot algorithm; and see [40] for citations of older work.

*Branched polymers.* In recent years much attention has been devoted to the theoretical and experimental study of branched polymers, whose behavior is quite different from that of linear or ring polymers. The simplest case is that of branched polymers with *fixed* topology, such as star or comb polymers. Many of the algorithms for linear polymers can be adapted to this case, although both the ergodicity and the efficiency are nontrivial problems. See [40, Section 2.4.3] for some references, and see [169, 170, 171, 172, 174, 176] for recent work using the pivot algorithm.

A more difficult problem is that of branched polymers with *variable* topology, such as arbitrarily-branched lattice trees. Early works used simple sampling [239], while more recent works have used variants of the slithering-tortoise or incomplete-enumeration algorithms [240]. However, in all these cases it has been difficult to generate branched polymers of more than $\approx 50$ segments, because of the critical slowing-down. On the other hand, the need for large polymers is even more acute here than for linear or ring polymers, since one needs a rather large number of segments in order to feel the full effects of the branching. (That is, one expects large corrections to scaling in which the effective exponents for small $N$ are biased toward the unbranched-polymer values.) A very promising non-local algorithm was devised recently by Janse van Rensburg and Madras [241].

*Multi-chain systems.* Much current work in polymer science, both theoretical and experimental, focusses on semidilute and concentrated solutions and on melts. The simulation of multi-chain systems poses very difficult problems: for example, it is an open question whether *any* known algorithm is ergodic at *any* nonzero density! In the semidilute case, many of the algorithms described here can be applied with minor modification, and their performance in practice (if one disregards the ergodicity problem) will probably be similar to that for single chains. On the other hand, dense solutions and melts constitute a much more difficult problem, due to the possibility of “gridlock”. Several algorithms have been proposed [242], involving both local-deformation and subunit-exchange moves, but the relaxation seems in general to be very slow. Progress in this area will probably require new physical and algorithmic ideas. See [40, Section 3] for some references.

## Acknowledgments

Many of the ideas reported here have grown out of joint work with my colleagues Alberto Berretti, Sergio Caracciolo, Tony Guttmann, Greg Lawler, Bin Li, Neal Madras, Andrea Pelissetto and Larry Thomas. I thank them for many pleasant and fruitful collaborations. I have also learned much from discussions with Carlos Aragão de Carvalho, Jim Barrett, Marvin Bishop, Bertrand Duplantier, Roberto Fernández, Michael Fisher, Jürg Fröhlich, Paul Gans, Takashi Hara, Mal Kalos, Bernie Nickel, Gordon Slade, Stu Whittington and too many others to thank here. Finally, I wish to thank Kurt Binder and Neal Madras for many helpful comments on early drafts of this manuscript.

The author's research is supported in part by U.S. National Science Foundation grant DMS–9200719, U.S. Department of Energy contract DE-FG02-90ER40581 and NATO Collaborative Research Grant CRG 910251, as well as by a New York University Research Challenge Fund grant. Acknowledgment is also made to the donors of the Petroleum Research Fund, administered by the American Chemical Society, for partial support of this research.